\documentclass[12pt,a4paper]{article}
\usepackage[T2A]{fontenc}
\usepackage[cp1251]{inputenc}
\usepackage[russian]{babel}
\usepackage{cite}
\usepackage{mathtext}
\usepackage{amssymb,amsthm,amsmath}
\usepackage{array}
\usepackage[dvips]{epsfig}

\begin{document}
\author{\bf Yu.A. Markov$\!\,$\thanks{e-mail:markov@icc.ru}
$\,$, M.A. Markova$\!\,$\thanks{e-mail:markova@icc.ru} $\,$, A.A. Shishmarev,$\!\,$\thanks{e-mail:a.a.shishmarev@mail.ru}\\
\bf and A.N. Vall$\!\,$\thanks{e-mail:vall@irk.ru}}
\title{Equations of motion for a classical color particle\\
in background non-Abelian fermionic and bosonic fields: Inclusion of pseudoclassical spin}
\date{\it Institute for System Dynamics and Control Theory Siberian Branch\\
of Academy of Sciences of Russia, P.O. Box 1233, 664033 Irkutsk, Russia\\
\vspace{0.4cm}
$^{\S}$Irkutsk State University, Department of Theoretical Physics,\\
664003, Gagarin blrd, 20, Irkutsk, Russia}

\thispagestyle{empty}
\maketitle{}


\def\theequation{\arabic{section}.\arabic{equation}}

\[
{\bf Abstract}
\]

A generalization of the Lagrangian introduced earlier in [$\hspace{0.015cm}$2011 {\it J. Phys. G} ${\bf 37}$ 105001]
for a classical color spinning particle interacting with background non-Abelian gauge and fermion
fields for purpose of considering a change in time of the spin particle degree of freedom, is suggested.
In the case under consideration the spin degree of freedom is described by a commuting $c$-number Dirac
spinor $\psi_{\alpha}$. A mapping of this spinor into new variables: anticommu\-ting pseudovector $\xi_{\mu}$
and pseudoscalar $\xi_5$ commonly used in a description of the spin degree of freedom of a massive
spin-$\frac{1}{2}$ particle, is constructed. An analysis of one-to-one correspondence of this mapping is given.
It is shown that for the one-to-one correspondence it is necessary to extend a class of real tensor quantities
including besides $\xi_{\mu}$ and $\xi_5$, also odd vector $\hat{\xi}_{\mu}$, scalar $\hat{\xi}_5$, and (dual)
pseudotensor $^{\ast}\zeta_{\mu \nu}$. In addition, it is shown that it is necessary either to restrict
a class of the initial spinor $\psi_{\alpha}$ to Majorana one or to double the number of variables in the
tensor aggregate $(\xi_{\mu},\,\xi_5,\!\, ^{\ast}\zeta_{\mu \nu},\,\hat{\xi}_{\mu},\,\hat{\xi}_5)$. Various
special cases of the desired mapping are considered. In particular, a connection with the Lagrangian suggested
by A.M. Polyakov, is studied. It is also offered the way of obtaining the supersymmetric Lagrangian in terms
of the even $\psi_{\alpha}$ and odd $\theta_{\alpha}$ spinors. The map of the Lagrandian leads to local SUSY
Lagrangian in terms of $\xi_{\mu}$ and $\xi_5$.

{}


\newpage

\section{Introduction}
\setcounter{equation}{0}

In our previous paper \cite{markov_J_Phys_G_2010} a model Lagrangian describing interaction of a relativistic spinning color-charged classical particle with background
non-Abelian gauge and fermion fields was suggested. The spin degree of freedom has been presented in \cite{markov_J_Phys_G_2010} by a $c$-number Dirac spinor
$\psi_{\alpha}$, $\alpha = 1,\ldots,4$.
By virtue of that the background fermion field $\Psi_{\alpha}^i(x)$, which within a classical description is considered as Grassmann-odd one, has by the definition spinor index,
the description of the spin degree of freedom of the particle in terms of the spinor $\psi_{\alpha}$ is a very natural and simplest in technical respect. There is some vagueness
with respect to Grassmann evenness of this spinor. In our papers \cite{markov_NPA_2007, markov_IJMPA_2010} in the application to analysis of dynamics of the spinning color particle
moving in a hot quark-gluon plasma, the spinor $\psi_{\alpha}$ was thought as Grassmann evenness one (although it is improbably that for a complete description of the spin dynamics
in external fields of different statistics it demands simultaneously to take into consideration spinors of the different Grassmann evenness, i.e. in other words, it requires
introducing a superspinor, see Conclusion). Furthermore, for simplicity throughout our previous works, we neglected a change of a spin state of the particle, i.e. we believed
$\psi_{\alpha}$ to be a spinor independent of the evolution parameter $\tau$. As a consequence we completely neglected an influence of the particle spin on the general dynamics
of interaction of the particle with background fields. However, for a more detailed study of particle motion in external fields of different statistics and comparing the suggested
model with the other approaches known in the literature, it is necessary to consider the change in time of the spin variable. At present there exist a few approaches
to the description of the spin degree of freedom of a particle within (semi)classical approximation. Below we briefly consider only one approach most close to the subject
of investigation in this work.\\
\indent
Notice that a description of the spin degree of freedom by means of classical commuting spinor is not new. Such a way of the description arises naturally in determination of
a connection of relativistic quantum mechanics of electron with relativistic classical mechanics \cite{pauli_1932}. In particu\-lar, it was shown \cite{akhiezer_1969}
that within the WKB-method extended to the relativistic case, the relativistic wave Dirac equation results in a system of equations incorporating not only classical canonical
equation of motion, but also a further equation for the spin degree of freedom. This equation is connected directly with the Schr\"odinger equation
\begin{equation}
i\hbar\,\frac{d\psi(\tau)}{d\tau} =
-\frac{e\hbar}{4m}\,\sigma^{\mu \nu} F_{\mu \nu}(x)\psi(\tau),
\quad \sigma^{\mu\nu}\equiv\frac{1}{2i}\,[\gamma^{\mu},\gamma^{\nu}]
\label{eq:1q}
\end{equation}
for a spinor function $\psi_{\alpha}$ (we put throughout $c=1$ for the speed of light). This equation describes a motion of electron spin in external electromagnetic
field $F_{\mu \nu}(x)$. The field in (\ref{eq:1q}) is defined along the path of particle $x_{\mu} = x_{\mu} (\tau,x_0,\tau_0)$.\\
\indent
Further, a similar way of the description of the spin degree of freedom of elementary particle has been used extensively by A.O. Barut et al. \cite{barut_1984, barut_1989}
(see also \cite{kar_2011}). In these works it has been proposed the following Lagrangian which in our notions is
\begin{equation}
L = \frac{1}{2}\,i\lambda\biggl(\frac{d\bar{\psi}}{d\tau}\,\psi - \bar{\psi}\,\frac{d\psi}{d\tau}\biggr)+\,
p_{\mu}\bigl(\dot{x}^{\mu} - \bar{\psi} \gamma^{\mu} \psi\bigr)+\,eA_{\mu}(x)\bigl(\bar{\psi}\gamma^{\mu}\psi\bigr).
\label{eq:1w}
\end{equation}
Here, $\lambda$ is a constant with the dimension of action and these authors consider the $p_{\mu}$ momentum as Lagrangian multiplier for the constraint
\begin{equation}
\dot{x}_{\mu} = \bar{\psi} \gamma_{\mu}\psi,
\label{eq:1e}
\end{equation}
where the dot denotes the differentiation with respect to $\tau$. Within the classical model \cite{barut_1984, barut_1989} the whole phase consists of the usual pair
of conjugate variables $(x_{\mu}, p_{\mu})$ and another pair of conjugate classical spinor variables $(\psi, -i \bar{\psi})$ represen\-ting internal degrees of freedom.
The configuration space is thus ${\rm M}_4\otimes \mathbb{C}_4,\,\psi\in \mathbb{C}_4$ and the Lagrangian (\ref{eq:1w}) describes a symplectic system. In our approach
we depart from the constraint (\ref{eq:1e}). Next we define an interaction term with external (non-Abelian) gauge field such that it will be consistent with the equation
of motion (\ref{eq:1q}). Under these circumstances we suggest the following model Lagrangian which account for a change in both color and spinning degrees of freedom of
a classical particle propaga\-ting in background non-Abelian bosonic and fermionic fields
\begin{equation}
L = L_0  + L_m + L_{\theta} + L_{\vartheta} + L_{\Psi},
\label{eq:1r}
\end{equation}
where
\begin{equation}
\hspace{0.6cm}
L_0=-\frac{1}{2e}\,\dot{x}_{\mu\hspace{0.02cm}}\dot{x}^{\mu} + \frac{1}{2\hspace{0.03cm}i}\!\left(\frac{d\bar{\psi}}{d\tau}\,\psi -
\bar{\psi}\,\frac{d\psi}{d\tau}\right),
\label{eq:1t}
\end{equation}
\begin{equation}
L_m=-\frac{e}{2}\,m^2,
\hspace{3.85cm}
\label{eq:1y}
\end{equation}
\begin{equation}
\hspace{0.5cm}
L_{\theta}=i(\theta^{\dagger{i}}D^{ij}\theta^{j}) - e\,\frac{g}{4}\,Q^aF^{a}_{\mu\nu}(\bar{\psi}\sigma^{\mu\nu}\psi),
\label{eq:1u}
\end{equation}
\begin{equation}
\hspace{0.45cm}
L_{\vartheta}=\frac{i}{2}\,\vartheta^aD^{ab}\vartheta^b - e\,\frac{g}{4}\,{\cal Q}^aF^{a}_{\mu\nu}(\bar{\psi}\sigma^{\mu\nu}\psi),
\label{eq:1i}
\end{equation}
\begin{equation}
L_{\Psi}=-\frac{e}{\sqrt{2}}\,g\left\{\theta^{\dagger{i}}(\bar{\psi}_{\alpha}\Psi^{i}_{\alpha})
+(\bar{\Psi}^{i}_{\alpha}\psi_{\alpha})\theta^{i}\right\}
\hspace{7.1cm}
\label{eq:1o}
\end{equation}
\[
+\biggl[\,\frac{e}{\sqrt{2}}\,g\biggl(\frac{C_F}{2T_F}\biggr)
Q^a\Bigl\{\theta^{\dagger{j}}(t^a)^{ji}(\bar{\psi}_{\alpha}\Psi^{i}_{\alpha})
+ (\bar{\Psi}^{i}_{\alpha}\psi_{\alpha})(t^a)^{ij}\theta^{j}\Bigr\}+(Q^a\rightarrow{\cal Q}^a)\biggr]
\]
\[
+\biggl[\,\frac{e}{\sqrt{2}}\,g\biggl(\frac{C_F}{2T_F}\biggr)Q^aF^{a}_{\mu\nu}
\Bigl\{\theta^{\dagger{i}}(\bar{\psi}\sigma^{\mu\nu}\Psi^{i})
+ (\bar{\Psi}^{i}\sigma^{\mu\nu}\psi)\theta^{i}\Bigr\}+(Q^a\rightarrow{\cal Q}^a)\biggr]
\hspace{0.6cm}
\]
\[
+\, \frac{e}{\sqrt{2}}\,g\biggl(\frac{C_F}{2T_F}\biggr)F^{a}_{\mu\nu}\Bigl\{\theta^{\dagger{j}}(t^a)^{ji}(\bar{\psi}\sigma^{\mu\nu}\Psi^{j})
+ (\bar{\Psi}^{j}\sigma^{\mu\nu}\psi)(t^a)^{ij}\theta^{j}\Bigr\} + \,\dots\,.
\hspace{0.25cm}
\]
Here, $e$ is the (one-dimensional) vierbein field (it should not be confused with electric charge $e$ in (\ref{eq:1q})); $D^{ij}=\delta^{ij}\partial/\partial\tau+ig\hspace{0.02cm}\dot{x}^{\mu\!}A^{a}_{\mu}(t^{a})^{ij}$ is the covariant
derivative along the direction of motion; self-conjugate pair $(\theta^{\dagger{i}},\hspace{0.01cm}\theta^i)$ and real $\vartheta^a$ are a set of Grassmann variables
belonging to the fundamental and adjoint representations of the $SU(N_c)$ color group, respectively\footnote{Here, one can draw some interesting analogy to
(super)string theory for the interacting terms, for example in (\ref{eq:1u}). In our case the term $\dot{x}^{\mu} A_{\mu}^a (\theta^+ t^a \theta)$ is similar to one
$j^a \bar{\partial}x^{\mu}A_{\mu}^a(x)$ defining interaction with so-called the Neveu-Schwarz $(N\!S\hspace{0.01cm}N\!S)$ gauge fields \cite{callan_1985}.
Another term of a form $F_{\mu\nu}(\bar{\psi}\sigma^{\mu\nu}\psi)$ represents analog of a term $\bar{S}\Gamma^{[\mu_1 \ldots} \Gamma^{\mu_n ]} S F_{\mu_1 \ldots \mu_n}$ for $n=2$,
where $\bar{S}_{\alpha}$ and $S_{\alpha}$ are the spin fields. This term in a string theory defines interaction with Ramond-Ramond $(RR)$ gauge fields \cite{polchinski_1995}.
$N\!S\hspace{0.01cm}N\!S$ and $RR$ gauge fields are quite different in string theory in contrast to the theory of point particles.}, i.e.
$i,\,j,\ldots = 1,\ldots, N_c$, and $a,\,b,\ldots =1,\ldots, N^2_c-1$; the commuting color charges $Q^a$ and ${\cal Q}^a$ are defined by
\[
\quad
Q^a\equiv\theta^{\dagger{i}}(t^{a})^{ij}\theta^{j},\quad
{\cal Q}^a\equiv\frac{1}{2}\,\vartheta^{b}(T^{a})^{bc}\vartheta^{c}.
\]
The last contribution $L_{\Psi}$, Eq.\,(\ref{eq:1o}), describes interaction with external non-Abelian fermionic field. As was shown in \cite{markov_J_Phys_G_2010} in principle this
contribution includes an infinite number of interaction terms. In the work mentioned just above we have given the concrete examples of such terms.\\
\indent
An alternative approach most generally employed for a spin description of a massive particle is connected with introduction into consideration of the pseudovector and pseudoscalar dynamical
variables $\xi_{\mu}, \mu=1,\ldots, 4$ and $\xi_5$ that are elements of the Grassmann algebra \cite{berezin_1975, barducci_1976, brink_1976, balachandran_1977}. For this variables an appropriate
Lagrangian of the first order time derivative, is defined. In view of its great importance for the subsequent discussion and for convenience of further references we give in Appendix A
a complete form of this Lagrangian. It is these Grassmann-valued variables that appear in
representation of the one-loop effective action in quantum chromodynamics in terms of the path integral over world lines of hard particle in external field
\cite{borisov_1982, strassler_1992, d_hoker_1996}. We notice also that these variables in two-dimensional case are used in a description of a spinning string
within the Ramond-Neveu-Schwarz formalism and essentially in construction of the covariant string vertex operator (it is the most important notion of string theory) describing
emission (absorption) of a fermion by string.\\
\indent
The description of the spin degree of freedom in terms of the odd pseudovector and pseudosca\-lar quantities is to some extent more fundamental in comparison
with the description in terms of even spinor $\psi_{\alpha}$. This raises the interesting question of defining relations (mapping) between these variables,
and, finally, a possibility of constructing the mapping between the Lagrangian (\ref{eq:1r}) (without interaction term $L_{\Psi}$) and Lagrangian (A.1).
The construction of such a mapping in an explicit form is very important for us. The reason is that counterparts of the interaction terms (\ref{eq:1o})
in the Lagrangian (A.1) is unknown. Thus having understood the connection between the Lagrangians without an external fermion field, one can define an explicit form
of interaction terms with the background $\Psi$-fields in terms of the Grassmann pseudovector and pseudoscalar variables $\xi_{\mu}$ and $\xi_5$ merely by means of
an appropriate replacement of the $\psi_{\alpha}$ spinor by the mapping  $\psi_{\alpha}=\psi_{\alpha}(\xi_{\mu},\,\xi_5)$ in (\ref{eq:1o}).\\
\indent
It is pertinent at this point to make one remark, which is completely analogous to that made in Introduction in the paper \cite{markov_J_Phys_G_2010}. This remark
concerns with introducing the Grassmann color charges $\theta^{+i}$ and $\theta^{i}$ into consideration. If we carefully look at the equations of motion, the constraints
(A.7)\,--\,(A.11) and the expression for color current (A.12), then we may notice that the odd pseudovector $\xi_{\mu}$ enters these equations only in the following
even combination
\begin{equation}
S^{\mu\nu}\!\equiv -i\,\xi^{\mu}\xi^{\nu},
\label{eq:1p}
\end{equation}
as well as the Grassmann color charges enter these equations in the combination $\theta^{\dagger}t^a\theta\,(\equiv Q^a)$. By virtue of (A.8) the function $S^{\mu \nu}$
obeys the equation
\begin{equation}
\frac{dS^{\mu\nu}}{d\tau}=\frac{g}{m}\,Q^a(F^{a\mu}_{\;\;\;\;\lambda}S^{\lambda\nu} - F^{a\nu}_{\;\;\;\;\lambda}S^{\lambda\mu}).
\label{eq:1a}
\end{equation}
We notice that a similar tensor of spin can be defined also in terms of the $\psi_{\alpha}$ spinor if one sets
\begin{equation}
S^{\mu\nu} = \frac{1}{2}\,\bar{\psi}\sigma^{\mu\nu}\psi.
\label{eq:1s}
\end{equation}
By virtue of (\ref{eq:1u}) this tensor of spin obeys the same equation (\ref{eq:1a}). The anti-symmetric tensor $S^{\mu\nu}$ can be considered as
semiclassical limit of quantum-mechanical average of $\sigma_{\mu\nu}$, i.e.
\[
S^{\mu\nu}=\frac{1}{2}\,\langle\sigma^{\mu\nu}\rangle_{\hbar\rightarrow 0},
\]
where $\langle\cdot\rangle$ assigns an average operation.\\
\indent
Thus in actual dynamics of a classical color spinning particle introduction into consideration of the Grassmann pseudovector $\xi_{\mu}$ is reveal by no means
(and one can quite get by with the usual commuting function $S^{\mu\nu}$). The odd variables give merely the possibility of an Lagrangian formulation with subsequent
quantization of the model. One can expect that the situation can qualitatively changes only if the system is subjected to a background non-Abelian fermion field, which as it
were splits the combinations $-i\xi_{\mu\,}\xi_{\nu}$ into two independent (Grassmann-odd) parts. Here, the necessity of introducing Grass\-mann
pseudovector $\xi_{\mu}$ as dynamical variable enjoying full rights should be manifested in full.\\
\indent
Furthermore, by virtue of the fact that we have the even spinor $\psi_{\alpha}$ on the one hand and odd pseudovector $\xi_{\mu}$ (and pseudoscalar $\xi_5$)
on the other hand, for construction of the desired mapping inevitably we must introduce some auxiliary odd spinor $\theta_{\alpha}$. The idea of construction
of such a mapping is not new. In due time this problem has been studied extensively in view of analysis of a classical correspondence of theories of
relativistic massless spin-$\frac{1}{2}$ particles \cite{berezin_1975, barducci_1976, brink_1976} and superparticles
\cite{casalbuoni_1976, volkov_1980, brink_1_1981, brink_2_1981} and in a more general context between spinning string and superstring.
In paper by Sorokin et al. \cite{sorokin_1989} within the superfield formalism it was noted that such a classical correspondence can be defined by the following
relation
\begin{equation}
\xi_{\mu} \sim \bar{\theta}\gamma_{\mu}\psi + (\mbox{conj.\,part}).
\label{eq:1d}
\end{equation}
In \cite{sorokin_1989} commuting spinor $\psi_{\alpha}$ has played the role of a twistor-like variable, which is not dynamical one. In our paper (see the next section) we use
the relation (\ref{eq:1d}). The only difference is that by virtue of initial setting of the problem the anticommuting spinor $\theta_{\alpha}$ will play a role
of auxiliary variable rather than $\psi_{\alpha}$.\\
\indent
The paper is organized as follows. In Section\,2 as the first example a detailed analysis of the simplest relation between the even Dirac spinor $\psi_{\alpha}$
and odd pseudovector and pseudoscalar variables $\xi_{\mu}$, $\xi_5$ is carried out and the mapping between individual terms of Lagrangians (\ref{eq:1r}) and
(A.1) is considered.
Section\,3 is devoted to a discussion of the conditions for the one-to-one correspondence $(\psi,\bar{\psi}) \leftrightarrow (\xi_{\mu}, \xi_5)$ suggested in
the previous section. It is shown that the requirement of the one-to-one correspondence results in the necessity of generalization of the desired mapping by considering not
only the pseudovari\-ab\-les $\xi_{\mu}$ and $\xi_5$, but also Grassmann vector $\hat{\xi_{\mu}}$, scalar $\hat{\xi_5}$, and pseudotensor $^{*}\zeta_{\mu\nu}$ variables.
In Section\,4 a general analysis of the mapping between pair $(\psi_{\alpha}, \theta_{\alpha})$ and the real tensor aggregate ($\xi_{\mu}$,
$\xi_{5}$, $\zeta_{\mu\nu}$, $\hat{\xi}_{\mu}$, $\hat{\xi}_5$) is proposed. On the basis of this analysis it is concluded that for the existence of the one-to-one correspondence
it is necessary either restrict the $(\psi_{\alpha}$  and $\theta_{\alpha})$ spinors to Majorana ones or double a number of real tensor quantities:
$(\xi_{\mu}^{(i)}\!\!,\,\xi_{5}^{(i)}\!\!\!,\ldots)$, $i=1, 2$.
Section\,5 is concerned with an analysis similar to one in Section\,2, where instead of the pseudoscalar contribution $\xi_5$ the pseudotensor contribution $^{*}\zeta_{\mu\nu}$ is considered.
In the subsequent Section\,6 we compare thus obtained Lagrangian with Lagrangian suggested by A.M. Polyakov. In Section\,7 a qualitative consideration of supersym\-metric
generaliza\-tion of our initial Lagrangian (\ref{eq:1r}) is performed.
In Section\,8 it is considered a problem of construction in an explicit
form of the auxiliary odd spinor $\theta_{\alpha}$ in terms of known functions of the problem under consideration, namely, in terms of external non-Abelian fermionic
field $\Psi_{\alpha}^i(x)$ defined along world line of a particle $x_{\mu} = x_{\mu}(\tau)$ and color charges $\theta^i$, $\vartheta^a$ in the fundamental and adjoin
representations of the gauge group. In the concluding section we briefly discuss the question of a further generalization of the ideas of this work connected with
a construction of hybrid description of both spinning and color degrees of freedom of a relativistic particle within the models possessing double supersymmetry
(the doubly-graded models).\\
\indent
In Appendix\,A a complete form of the Lagrangian for spin-$\frac{1}{2}$ color particle, is given and local SUSY $n=1$ transformations, constraints and equations of the
motion are written out. In Appendix\,B all required formulas of spinor algebra are listed.

\section{\bf Mapping of the even spinor variables $(\psi_{\alpha},\,\bar{\psi}_{\alpha})$ into odd pseudovector and pseudoscalar ones $(\xi_{\mu},\,\xi_{5})$}
\setcounter{equation}{0}

Let us consider our initial Lagrangian (\ref{eq:1r}) without two last contributions  $L_{\vartheta}$ and $L_{\Psi}$. We will try to construct such an explicit mapping of spin
variables $\psi_{\alpha}$ and $\bar{\psi}_{\alpha}$ to spin variables $\xi_{\mu}$ and $\xi_{5}$ that reproduce Lagrangian (A.1).  A sum of terms (\ref{eq:1t}), (\ref{eq:1y})
and (\ref{eq:1u}) is devoid of any supersymmetry whereas a sum of terms (A.2), (A.3) and (A.4) possesses the local $n=1$ SUSY. The terms providing the local  supersymmetry
of expression (A.1) are connected with the last terms in (A.2) and (A.3) containing the one-dimensional gravitino field $\chi(\tau)$. By virtue of the fact that there is
no any analogy to these terms in (\ref{eq:1r}), we drop them for a moment. A question concerning the possibility of their arising in the mapping between Lagrangians (\ref{eq:1r})
and (A.1) will be discussed in section\,7.\\
\indent
Let us consider a linear mapping of the following form
\begin{equation}
\psi=\kappa\,\xi_{\mu}(\gamma^{\mu}\gamma_{5\hspace{0.02cm}}\theta) +\, \alpha\hspace{0.02cm}\xi_5(\gamma_{5\hspace{0.02cm}}\theta),
\hspace{0.2cm}
\label{eq:2q}
\end{equation}
and, correspondingly, for the conjugate function we have
\begin{equation}
\hspace{0.5cm}
\bar\psi=-\kappa^{*}(\bar\theta\gamma_5\gamma^{\mu})\xi_{\mu} - \alpha^{*}(\bar\theta\gamma_5)\xi_5.
\label{eq:2w}
\end{equation}
Here, $\kappa,\,\alpha$ are unknown (complex) functions, $\theta=(\theta_{\alpha}),\,\alpha=1,\ldots,4$, is some auxiliary Grassmann-odd Dirac spinor and the symbol $^{\ast}$ is a
complex conjugation sign.\\
\indent
Let us assume further that an inverse mapping will have the following structure
\begin{equation}
\xi_{\mu} = \frac{1}{2}\,\Bigl\{\beta(\bar\theta\gamma_{\mu}\gamma_{5\hspace{0.02cm}}\psi) - \beta^{*}(\bar\psi\gamma_{5}\gamma_{\mu}\theta)\Bigr\},
\label{eq:2e}
\end{equation}
\begin{equation}
\xi_5 = \frac{1}{2}\left\{\tilde\beta(\bar\theta\gamma_{5\hspace{0.02cm}}\psi) - \tilde\beta^{*}(\bar\psi\gamma_{5\hspace{0.02cm}}\theta)\right\}\!,
\label{eq:2r}
\hspace{0.74cm}
\end{equation}
where $\beta$ and $\tilde{\beta}$ are some new unknown coefficient functions. Certain restrictions for these unknown functions can be obtained if one requires that
upon substitution (\ref{eq:2q}) and (\ref{eq:2w}) into (\ref{eq:2e}) and (\ref{eq:2r}) we get the identities. For example, substituting (\ref{eq:2q}) and (\ref{eq:2w})
into (\ref{eq:2e}), we find
\[
\xi_{\mu}=\frac{1}{2}\,(\beta\kappa+\beta^{*}\kappa^{*})(\bar\theta\theta)\,\xi_{\mu} +
\frac{i}{2}\,(\beta\kappa - \beta^{*}\kappa^{*})(\bar\theta\sigma_{\mu\nu}\theta)\,\xi^{\nu}
- \frac{1}{2}\,(\beta\alpha + \beta^{*}\alpha^{*})(\bar\theta\gamma_{\mu}\theta)\,\xi_{5},
\]
The requirement of identical coincidence of left- and right-hand sides leads to the following system of algebraic equations
\begin{equation}
\begin{split}
\frac{1}{2}\,(\beta\kappa+\beta^{*}\kappa^{*})(\bar\theta\theta)=1,\\
\beta\kappa-\beta^{*}\kappa^{*}=0,\\
\beta\alpha+\beta^{*}\alpha^{*}=0.\\
\end{split}
\label{eq:2t}
\end{equation}
A similar requirement for the second relation (\ref{eq:2r}) leads still to the additional equations
\begin{equation}
\begin{split}
-\frac{1}{2}\,(\tilde\beta\alpha + \tilde\beta^{*}\alpha^{*})(\bar\theta\theta)=1,\\
(\tilde\beta\kappa + \tilde\beta^{*}\kappa^{*})=0.\\
\end{split}
\label{eq:2y}
\end{equation}
\indent
Let us now turn to analysis of separate terms in Lagrangian (\ref{eq:1r}). At first we consider a mapping of the kinetic term in (\ref{eq:1t}), more exactly, the term
\begin{equation}
\frac{1}{2\hspace{0.01cm}i}\biggl(\frac{d\bar\psi}{d\tau}\,\psi-\bar\psi\,\frac{d\psi}{d\tau}\biggr).
\label{eq:2u}
\end{equation}
Rather cumbersome, but straightforward calculations give the following expression for (\ref{eq:2u}) in terms of new variables $\xi_{\mu}$ and $\xi_5$
\begin{equation}
\begin{split}
&-i|\kappa|^{2\,}(\bar\theta\theta)\!\left(\xi_{\mu}\,\frac{d\xi^{\mu}}{d\tau}\right) - i|\alpha|^{2}(\bar\theta\theta)\!\left(\xi_{5}\,\frac{d\xi_{5}}{d\tau}\right)\\
&-\,\frac{1}{2}\,\biggl\{\frac{d\kappa^{*}}{d\tau}\,\kappa - \kappa^{*}\,\frac{d\kappa}{d\tau}\biggl\}(\bar{\theta}\sigma^{\mu\nu}\theta)\,\xi_{\mu}\xi_{\nu}
-\,\frac{1}{2}\,|\kappa|^{2\!}\left\{\left(\frac{d\bar{\theta}{d\tau}}\,\sigma^{\mu\nu}\theta\right)-\left(\bar\theta\,\sigma^{\mu\nu}\frac{d\theta}{d\tau}\right)\right\}\!\xi_{\mu\,}\xi_{\nu}\\
&+\,\frac{1}{2\hspace{0.01cm}i}\,\biggl\{\biggl(\kappa\,\frac{d\alpha^{*}}{d\tau} - \frac{d\kappa}{d\tau}\,\alpha^{*}\biggl)
+\, \biggl(\kappa^{*}\,\frac{d\alpha}{d\tau} - \frac{d\kappa^{*}}{d\tau}\,\alpha\biggl)\biggl\}
(\theta\gamma^{\mu}\theta)\,\xi_{5\,}\xi_{\mu}\\
&+\,\frac{1}{2\hspace{0.01cm}i}\,(\alpha^{*}\kappa + \alpha\kappa^{*})(\bar\theta\gamma^{\mu}\theta)\!\left\{\frac{d\xi_{5}}{d\tau}\,\xi_{\mu}-\xi_{5}\,\frac{d\xi_{\mu}}{d\tau}\right\}\\
&+\,\frac{1}{2\hspace{0.01cm}i}\,(\alpha^{*}\kappa - \alpha\kappa^{*})\left\{\left(\frac{d\bar\theta}{d\tau}\,\gamma^{\mu}\hspace{0.02cm}\theta\right) - \left(\bar\theta\hspace{0.02cm}\gamma^{\mu}\,\frac{d\theta}{d\tau}\right)\right\}\xi_{5\,}\xi_{\mu}.
\end{split}
\label{eq:2i}
\end{equation}
Here, the first two terms have in exact the same structure as the terms
\[
-\frac{i}{2}\left(\xi_{\mu}\,\frac{d\xi^{\mu}}{d\tau}\right)\quad \mbox{and} \quad  +\frac{i}{2}\left(\xi_{5}\,\frac{d\xi_{5}}{d\tau}\right)
\]
in (A.2) and (A.3), respectively. The requirement of literal coincidence of these two terms results in the algebraic equations additional to (\ref{eq:2t}) and (\ref{eq:2y})
\begin{equation}
+\,|\kappa|^2(\bar\theta\theta)=\frac{1}{2}\,,
\label{eq:2o}
\end{equation}
\begin{equation}
-\,|\alpha|^2(\bar\theta\theta)=\frac{1}{2}\,.
\label{eq:2p}
\end{equation}
Here, a minus sign in the second equation casts some doubt upon correctness of such a direct approach. Actually, there exist a little subtlety concerning the terms in (A.3)
with pseudoscalar $\xi_5$. As was mentioned above the Lagrangian (A.1) possesses local $n=1$ SUSY, while (\ref{eq:1r}) does not. For their mutual comparison
among themselves we dropped all terms in (A.1) containing the one-dimensional gravitino field $\chi$. However such a direct the $\chi$-term truncation is not strictly accurate.
It is more correct in this situation to use appropriate constraint equation in (A.6) for elimination of this variables from Lagrangian (A.1). In other words, before comparison
the second term in (\ref{eq:2i}) with the second term in (A.3), it is necessary preliminary to eliminate the $\chi$-field by using of the equation of motion for $\xi_5$
\[
2\hspace{0.04cm}\dot{\xi}_{\hspace{0.01cm}5} - m\chi = 0.
\]
After such an elimination the kinetic term $\xi_{5\hspace{0.02cm}}\dot{\xi}_{5}$ in (A.3) changes its sign to opposite one and comparing with appropriate terms in (\ref{eq:2i})
results in the algebraic equation with the correct sign, instead of (\ref{eq:2p})
\begin{equation}
+\,|\alpha|^2(\bar\theta\theta)=\frac{1}{2}\,.
\label{eq:2a}
\end{equation}
\indent
Further, it is evident that we cannot simply set the coefficients before the other terms in (\ref{eq:2i}) equal to zero, since this leads immediately to degeneracy of
our initial mapping. Formally, we must at first solve a system of algebraic equations (\ref{eq:2t}), (\ref{eq:2y}), (\ref{eq:2o}) and (\ref{eq:2a}) and then substitute
the solution obtained in the relevant coefficient functions in (\ref{eq:2i}). It will be made just below.\\
\indent
Let us consider the mapping of the term in Lagrangian (\ref{eq:1u}) defining interaction of the spin of a particle with background non-Abelian gauge field, more exactly a mapping
of the tensor of spin $\frac{1}{2} (\bar{\psi}\sigma^{\mu\nu}\psi)$ into the quadratic combination $\xi^{\mu}\xi^{\nu}$. A direct substitution of (\ref{eq:2q}) and (\ref{eq:2w})
and using formulas of the spinor algebra (B.2) gives the following expression
\begin{equation}
\begin{split}
\frac{1}{2}\bigl(\bar\psi\sigma^{\mu\nu}\psi\bigr)=\;\,
&i\hspace{0.02cm}|\kappa|^{2}(\bar\theta\theta)\,\xi^{\mu}\xi^{\nu} - \frac{1}{2}\,|\kappa|^{2}(\bar\theta\gamma_5\theta)\,\epsilon^{\mu\nu\lambda\sigma}\xi_{\lambda}\xi_{\sigma}\\
-\,&\frac{i}{2}\,(\alpha\kappa^{*}+\alpha^{*}\kappa)\hspace{0.02cm}\bigl\{(\bar\theta\hspace{0.02cm}\gamma^{\nu}\theta)\hspace{0.02cm}\xi^{\mu} - (\bar\theta\hspace{0.02cm}\gamma^{\mu}\theta)\hspace{0.02cm}\xi^{\nu}\bigr\}\hspace{0.02cm}\xi_{5}\\
-\,&\frac{1}{4}\,(\alpha\kappa^{*}-\alpha^{*}\kappa)\,\epsilon^{\mu\nu\lambda\sigma}\bigl\{(\bar\theta\gamma_{\sigma}\gamma_{5\hspace{0.02cm}}\theta)\hspace{0.02cm}\xi_{\lambda}
- (\bar\theta\gamma_{\lambda}\gamma_{5\hspace{0.02cm}}\theta)\hspace{0.02cm}\xi_{\sigma}\bigr\}\hspace{0.01cm}\xi_{5}.
\end{split}
\label{eq:2s}
\end{equation}
The requirement of coincidence of the force terms in different representations
\begin{equation}
- \frac{eg}{4}\,Q^aF^{a}_{\mu\nu}(\bar{\psi}\hspace{0.02cm}\sigma^{\mu\nu}\psi)\sim
\frac{ieg}{2}\,Q^aF^{a}_{\mu\nu\,}\xi^{\mu}\xi^{\nu} +\, \ldots
\label{eq:2d}
\end{equation}
leads to the following condition
\begin{equation}
-\,|\kappa|^2(\bar\theta\theta)=1\,.
\label{eq:2f}
\end{equation}
The algebraic equation obtained contradicts a similar equation (\ref{eq:2o}). It is the more so surprising that the spin tensor $S_{\mu\nu}$ in the form (\ref{eq:1p})
and (\ref{eq:1s}) obeys the same dynamical equation (\ref{eq:1a}). This suggests that the initial naive mapping (\ref{eq:2q}) is not complete. An extended discussion of this subject
will be considered in the subsequent sections and a possible way of overcoming the contradiction between (\ref{eq:2o}) and (\ref{eq:2f}) will be given in section\,5. For the reminder
of this section we will solve algebraic system (\ref{eq:2t}), (\ref{eq:2y}), (\ref{eq:2o}) and (\ref{eq:2a}) putting aside the equation (\ref{eq:2f}).\\
\indent
We shall seek a solution of the system of algebraic equations in the form $\kappa = a\hspace{0.02cm}e^{i\varphi}$, $a$ is some real function etc. Simple calculations leads
to the general solution
\begin{equation}
\begin{split}
\kappa &= (\pm)_{\kappa}\,\frac{1}{\sqrt{2}\hspace{0.01cm}(\bar\theta\theta)^{1/2}}\;e^{i\varphi},\quad
\alpha = (\pm)_{\alpha}\,\frac{i\hspace{0.02cm}(-1)^n}{\sqrt{2}\hspace{0.01cm}(\bar\theta\theta)^{1/2}}\;e^{i\varphi},\\
\beta &= (\pm)_{\kappa}\,\frac{\sqrt{2}\hspace{0.01cm}}{(\bar\theta\theta)^{1/2}}\;e^{-i\varphi},\quad\;\;\;
\tilde{\beta} = (\pm)_{\alpha}\,\frac{i\sqrt{2}\hspace{0.03cm}(-1)^n}{(\bar\theta\theta)^{1/2}}\;e^{-i\varphi}.\\
\end{split}
\label{eq:2g}
\end{equation}
Here, $n = 0,\pm\,1,\pm\,2,\dots\,$; $\varphi$ is an arbitrary phase that generally speaking, is a function of the evolution parameter $\tau$ and the symbols $(\pm)_{\kappa}$, $(\pm)_{\alpha}$
denote arbitrariness in choose of signs independent for functions $(\kappa,\hspace{0.02cm}\beta)$ and $(\alpha,\hspace{0.02cm}\tilde{\beta})$. The solutions (\ref{eq:2g}) should be considered
as formal ones because strictly speaking, by virtue of Grassmann nature of the spinor $\theta_{\alpha}$ there is no inverse function to $(\bar{\theta} \theta)$ one \cite{dewitt_book}.
The expressions (\ref{eq:2g}) and all similar expressions should be considered as multiplied by $(\bar{\theta} \theta)$ in appropriate degree.\\
\indent
If we now substitute the obtained solutions (\ref{eq:2g}) in the remainder of the untapped coefficient functions in (\ref{eq:2i}), then we find quite compact expressions for them
\[
\begin{split}
&\frac{1}{2}\,\biggl\{\frac{d\kappa^{*}}{d\tau}\,\kappa - \kappa^{*}\,\frac{d\kappa}{d\tau}\biggl\} = \frac{1}{2\hspace{0.01cm}i}\,\frac{1}{(\bar\theta\theta)}\,
\frac{d\varphi}{d\tau}\,,\\
&\frac{1}{2\hspace{0.03cm}i}\,\biggl\{\biggl(\kappa\frac{d\alpha^{*}}{d\tau} - \frac{d\kappa}{d\tau}\,\alpha^{*}\biggl)\,
+\; \biggl(\kappa^{*}\frac{d\alpha}{d\tau} - \frac{d\kappa^{*}}{d\tau}\,\alpha\biggl)\biggl\}\, =
\pm\,i\,\frac{1}{(\bar\theta\theta)}\,\frac{d\varphi}{d\tau}\,,\\
&\frac{1}{2\hspace{0.03cm}i}\,(\alpha^{*}\kappa + \alpha\kappa^{*})=0,\quad
\frac{1}{2\hspace{0.03cm}i}\,(\alpha^{*}\kappa - \alpha\kappa^{*}) = \mp\,\frac{1}{2(\bar\theta\theta)}\,.
\end{split}
\]
A special feature of the first two expressions involving derivatives of coefficient functions is that they reduce to the derivative of arbitrary phase $\varphi$.
If we set $\varphi=\mbox{const}$, then noticeable part of redundant terms in (\ref{eq:2i}) will be vanishing without any additional assumptions. The terms containing
derivatives of the odd spinor $\theta_{\alpha}$ will not to be zero under no circumstances. Another feature is the vanishing of the last coefficient function but one.
This function is connected with the term in (\ref{eq:2i}) proportional to the following mixed expression
\[
\dot{\xi}_{5}\hspace{0.03cm}\xi_{\mu} - \xi_{5}\hspace{0.03cm}\dot{\xi}_{\mu}.
\]
A similar contribution to the Lagrangian of a spin particle was first considered in paper \cite{barducci_1976} and also in several others
\cite{brink_1976, casalbuoni_2008, casalbuoni_2011}. In our case for the simplest mapping (\ref{eq:2q}), (\ref{eq:2w}) this mixed contribution exactly turns to zero.\\
\indent
Thus we see that a simple choice of (\ref{eq:2q}), (\ref{eq:2w}) leads to sufficiently reasonable and visible expressions for the one-sided mapping
$(\psi,\hspace{0.02cm}\bar{\psi}) \rightarrow (\xi_{\mu},\hspace{0.02cm}\xi_5)$. We can substitute (\ref{eq:2q}), (\ref{eq:2w}) with coefficients (\ref{eq:2g}) into the
Lagrangian $L_{\Psi}$, Eq.\,(\ref{eq:1o}) and thereby to obtain an explicit form of interacting terms of a color spinning particle with the background fermionic field
$\Psi_{\alpha}^i(x)$ expressed in terms of the pseudovariables $\xi_{\mu},\hspace{0.02cm}\xi_5$ and auxiliary anticommuting spinor $\theta_{\alpha}$. The only problem
we are faced with is evident contradiction between the mapping of kinetic term (\ref{eq:2u}) and the mapping of the spin interaction term of a particle with external gauge
field, Eq.\,(\ref{eq:2d}). Furthermore, we leave delicate problem of the one-to-one correspondence of our mapping out of the consideration.
In two forthcoming sections we consider in detail the conditions and requirements under which it is possible to obtain one-to-one mapping.

\section{\bf Requirement of one-to-oneness of the map $(\psi,\bar{\psi})\rightleftarrows (\xi_{\mu},\hspace{0.02cm}\xi_5)$}
\setcounter{equation}{0}

In the preceding section in the analysis of the mapping $(\psi,\hspace{0.02cm}\bar{\psi}) \rightarrow (\xi_{\mu},\hspace{0.02cm}\xi_5)$ we require that
substitution (\ref{eq:2q}), (\ref{eq:2w}) into (\ref{eq:2e}) and (\ref{eq:2r}) leads to identities. Let us consider now what we will have for an inverse
mapping $(\xi_{\mu},\hspace{0.02cm}\xi_5) \rightarrow (\psi,\hspace{0.02cm}\bar{\psi})$. For this purpose we substitute (\ref{eq:2e}), (\ref{eq:2r})
back into (\ref{eq:2q}) and collect like terms
\[
\begin{split}
\psi_{\alpha} =\; &\frac{1}{2}\, \Bigl\{\kappa\hspace{0.02cm}\beta(\bar\theta\gamma^{\mu}\gamma_{5})_{\beta}(\gamma_{\mu}\gamma_5\theta)_{\alpha}
+\hspace{0.02cm} \alpha\hspace{0.02cm}\tilde{\beta}(\bar\theta\gamma_{5})_{\beta}(\gamma_5\theta)_{\alpha}\Bigr\}\psi_{\alpha}\\
+\,&\frac{1}{2}\,\bar{\psi}_{\beta}\Bigl\{\kappa\hspace{0.02cm}\beta^{*}(\gamma_{5}\gamma^{\mu}\theta)_{\beta}(\gamma_{\mu}\gamma_5\theta)_{\alpha}
+\hspace{0.02cm} \alpha\hspace{0.02cm}\tilde{\beta}^{*}(\gamma_{5}\theta)_{\beta}(\gamma_5\theta)_{\alpha}\Bigr\}.
\end{split}
\]
Here, the requirement of identical coincidence of the left- and right-hand sides leads to the following equations
\begin{equation}
\bar{\theta}_{\gamma}\Bigl\{\kappa\hspace{0.02cm}\beta(\gamma^{\mu}\gamma_{5})_{\gamma\beta}(\gamma_{\mu}\gamma_5)_{\alpha\delta}
+ \alpha\hspace{0.02cm}\tilde{\beta}(\gamma_{5})_{\gamma\beta}(\gamma_5)_{\alpha\delta}\Bigr\}\theta_{\delta} = 2\delta_{\alpha\beta}
\label{eq:3q}
\end{equation}
and
\begin{equation}
\hspace{0.6cm}
\Bigl\{\kappa\hspace{0.02cm}\beta^{*}(\gamma^{\mu}\gamma_{5})_{\beta\gamma}(\gamma_{\mu}\gamma_5)_{\alpha\delta}
- \alpha\hspace{0.02cm}\tilde{\beta}^{*}(\gamma_{5})_{\beta\gamma}(\gamma_5)_{\alpha\delta}\Bigr\}\theta_{\gamma}\theta_{\delta} = 0.
\label{eq:3w}
\end{equation}
At the beginning, we analyze the first equation. In the left-hand side of the expression in braces we pick out a contribution proportional to $\delta_{\alpha \beta}$.
For this purpose we use the Fierz identities (see, for example, appendix in the book by Okun \cite{okun_book}). The Fierz identities enable us to result (\ref{eq:3q}) in the form
\[
\begin{split}
&\Bigl\{-\kappa\beta + \frac{1}{4}\,\alpha\tilde{\beta}\Bigr\}\,(\bar{\theta}\theta)\delta_{\alpha\beta} -
\frac{1}{2}\,\Bigl\{\kappa\beta + \frac{1}{2}\,\alpha\tilde{\beta}\Bigr\}\,(\bar{\theta}\gamma_{\mu}\theta)(\gamma^{\mu})_{\alpha\beta}\\
&+\,\frac{1}{8}\,\alpha\tilde{\beta}\,(\bar{\theta}\sigma_{\mu\nu}\theta)(\sigma^{\mu\nu})_{\alpha\beta}\,+\\
&\Bigl\{\kappa\beta + \frac{1}{4}\,\alpha\tilde{\beta}\Bigr\}\,(\bar{\theta}\gamma_{5}\theta)(\gamma_{5})_{\alpha\beta}
-\,\frac{1}{2}\,\Bigl\{\kappa\beta - \frac{1}{2}\,\alpha\tilde{\beta}\Bigr\}\,(\bar{\theta}\gamma_{\mu}\gamma_{5}\theta)(\gamma^{\mu}\gamma_{5})_{\alpha\beta}
= 2\delta_{\alpha\beta}.
\end{split}
\]
The requirement of single-valuedness of the inverse mapping is reduced here to a system of algebraic equations for the coefficient functions
\[
\begin{split}
&\underline{\mbox{S\hspace{0.02cm}-\hspace{0.02cm}channel:}}\qquad\qquad \Bigl\{-\kappa\beta + \frac{1}{4}\,\alpha\tilde{\beta}\Bigr\}\,(\bar{\theta}\theta) = 2,\\
&\underline{\mbox{V-\hspace{0.02cm}channel:}}\qquad\qquad \kappa\beta + \frac{1}{2}\,\alpha\tilde{\beta} = 0,\\
&\underline{\mbox{T-\hspace{0.02cm}channel:}}\qquad\qquad \alpha\tilde{\beta} = 0,\\
&\underline{\mbox{A-\hspace{0.02cm}channel:}}\qquad\qquad \kappa\beta - \frac{1}{4}\,\alpha\tilde{\beta} = 0,\\
&\underline{\mbox{P-\hspace{0.02cm}channel:}}\qquad\qquad \kappa\beta + \frac{1}{4}\,\alpha\tilde{\beta} = 0.
\end{split}
\]
It is easy to see that in the presence of only pseudovector $\xi_{\mu}$ and pseudoscalar $\xi_5$ contributions to the mapping (\ref{eq:2q}) this algebraic system has no solutions.\\
\indent
Let us extent the mapping (\ref{eq:2q}) by adding to it the vector $\hat{\xi}_{\mu}$ and scalar $\hat{\xi}_5$ anticommuting parts, i.e. we set
\[
\psi=\; \ldots\; + \,\hat{\kappa}\,(\hat{\xi}_{\mu}\gamma^{\mu})\,\theta \,+\, \hat{\alpha}\,\hat{\xi}_{5\,}\theta,
\]
where the dots refer to the terms from the right-hand side of (\ref{eq:2q}) and
\begin{equation}
\begin{split}
&\hat{\xi}_{\mu} = \frac{1}{2}\,\Bigl\{\hat{\beta}(\bar\theta\gamma_{\mu}\psi) + \hat{\beta}^{*}(\bar\psi\gamma_{\mu}\theta)\Bigr\},\\
&\hat{\xi}_5 = \frac{1}{2}\left\{\hat{\tilde\beta}(\bar{\theta}\hspace{0.02cm}\psi) + \hat{\tilde\beta}^{*}(\bar{\psi}\hspace{0.02cm}\theta)\right\}.
\end{split}
\label{eq:3e}
\end{equation}
Here, $\hat{\kappa},\, \hat{\alpha},\, \hat{\beta}$ and $\tilde{\hat{\beta}}$ are some new unknown coefficients. To avoid large amount of notations, for the vector
and scalar contributions we use the same notations just like the pseudovector and pseudoscalar contributions with hat above only. The consideration of these additional
contributions appreci\-able improves situation with solvability of the algebraic system for coeffi\-ci\-ents, but eventually results in contradiction anyway.\\
\indent
Finally, as the last step we can add a tensor contribution $\zeta_{\mu\nu}$ (in our case it is more convenient to take pseudotensor $^{\ast\!}\zeta_{\mu\nu}$
dual to $\zeta_{\mu\nu}$) to construction of the one-to-one mapping and thus write eventually, instead of (\ref{eq:2q})
\begin{equation}
\psi = \Bigl[\kappa\,\xi_{\mu}(\gamma^{\mu}\gamma_{5\hspace{0.02cm}}\theta) + \alpha\,\xi_5(\gamma_{5\hspace{0.02cm}}\theta)\Bigr] +\,
\rho\,^{\ast\!}\zeta_{\mu\nu}(\sigma^{\mu\nu}\gamma_{5\hspace{0.02cm}}\theta)
\,+\, \Bigl[\hat{\kappa}\,\hat{\xi}_{\mu}(\gamma^{\mu}\theta) + \hat{\alpha}\,\hat{\xi}_{5\hspace{0.02cm}}\theta\Bigr],
\label{eq:3r}
\end{equation}
where
\begin{equation}
\,^{\ast\!}\zeta_{\mu\nu} = \frac{1}{2}\,\Bigl\{\!\hspace{0.01cm}s\hspace{0.02cm}(\bar{\theta}\sigma_{\mu\nu}\gamma_5\psi) -
s^{\ast}\hspace{0.02cm}(\bar{\psi}\hspace{0.01cm}\gamma_{5\,}\sigma_{\mu\nu}\theta)\Bigr\},
\label{eq:3t}
\end{equation}
$\rho$ and $s$ are the others unknown coefficients. Substituting (\ref{eq:2e}), (\ref{eq:2r}), (\ref{eq:3e}) and (\ref{eq:3t}) into (\ref{eq:3r}) and using
the Fierz identities we result in the following system of algebraic equations
\begin{equation}
\begin{split}
&\underline{\mbox{S\hspace{0.02cm}-\hspace{0.02cm}channel:}}\qquad\qquad \Bigl\{-\Bigl(\kappa\beta - \hat{\kappa}\hat{\beta}\Bigr) + \frac{1}{4}\,\Bigl(\alpha\tilde{\beta} + \hat{\alpha}\hat{\tilde{\beta}}\Bigr)+ 3\rho s\Bigr\}(\bar{\theta}\theta) = 2\,,\\
&\underline{\mbox{V-\hspace{0.02cm}channel:}}\qquad\qquad \Bigl(\kappa\beta + \hat{\kappa}\hat{\beta}\Bigr) + \frac{1}{2}\,\Bigl(\alpha\tilde{\beta} - \hat{\alpha}\hat{\tilde{\beta}}\Bigr) = 0,\\
&\underline{\mbox{T-\hspace{0.02cm}channel:}}\qquad\qquad \alpha\tilde{\beta} + \hat{\alpha}\hat{\tilde{\beta}} = 4\rho s,\\
&\underline{\mbox{A-\hspace{0.02cm}channel:}}\qquad\qquad \Bigl(\kappa\beta + \hat{\kappa}\hat{\beta}\Bigr) - \frac{1}{2}\,\Bigl(\alpha\tilde{\beta} - \hat{\alpha}\hat{\tilde{\beta}}\Bigr) = 0,\\
&\underline{\mbox{P-\hspace{0.02cm}channel:}}\qquad\qquad \Bigl(\kappa\beta - \hat{\kappa}\hat{\beta}\Bigr) + \frac{1}{4}\,\Bigl(\alpha\tilde{\beta} + \hat{\alpha}\hat{\tilde{\beta}}\Bigr)
+ 3\rho s = 0.
\end{split}
\label{eq:3y}
\end{equation}
The system obtained is consistent and it defines the following simple relations between the coefficient functions
\begin{equation}
\begin{split}
\alpha\tilde{\beta} = \,&\hat{\alpha}\hat{\tilde{\beta}},\quad \kappa\beta = - \hat{\kappa}\hat{\beta},\\
\kappa\beta = -\frac{1}{2(\bar{\theta}\theta)}\,,\quad &\alpha\tilde{\beta}= \frac{1}{2(\bar{\theta}\theta)}\,,\quad
\rho s= \frac{1}{4(\bar{\theta}\theta)}\,.
\end{split}
\label{eq:3u}
\end{equation}
We notice that from solutions (\ref{eq:2g}) in the preceding section follows
\[
\kappa\beta = \frac{1}{(\bar{\theta}\theta)}\,,\quad \alpha\tilde{\beta}= -\frac{1}{(\bar{\theta}\theta)}\,.
\]
Comparing the last equalities with appropriate expressions in (\ref{eq:3u}), we see that they differ from one another by the factor $(-1/2)$. Thus the approach stated in section 2
is quite good approximation of more subtle approach suggested in this section.\\
\indent
Let us consider now the second equation (\ref{eq:3w}), which must be also satisfied. Analysis of this equation is a slightly different. The distinctive feature of the equation
is that it contains a product of two anticommuting spinors $\theta_{\gamma}$ and $\theta_{\delta}$. Thus in order for this expression will to be zero, it is necessary
and sufficient that the expression in braces will be {\it symmetric} with respect to spinor indices $\gamma$ and $\delta$. It is well known \cite{okun_book} that there exist
only two combinations symmetric with respect to the last spinor indices, which can be symbolically written in the form:
\begin{equation}
3\hspace{0.02cm}({\cal S} + {\cal P}) + \frac{1}{2}\,{\cal T},\qquad 2\hspace{0.02cm}({\cal S} - {\cal P}) + {\cal V} -{\cal A},
\label{eq:3i}
\end{equation}
where ${\cal S}$ designates the contribution of the type $I\otimes I$, ${\cal P}\sim\gamma_{5}\otimes\gamma_{5}$, ${\cal V}\sim\gamma^{\mu}\otimes\gamma_{\mu}$, and so on.\\
\indent
From (\ref{eq:3i}) and (\ref{eq:3w}) one can see at once that it is impossible to form a symmetric combination of structures $A$ and $P$ only. Similar to the previous case,
here it is necessary to take into consideration all permissible contributions, i.e. to use expression (\ref{eq:3r}) as initial one. Then instead of (\ref{eq:3w}) we obtain
the equation in the form
\begin{equation}
\begin{split}
\Bigl\{\kappa\beta^{*}(\gamma_{\mu}\gamma_{5})_{\beta\gamma}&(\gamma^{\mu}\gamma_5)_{\alpha\delta}
- \alpha\tilde{\beta}^{*}(\gamma_{5})_{\beta\gamma}(\gamma_5)_{\alpha\delta}
-\frac{1}{2}\,\rho s^{\ast}(\sigma_{\mu\nu})_{\beta\gamma}(\sigma^{\mu\nu})_{\alpha\delta}\\
+ \;&\hat{\kappa}\hat{\beta}^{*}(\gamma_{\mu})_{\beta\gamma}(\gamma^{\mu})_{\alpha\delta}
\,+\,\hat{\alpha}\hat{\tilde{\beta}}^{*}\delta_{\beta\gamma}\delta_{\alpha\delta}
\Bigr\}\hspace{0.02cm}\theta_{\gamma}\theta_{\delta} = 0.
\end{split}
\label{eq:3o}
\end{equation}
\indent
Let us combine two expressions in (\ref{eq:3i}) preliminary multiplied them by arbitrary constants $\mu$ and $\nu$, respectively
\[
(3\mu + 2\nu)\hspace{0.02cm}{\cal S} + (3\mu - 2\nu)\hspace{0.02cm}{\cal P} + \frac{1}{2}\,\mu\hspace{0.01cm}{\cal T} + \nu\hspace{0.01cm}{\cal V} - \nu{\cal A}.
\]
Comparing the last expression with (\ref{eq:3o}), we find the following system of algebraic equations for the coefficient functions
\begin{equation}
\begin{split}
&\hat{\alpha}\hat{\tilde{\beta}}^{*} - \alpha\tilde{\beta}^{*} = -\hspace{0.03cm}6\rho s^{\ast},\\
\hat{\kappa}\hat{\beta}^{*} &= - \kappa\beta^{*} = \frac{1}{4}\,(\hat{\alpha}\hat{\tilde{\beta}}^{*} + \alpha\tilde{\beta}^{*})
\end{split}
\label{eq:3p}
\end{equation}
and the expressions for $\mu$ and $\nu$
\[
\begin{split}
&\mu = \frac{1}{6}\,(\hat{\alpha}\hat{\tilde{\beta}}^{*} - \alpha\tilde{\beta}^{*}),\\
&\nu = \frac{1}{4}\,(\hat{\alpha}\hat{\tilde{\beta}}^{*} + \alpha\tilde{\beta}^{*}).
\end{split}
\]
Incidentally, the system (\ref{eq:3p}) can be directly obtained by not appealing to (\ref{eq:3i}) by using the Fierz identities only. We notice also that as distinct from (\ref{eq:3y})
the equation (\ref{eq:3p}) in fact represents a system of equations for phases of the desired coefficient functions. To see that we express
variables $\rho$, $\alpha$ and $\hat{\alpha}$ in terms of $s$, $\tilde{\beta}$ and $\hat{\tilde{\beta}}$ with the help of (\ref{eq:3u}) and substitute them into (\ref{eq:3p}). Finally, we have
\[
\biggl(\frac{\hat{\tilde{\beta}}^{*}}{\!\hat{\tilde{\beta}}} \,-\, \frac{\tilde{\beta}^{*}}{\!\tilde{\beta}}\biggr)
= -\hspace{0.01cm}3\,\frac{s^{*}}{\!s},\qquad
\frac{\beta^{*}}{\!\beta} = \frac{\hat{\beta}^{*}}{\!\hat{\beta}} =
\frac{1}{4}\biggl(\frac{\hat{\tilde{\beta}}^{*}}{\!\hat{\tilde{\beta}}}\hspace{0.03cm} + \hspace{0.03cm}\frac{\tilde{\beta}^{*}}{\!\tilde{\beta}}\biggr).
\]
If one formally sets
\[
\tilde{\beta} = |\tilde{\beta}|\,{\rm e}^{-i\tilde{\Phi}},\quad
\hat{\tilde{\beta}} = |\hat{\tilde{\beta}}|\,{\rm e}^{-i\hat{\tilde{\Phi}}},\quad
s = |s|\,{\rm e}^{-i\Psi},\quad
\beta = |\beta|\,{\rm e}^{-i\Phi},\quad
\hat{\beta} = |\hat{\beta}|\,{\rm e}^{-i\hat{\Phi}},
\]
then instead of two equations written out just above, we obtain
\[
{\rm e}^{2i\hat{\tilde{\Phi}}} -\, {\rm e}^{2i\tilde{\Phi}} = -3\hspace{0.02cm}{\rm e}^{2i\Psi},\qquad
{\rm e}^{2i\Phi} = {\rm e}^{2i\hat{\Phi}} = \frac{1}{4}\,({\rm e}^{2i\hat{\tilde{\Phi}}} + {\rm e}^{2i\tilde{\Phi}}).
\]
It is evident that the first equation has no solutions in no values of angles $\hat{\tilde{\Phi}}$, $\tilde{\Phi}$ and $\Psi$. Thus we lead to negative conclusion: the expression in braces in (\ref{eq:3o}) cannot be symmetric with respect to spinor indices $\gamma$ and $\delta$ in no values of the coefficient functions compatible with relations (\ref{eq:3u}). Hence, the equation (\ref{eq:3o}) is not identical fulfilment.\\
\indent
For indirect check of this result let us give a similar brief analysis for antisymmetric case of the spinor indices and show that here, there exist non-trivial solutions.
For the antisymmetric case there exist three antisymmetric combinations
\[
{\cal V} + {\cal A},\qquad {\cal S} + {\cal P} - \frac{1}{2}\,{\cal T},\qquad  2\,(\,{\cal S} - {\cal P}) - (\,{\cal V} - {\cal A}).
\]
The requirement of antisymmetry of the expression in braces in (\ref{eq:3o}) with respect to permutation of spinor indices $\gamma$ and $\delta$ results in the following
algebraic equation
\begin{equation}
\begin{split}
&\hat{\alpha}\hat{\tilde{\beta}}^{*} - \alpha\tilde{\beta}^{*} = 2\hspace{0.02cm}\rho s^{\ast},\\
&\hat{\alpha}\hat{\tilde{\beta}}^{*} + \alpha\tilde{\beta}^{*} = -2\hspace{0.01cm}(\hat{\kappa}\hspace{0.03cm}\hat{\beta}^{*} - \kappa\hspace{0.03cm}\beta^{*}),
\end{split}
\label{eq:3a}
\end{equation}
or in terms of the phases, we get

\[
{\rm e}^{2i\hat{\tilde{\Phi}}} -\, {\rm e}^{2i\tilde{\Phi}} = {\rm e}^{2i\Psi},\qquad
{\rm e}^{2i\hat{\tilde{\Phi}}} +\, {\rm e}^{2i\tilde{\Phi}} = -2\hspace{0.01cm}({\rm e}^{2i\Phi} + {\rm e}^{2i\hat{\Phi}}).
\]
A solution of this system exists and contains one arbitrary phase. If we take the function $\tilde{\Phi}$ as such a phase, then we will have
\begin{equation}
\begin{split}
&\hat{\tilde{\Phi}} = \tilde{\Phi} \,\mp\, \frac{\pi}{6} + 2\pi n,\quad n = 0,\,\pm 1,\ldots,\\
&\Psi = \tilde{\Phi} \,\mp\, \frac{\pi}{3} + \pi m,\quad m = 0,\,\pm 1,\ldots,\\
&\hat{\Phi} = -\Phi + 2\tilde{\Phi} \,\mp\, \frac{\pi}{6} + \pi + 2\pi m,
\end{split}
\label{eq:3s}
\end{equation}
and the angle $\Phi$ is expressed through the angle $\tilde{\Phi}$ from the relation
\[
\cos\Bigl[2(\Phi - \tilde{\Phi})\,\pm\,\frac{\pi}{6}\,\Bigr] = \mp\,\frac{\sqrt{3}}{4}\,.
\]
\indent
The map (\ref{eq:3r}) seems sufficiently complete, but here, additional physical variables appear. The pair of variables $(\hat{\xi}_{\mu}, \hat{\xi}_5)$ and also
the pseudotensor $^{\ast}\!\hspace{0.02cm}\zeta_{\mu \nu}$ were added to pair of the initial variables $(\xi_{\mu}, \xi_5)$. As was mentioned by Berezin and Marinov
\cite{berezin_1975}, generally speaking, before the procedure of quantization either of the pairs $(\xi_{\mu}, \xi_5)$ and $(\hat{\xi}_{\mu}, \hat{\xi}_5)$ can
be chosen as main dynamical variables. Since as the basic variables the first pair is chosen, then the second one must play the auxiliary role. It is possible
to reduce the number of auxiliary functions to one function $\hat{\xi}_{5}$, if to impose the condition in the form
\[
\hat{\xi}_{\mu} \sim\hspace{0.01cm} \dot{x}_{\mu\hspace{0.01cm}} \hat{\xi}_{5}.
\]
However, a physical interpretation of the odd scalar $\hat{\xi}_{5}$ remains unclear.\\
\indent
In regard to the pseudotensor $^{\ast}\!\hspace{0.02cm}\zeta_{\mu \nu}$, it is naturally to consider that
\begin{equation}
^{\ast}\!\hspace{0.02cm}\zeta_{\mu \nu} \sim\hspace{0.01cm} \dot{x}_{\mu\hspace{0.01cm}} \xi_{\nu} - \dot{x}_{\nu\hspace{0.01cm}} \xi_{\mu}
\label{eq:3d}
\end{equation}
up to possible term $\varepsilon_{\mu\nu\lambda\sigma\hspace{0.02cm}}\dot{x}^{\lambda}\xi^{\sigma}$. The tensor variables of type $^{\ast}\hspace{0.03cm}\!\zeta_{\mu \nu}$
(together with $\xi_{\mu}$ and $\xi_{5}$) was for the first time considered in \cite{casalbuoni_2011} in a context of construction of a pseudoclassical particle model
associated to the twisted ${\cal N} = 2$ SUSY algebra in the euclidian $4D$ space. In this work it was shown that in a certain choice of parameters of the particle
model it is possible to eliminate the odd tensor variable and ipso facto to obtain the vector SUSY particle model suggested early by the same authors \cite{casalbuoni_2008}.
However, we would like to show that a choice of form (\ref{eq:3d}) results in more nontrivial results, appearance of new interaction terms, especially in the
presence of the background fermion field in the system. This question will be considered in sections\,5 and 6. The constraint (\ref{eq:3d}) actually contains
in the above-mentioned paper \cite{casalbuoni_2011} in equality (21) for the generators of the twisted SUSY algebra.

\section{\bf General analysis of a connection of $\psi$ and $\theta$ spinors with real tensor quantities}
\setcounter{equation}{0}

The above analysis has shown that the most general mapping suggested in section\,3 with {\it real} tensor quantities (\ref{eq:2e}), (\ref{eq:2r}),
(\ref{eq:3e}) and (\ref{eq:3t}) is not one-to-one mapping. Substituting the tensor structures into (\ref{eq:3r}) we do not come to identity
in no values of the coefficient functions. In this section we would like to analyze a reason for this circumstance by using to some extent more general
approach. Answer the question of one-to-oneness of the inverse mapping will be the additional result of this analysis,
i.e. a result of a substitution of the mapping (\ref{eq:3r}) back into tensor structures (\ref{eq:2e}), (\ref{eq:2r}), (\ref{eq:3e}) and (\ref{eq:3t}).\\
\indent
The problem of defining the mapping $(\psi, \bar{\psi}) \rightleftarrows (\xi_{\mu},\,\xi_5,\,\ldots)$ stated in this paper is in fact a part of more
general analysis of a connection between spinors (Dirac, Majorana or Weyl ones) and Lorentz-invariant real or complex tensor aggregates. In the case
of one commuting c-number Dirac spinor and 16 real commuting bilinear quantities that are formed by the given spinor, such a problem has been studied by Takahashi
\cite{takahashi_1982}, Kaempffer \cite{kaempffer_1981} and from a different viewpoint by Zhelnorovich \cite{zhelnorovich_1970, zhelnorovich_1972, zhelnorovich_book}.
The latter considers also more special cases of Majorana and Weyl spinors, and most important for us the problem of a connection of {\it two} commuting spinors with
appropriate tensor set. In subsequent discussion we will follow essentially by Zhelnorovich \cite{zhelnorovich_1972, zhelnorovich_book}.\\
\indent
In the problem under consideration we also have at hand two (Dirac) spinors $\psi_{\alpha}$ and $\theta_{\alpha}$ (although in our case the latter plays an auxiliary role).
However, the second spinor as distinct from the works \cite{zhelnorovich_1972, zhelnorovich_book} is classical anticommuting one. For the convenience of further references
we written out once more all required formulas. We consider that a connection of even spinor $\psi_{\alpha}$ with real tensor quantities is defined
by the formula
\begin{equation}
\psi = \Bigl[\kappa\,\xi_{\mu}(\gamma^{\mu}\gamma_{5\hspace{0.02cm}}\theta) + \alpha\,\xi_5(\gamma_{5\hspace{0.02cm}}\theta)\Bigr]\! +
\rho\,^{\ast\!}\zeta_{\mu\nu}(\sigma^{\mu\nu}\gamma_{5\hspace{0.02cm}}\theta)
+ \Bigl[\hat{\kappa}\,\hat{\xi}_{\mu}(\gamma^{\mu}\theta) + \hat{\alpha}\,\hat{\xi}_{5\hspace{0.02cm}}\theta\Bigr],
\label{eq:4q}
\end{equation}
where in turn a connection of the tensor quantities with the original spinor $\psi$ is given by formulas
\begin{equation}
\begin{split}
&\xi_{\mu} = \frac{1}{2}\,\Bigl\{\beta(\bar\theta\hspace{0.02cm}\gamma_{\mu}\gamma_{5}\psi) - \beta^{*}(\bar\psi\hspace{0.02cm}\gamma_{5}\gamma_{\mu\hspace{0.01cm}}\theta)\Bigr\},\\
&\xi_5 = \frac{1}{2}\left\{\tilde\beta(\bar\theta\hspace{0.02cm}\gamma_{5}\psi) - \tilde\beta^{*}(\bar\psi\hspace{0.02cm}\gamma_{5\hspace{0.02cm}}\theta)\right\},\\
\,^{\ast\!}&\zeta_{\mu\nu} = \frac{1}{2}\,\Bigl\{\!\hspace{0.005cm}s\hspace{0.02cm}(\bar{\theta}\sigma_{\mu\nu}\gamma_5\psi) -
s^{\ast}\hspace{0.01cm}(\bar{\psi}\hspace{0.01cm}\gamma_{5\,}\sigma_{\mu\nu}\psi)\Bigr\},\\
&\hat{\xi}_{\mu} = \frac{1}{2}\,\Bigl\{\hat{\beta}(\bar\theta\hspace{0.02cm}\gamma_{\mu}\psi) + \hat{\beta}^{*}(\bar\psi\hspace{0.02cm}\gamma_{\mu\hspace{0.01cm}}\theta)\Bigr\},\\
&\hat{\xi}_5 = \frac{1}{2}\left\{\hat{\tilde\beta}(\bar{\theta}\hspace{0.02cm}\psi) + \hat{\tilde\beta}^{*}(\bar{\psi}\hspace{0.02cm}\theta)\right\}.
\end{split}
\label{eq:4w}
\end{equation}
Furthermore, we introduce by definition a (real) tensor aggregate that are formed only by the odd spinor $\theta$:
\begin{equation}
S\equiv\bar{\theta}\theta,\quad\; P\equiv\bar{\theta}\gamma_{5\hspace{0.02cm}}\theta,\quad\; V_{\mu}\equiv\bar{\theta}\gamma_{\mu}\hspace{0.02cm}\theta,\quad\;
A_{\mu}\equiv\bar{\theta}\gamma_{\mu}\gamma_{5\hspace{0.02cm}}\theta,   \quad\; \,^{\ast}T_{\mu\nu}\equiv\bar{\theta}\sigma_{\mu\nu}\gamma_{5\hspace{0.02cm}}\theta.
\label{eq:4e}
\end{equation}
A spinor of the second rank in the form $\bar{\theta}_{\beta} \theta_{\alpha}$ can be identically presented in terms of these tensor quantities
\begin{equation}
\bar{\theta}_{\beta}\theta_{\alpha} = \frac{1}{4}\,
\Bigl\{\!S\hspace{0.01cm}\delta_{\alpha\beta} + V_{\mu}(\gamma^{\mu})_{\alpha\beta} + \frac{1}{2}\,^{\ast}T_{\mu\nu}(\sigma^{\mu\nu}\gamma_{5})_{\alpha\beta} -
A_{\mu}(\gamma^{\mu}\gamma_{5})_{\alpha\beta} + P(\gamma_{5})_{\alpha\beta}\Bigr\}.
\label{eq:4r}
\end{equation}
Let us consider a spinor structure of a mixed type including both the even $\psi_{\alpha}$ and odd $\theta_{\alpha}$ spinors
\[
\lambda\,\bar{\theta\hspace{0.02cm}}_{\beta}\psi_{\alpha} + \lambda^{*}\bar{\psi}_{\beta\hspace{0.02cm}}\theta_{\alpha},
\]
where $\lambda$ is some complex number, which without loss of generality can be set equal to $e^{i\phi}$.
Let us define for this expression an expansion similar to (\ref{eq:4r}) making use the definitions (\ref{eq:4w}) as real tensor quantities
\begin{equation}
{\rm e}^{i\phi}\,\bar{\theta}_{\beta\hspace{0.02cm}}\psi_{\alpha} + \,{\rm e}^{-i\phi}\,\bar{\psi}_{\beta\hspace{0.02cm}}\theta_{\alpha}
\label{eq:4t}
\end{equation}
\[
= \frac{1}{4}\,
\Bigl\{\!a_{1}\hspace{0.02cm}\hat{\xi}_{5}\hspace{0.01cm}\delta_{\alpha\beta} +\hspace{0.02cm}a_{2}\hspace{0.04cm}\hat{\xi}_{\mu}(\gamma^{\mu})_{\alpha\beta}
+ \hspace{0.02cm}a_{3}\hspace{0.02cm}\!\,^{\ast\!}\zeta_{\mu\nu}(\sigma^{\mu\nu}\gamma_{5})_{\alpha\beta} +
a_{4}\hspace{0.02cm}\xi_{\mu}(\gamma^{\mu}\gamma_{5})_{\alpha\beta} + a_{5}\hspace{0.03cm}\xi_5(\gamma_{5})_{\alpha\beta}\Bigr\}.
\]
The constants $a_1, \ldots \,, a_5$ on the right-hand side are connected with the coefficients $\hat{\tilde{\beta}}, \hat{\beta}, \ldots, \beta$ in (\ref{eq:4w})
by the relations
\begin{equation}
\begin{split}
&a_{1} = a_{1}^{*} = 2\,\frac{\,{\rm e}^{i\phi}}{\hat{\tilde{\beta}}},\quad
a_{2} = a_{2}^{*} = 2\,\frac{\,{\rm e}^{i\phi}}{\hat{\beta}},\quad
a_{3} = a_{3}^{*} = \frac{\,{\rm e}^{i\phi}}{s},\\
&a_{4} = a_{4}^{*} = -2\,\frac{\,{\rm e}^{i\phi}}{\beta},\quad
a_{5} = a_{5}^{*} = 2\,\frac{\,{\rm e}^{i\phi}}{\tilde{\beta}}.
\end{split}
\label{eq:4y}
\end{equation}
The expression (\ref{eq:4t}) as well as (\ref{eq:4r}) is exact. Our question is as follows: is it possible to recover Dirac the commuting spinor $\psi_{\alpha}$
by using {\it only} tensor aggregate (\ref{eq:4w}), i.e. is it the expansion (\ref{eq:4q}) true? According to V.A. Zhelnorovich
the answer to this question lies in the expansion (\ref{eq:4t}). Let us contract expression (\ref{eq:4t}) with a spinor $\theta_{\beta}$
\[
{\rm e}^{i\phi}(\bar{\theta}\theta)\hspace{0.03cm}\psi_{\alpha} + \,{\rm e}^{-i\phi}(\bar{\psi}\hspace{0.03cm}\theta)\hspace{0.03cm}\theta_{\alpha}
\]
\[
= \frac{1}{4}\,\Bigl\{a_{1}\hspace{0.03cm}\hat{\xi}_{5}\theta_{\alpha} +\hspace{0.02cm} a_{2}\hspace{0.03cm}\hat{\xi}_{\mu}(\gamma^{\mu}\theta)_{\alpha}
+\hspace{0.02cm} a_{3}\hspace{0.03cm}\!\,^{\ast\!}\zeta_{\mu\nu}(\sigma^{\mu\nu}\gamma_5\theta)_{\alpha} +\hspace{0.02cm}
a_{4}\hspace{0.03cm}\xi_{\mu}(\gamma^{\mu}\gamma_{5}\hspace{0.02cm}\theta)_{\alpha}
+\hspace{0.02cm} a_{5}\hspace{0.03cm}\xi_{5}(\gamma_{5}\hspace{0.02cm}\theta)_{\alpha}\Bigr\}.
\]
Here, on the right-hand side we have the correct expression as it was written out in (\ref{eq:4q}). On the left-hand side the first term also gives us the correct
answer, but the second term spoils a pattern. It does not enable us to express uniquely the $\psi_{\alpha}$ spinor in terms of real tensor variables (\ref{eq:4w})
and (\ref{eq:4e}). Here, it is necessary to start from the more general form
\[
\bar{\theta}_{\beta}\psi_{\alpha} = \frac{1}{4}\,
\Bigl\{\!\hspace{0.02cm}b_{1}W^{({\cal S})}\delta_{\alpha\beta} +\hspace{0.02cm} b_{2}W_{\mu}^{({\cal V})}(\gamma^{\mu})_{\alpha\beta}
+\hspace{0.03cm} b_3\!\,^{\ast}W_{\mu\nu}^{({\cal T})}(\sigma^{\mu\nu}\gamma_{5})_{\alpha\beta} +
b_{4}W_{\mu}^{({\cal A})}(\gamma^{\mu}\gamma_{5})_{\alpha\beta} + b_{5}W^{({\cal P})}(\gamma_{5})_{\alpha\beta}\Bigr\}.
\]
Contracting the last expression with $\theta_{\beta}$ or with $(\gamma_{5}\hspace{0.03cm}\theta)_{\beta}$ and so on we can define $\psi_{\alpha}$ in terms of the real tensors
(\ref{eq:4e}) and {\it complex} tensor quantities $W^{({\cal S})}, W^{({\cal V})}_{\mu},\ldots\,$. The last ones are connected with (\ref{eq:4w}) by relations of the form
\[
\hat{\xi}_{5} = \frac{1}{2}\,
\Bigl\{b_1W^{({\cal S})} +\, b_1^{*}\hspace{0.03cm}\bigl(W^{({\cal S})}\bigr)^{*}\Bigr\},
\]
etc. We thus come to the important conclusion that the more general {\it Dirac} spinor $\psi_{\alpha}$ cannot be presented in the form of an expansion in the real
tensor quantities (\ref{eq:4w}) only. Therefore if we wish to obtain a complete one-to-oneness of the mapping $(\psi, \bar{\psi}) \rightleftarrows (\xi_5,\,\xi_{\mu},\,\ldots)$
we must use some additional suggestions. The simplest of them is to restrict a class of the spinor $\psi_{\alpha}$ as well as $\theta_{\alpha}$ to Majorana one, i.e. to require that
they satisfy the condition
\[
\psi = \psi^{\hspace{0.02cm}c},\qquad \theta = \theta^{\hspace{0.02cm}c},
\]
where $\psi^c$ and $\theta^c$ are the charge-conjugate spinors. In this case, instead of (\ref{eq:4w}) we will have
\[
\begin{split}
&\xi_{\mu} = i\, {\rm Im}\hspace{0.02cm}\beta \,(\bar{\theta}\gamma_{\mu}\gamma_{5}\psi),\\
&\xi_{5} = {\rm Re}\hspace{0.02cm}\tilde{\beta} \,(\bar{\theta}\gamma_{5}\psi),
\end{split}
\]
and so on. If we now substitute these expressions into (\ref{eq:4q}), we obtain
\[
\begin{split}
\psi =\, &\Bigl\{i\kappa\,{\rm Im}\hspace{0.02cm}\beta (\bar{\theta}\gamma_{\mu}\gamma_{5}\hspace{0.02cm}\psi)(\gamma^{\mu}\gamma_{5\hspace{0.02cm}}\theta) +\,
\alpha\,{\rm Re}\hspace{0.02cm}\tilde{\beta}\,(\bar{\theta}\gamma_{5}\psi)(\gamma_{5\hspace{0.02cm}}\theta)\Bigr\} +\hspace{0.02cm}
\rho\hspace{0.04cm}{\rm Re}\hspace{0.02cm}s\hspace{0.02cm}(\bar{\theta}\sigma_{\mu\nu}\gamma_{5}\hspace{0.02cm}\psi)(\sigma^{\mu\nu}\gamma_{5}\hspace{0.02cm}\theta)\\
+\, &\Bigl\{\hat{\kappa}\,{\rm Re}\hspace{0.02cm}\tilde{\hat{\beta}}\,(\bar{\theta}\gamma_{\mu}\psi)(\gamma^{\mu}\theta)
+\, i \hspace{0.02cm}\hat{\alpha}\,{\rm Im}\hspace{0.02cm}\hat{\tilde{\beta}}(\bar{\theta}\hspace{0.02cm}\psi)\hspace{0.04cm}\theta\Bigr\}.
\end{split}
\]
Here, the terms containing a product of two anticommiting spinors $\theta_{\gamma} \theta_{\delta}$ formally vanish and thus equation (\ref{eq:3o}) disappears. Recall that
it is this equation that gives no way of construction of the one-to-one mapping $(\psi, \bar{\psi}) \rightleftharpoons (\xi_{\mu},\,\xi_5,\,\ldots)$ with the Dirac spinor.
Thus for the Majorana spinors we are left only with the system (\ref{eq:3y}) with known modification of the coefficient functions. Instead of relations (\ref{eq:3u}) now
we will have
\[
\begin{split}
\alpha\,{\rm Re}\hspace{0.02cm}\tilde{\beta} = i\,&\hat{\alpha}\,{\rm Im}\hspace{0.02cm}\hat{\tilde{\beta}},\qquad
i\kappa\,{\rm Im}\hspace{0.02cm}\beta = - \hat{\kappa}\,{\rm Re}\hspace{0.02cm}\hat{\beta},\\
\kappa\,{\rm Im}\hspace{0.02cm}\beta = \frac{i}{4(\bar{\theta}\theta)}\,,\qquad &\alpha\,{\rm Re}\hspace{0.02cm}\tilde{\beta}= \frac{1}{4(\bar{\theta}\theta)}\,,\qquad
\rho \,{\rm Re}\hspace{0.02cm}s = \frac{1}{8(\bar{\theta}\theta)}\,.
\end{split}
\]
\indent
Certainly, in the case of Majorana spinors the four-component formalism is not technically  optimal. Here, it is more adequately to use the two-component Weyl formalism
\begin{equation}
\psi_{\rm M} =
\left(
\begin{array}{c}
\psi_{\alpha}\\
\bar{\psi}^{\dot{\alpha}}
\end{array}
\right),
\qquad
\theta_{\rm M} =
\left(
\begin{array}{c}
\theta_{\alpha}\\
\bar{\theta}^{\dot{\alpha}}
\end{array}
\right),
\label{eq:4u}
\end{equation}
where now $\alpha, \dot{\alpha} = 1,2$. In two-component notations the mapping (\ref{eq:4q}) takes a form
\begin{equation}
\psi_{\alpha} = \Bigl[\kappa\,\xi_{\mu}(\sigma^{\mu})_{\alpha\dot{\alpha}\,}\bar{\theta}^{\dot{\alpha}} - \alpha\,\xi_{5\hspace{0.02cm}}\theta_{\alpha}\Bigr]
-i\rho\;^{\ast\!}\zeta_{\mu\nu}(\sigma^{\mu\nu})_{\alpha\,\cdot}^{\,\cdot\,\beta\,}\theta_{\beta}
+ \Bigl[\hat{\kappa}\,\hat{\xi}_{\mu}(\sigma^{\mu})_{\alpha\dot{\alpha}\,}\bar{\theta}^{\dot{\alpha}} + \hat{\alpha}\,\hat{\xi}_{5\hspace{0.02cm}}\theta_{\alpha}\Bigr].
\label{eq:4i}
\end{equation}
Here, in the notation of the book \cite{bailin_love_book} we have
\[
\sigma^{\mu}\equiv (I,\,\boldsymbol{\sigma}), \quad \bar{\sigma}^{\mu}\equiv (I,\,-\boldsymbol{\sigma}),
\quad
\sigma^{\mu\nu}\equiv \frac{1}{4}\,(\sigma^{\mu}\bar{\sigma}^{\nu} - \sigma^{\nu}\bar{\sigma}^{\mu}),
\quad
\bar{\sigma}^{\mu\nu}\equiv \frac{1}{4}\,(\bar{\sigma}^{\mu}\sigma^{\nu} - \bar{\sigma}^{\nu}\sigma^{\mu}),
\]
and the coefficients obey the conditions
\[
\kappa = -\kappa^{*},\quad \alpha = -\alpha^{*},\quad \rho = \rho^{*},\quad \hat{\kappa} = \hat{\kappa}^{*},\quad \hat{\alpha} = \hat{\alpha}^{*}.
\]
The term defining interaction of spin with background gauge field in these variables takes a form\footnote{Here, it must be understood that on the left-hand side the $(4\times 4)$
$\sigma$-matrix stands and on the right-hand side $(2 \times 2)$ $\sigma$-matrix does.}
\[
- \,\frac{eg}{4}\,Q^aF^{a}_{\mu\nu}(\hspace{0.02cm}\bar{\psi}\hspace{0.01cm}\hspace{0.02cm}\sigma^{\mu\nu}\psi) =
\frac{ieg}{2}\,Q^aF^{a}_{\mu\nu}\!
\left(\hspace{0.02cm}\psi^{\alpha\!}\hspace{0.01cm}(\sigma^{\mu\nu})_{\alpha\,\cdot}^{\cdot\,\beta\,}\psi_{\beta}\right)
+\hspace{0.03cm} \frac{ieg}{2}\,Q^aF^{a}_{\mu\nu}
\bigl(\hspace{0.01cm}\bar{\psi}^{\dot{\alpha}\!}\hspace{0.01cm}(\bar{\sigma}^{\mu\nu})_{\dot{\alpha}\,\cdot}^{\cdot\,\dot{\beta}\,}\bar{\psi}_{\dot{\beta}}\bigr),
\]
and the kinetic term is
\[
\frac{1}{2\hspace{0.03cm}i}\left(\frac{d\bar{\psi}}{d\tau}\,\psi \,-\, \bar{\psi}\,\frac{d\psi}{d\tau}\right) =
\frac{1}{2\hspace{0.03cm}i}\left(\frac{d\psi^{\alpha}}{d\tau}\,\psi_{\alpha} - \psi^{\alpha}\,\frac{d\psi_{\alpha}}{d\tau}\right) +\hspace{0.03cm}
\frac{1}{2\hspace{0.03cm}i}\left(\frac{d\bar{\psi}_{\dot{\alpha}}}{d\tau}\,\bar{\psi}^{\dot{\alpha}} - \bar{\psi}_{\dot{\alpha}}\,\frac{d\bar{\psi}^{\dot{\alpha}}}{d\tau}\right).
\]
Ipso facto, if an external fermion field is absent in the system, then a part of Lagrangian (\ref{eq:1r}) responsible for a description of spin degree of freedom can be written
in terms of only one variable, that is by means of the two-component spinor $\psi_{\alpha}$:
\[
L_{\rm spin} = \frac{1}{2\hspace{0.03cm}i}\left(\frac{d\psi^{\alpha}}{d\tau}\,\psi_{\alpha} - \psi^{\alpha}\,\frac{d\psi_{\alpha}}{d\tau}\right) +\hspace{0.03cm}
\frac{ieg}{2}\,Q^aF^{a}_{\mu\nu}\!\left(\hspace{0.02cm}\psi^{\alpha\!}\hspace{0.01cm}(\sigma^{\mu\nu})_{\alpha\,\cdot}^{\cdot\,\beta\,}\psi_{\beta}\right),
\]
and the (one-to-one) mapping into real tensor variables can be defined by (\ref{eq:4i}).\\
\indent
The situation qualitatively changes in the presence of the external fermion field $\Psi_{\alpha}^i(x)$ that in the general case should be considered as the Dirac one.
It is clear that such a field inevitably violates the representations (\ref{eq:4u}) for Majorana spinors. Here one can follow the next line. As known, a general Dirac
spinor $\psi_{\rm D}$ can be always written in terms of two Majorana spinors
\begin{equation}
\psi_{\rm D} = \psi_{\rm M}^{(1)} + \psi_{\rm M}^{(2)},
\label{eq:4o}
\end{equation}
where
\[
\psi_{\rm M}^{(1)} = \frac{1}{2}\,(\psi_{\rm D} + \psi_{\rm D}^{\hspace{0.02cm}c}),\qquad
\psi_{\rm M}^{(2)} = \frac{1}{2\hspace{0.02cm}i}\,(\psi_{\rm D} - \psi_{\rm D}^{\hspace{0.02cm}c}).
\]
Such a decomposition can be performed both for the background fermion field $\Psi_{\alpha}^i(x)$ and for spinors $\psi_{\alpha}$ and $\theta_{\alpha}$. Then for each
of Majorana spinors $\psi_{\rm M}^{(i)},\,i = 1,\,2$ we define own set of odd real currents $\xi_{\mu}^{(i)},\,\xi_5^{(i)},\,\xi_{\mu \nu}^{(i)},\,\ldots,$ such that
\begin{equation}
\begin{split}
&\psi_{\rm M}^{(1)} = \Bigl[\kappa_{1}\,\xi_{\mu}^{(1)}(\gamma^{\mu}\gamma_{5\hspace{0.02cm}}\theta^{(1)}_{\rm M}) +
\alpha_{1}\,\xi_5^{(1)}(\gamma_{5\hspace{0.02cm}}\theta^{(1)}_{\rm M})\Bigr] +\,
\rho_{1}\,^{\ast\!}\zeta_{\mu\nu}^{(2)}(\sigma^{\mu\nu}\gamma_{5\hspace{0.02cm}}\theta^{(1)}_{\rm M}) \,+ \dots\,,\\
&\psi_{\rm M}^{(2)} = \Bigl[\kappa_{2}\,\xi_{\mu}^{(2)}(\gamma^{\mu}\gamma_{5\hspace{0.02cm}}\theta^{(2)}_{\rm M}) +
\alpha_{2}\,\xi_5^{(2)}(\gamma_{5\hspace{0.02cm}}\theta^{(2)}_{\rm M})\Bigr] +\,
\rho_{2}\,^{\ast\!}\zeta_{\mu\nu}^{(2)}(\sigma^{\mu\nu}\gamma_{5\hspace{0.02cm}}\theta^{(2)}_{\rm M}) \,+ \dots\,.
\end{split}
\label{eq:4p}
\end{equation}
Running ahead, notice that in section\,8 on the construction in an explicit form of the auxiliary odd spinor $\theta_{\alpha}$ in terms of the field $\Psi_{\alpha}^i(x)$ in footnote
\ref{foot_8} it is pointed that there exist two possibilities of the choice of $\theta_{\alpha}$ as Majorana spinor, Eq.\,(\ref{eq:8e}). Thus in formulas (\ref{eq:4p}) the first expression in
(\ref{eq:8e}) can be used as $\theta_{\rm M}^{(1)}$ and the second one can be used as $\theta_{\rm M}^{(2)}$. In doing so at the cost of doubling real tensor quantities we can construct
sufficiently consistent description of dynamics of a color particle with half-integer spin, moving in background non-Abelian {\it Dirac} fermionic field.\\
\indent
One additional remark is in order. As known \cite{gershun_1979, howe_1988, marnelius_1989, bastianelli_2007}, for a description of particles with spin $s=n/2$ it is necessary to introduce
$n$ real tensor structures. Thus for $n = 2$, as it takes place in Eq.\,(\ref{eq:4p}) we deal with particle with spin 1. In principle this is sufficiently reasonable if recall
that our particle due to the interaction with the background fermion field changes its statistics, namely, it turns from Dirac spinor into a vector boson and vice versa. A description of such a
particle with variable spin in terms of a single set of functions, for example $(\xi_{\mu},\xi_5,\,\ldots)$, will be most likely insufficient.\\
\indent
Concluding this section, for completeness let us consider the problem of inverse substitution of the expression (\ref{eq:4q}) into real tensor variables (\ref{eq:4w}). A direct analysis
of such a substitution is too cumbersome. Following Zhelnorovich \cite{zhelnorovich_1972, zhelnorovich_book}, we proceed as follows. Let us multiply together two exact structures (\ref{eq:4r}) and (\ref{eq:4t}). On the left-hand side we will have an expression of the form
\begin{equation}
\bar{\theta}_{\beta}\theta_{\alpha}
\left(\hspace{0.02cm}{\rm e}^{i\phi}\hspace{0.02cm}\bar{\theta}_{\delta\hspace{0.02cm}}\psi_{\gamma} + \,{\rm e}^{-i\phi}\hspace{0.02cm}\bar{\psi}_{\delta\hspace{0.02cm}}\theta_{\gamma}\right).
\label{eq:4a}
\end{equation}
One can obtain every possible bilinear equations that relate two systems of tensor aggregates (\ref{eq:4w}) and (\ref{eq:4e}), making cross-contraction of
(\ref{eq:4a}) with respect to spinor indices with different combinations of $16$ independent generators of the Clifford algebra: $I$, $\gamma_{5}$, $\gamma_{\mu}$, $i \gamma_{\mu}\gamma_{5}$ and $\sigma_{\mu\nu}$. Let us consider, for example, convolution with the simplest structure
\[
\delta_{\beta\gamma}\hspace{0.01cm}\delta_{\delta\alpha}.
\]
By virtue of (\ref{eq:4y}) we will have
\begin{equation}
-\frac{1}{2}\,a_{1}(\bar{\theta}\theta)\hspace{0.03cm}\hat{\xi}_{5} =
\label{eq:4s}
\end{equation}
\[
=
\frac{1}{4}\hspace{0.01cm}\Bigl\{a_{1}(\bar{\theta}\theta)\hspace{0.03cm}\hat{\xi}_{5} + \hspace{0.02cm}a_{2}\hspace{0.04cm}(\bar{\theta}\gamma_{\mu}\hspace{0.02cm}\theta)\hspace{0.03cm}\hat{\xi}^{\mu}
+ \hspace{0.02cm}a_{3}\hspace{0.02cm}(\bar{\theta}\sigma_{\mu\nu}\gamma_{5\hspace{0.02cm}}\theta)\hspace{0.03cm}\!\,^{\ast\!}\zeta^{\mu\nu}
+ \hspace{0.02cm}a_{4}\hspace{0.02cm}(\bar{\theta}\gamma_{\mu}\gamma_{5\hspace{0.02cm}}\theta)\hspace{0.03cm}\xi^{\mu}
+  \hspace{0.02cm}a_{5}\hspace{0.03cm}(\bar{\theta}\gamma_{5\hspace{0.02cm}}\theta)\hspace{0.03cm}\xi_{5}\Bigr\}.
\]
This expression is identically obeyed by virtue of the initial construction. From the other hand, let us substitute (\ref{eq:4q}) into expression $\hat{\xi}_5$ in (\ref{eq:4w})
\[
\hat{\xi}_{5} = -\frac{1}{2}\bigl(\hspace{0.02cm}\hat{\tilde{\beta}}\hat{\alpha} + \hat{\tilde{\beta}}^{\ast}\hat{\alpha}^{\ast}\bigr)(\bar{\theta}\theta)\hspace{0.03cm}\hat{\xi}_{5}
-\frac{1}{2}\bigl(\hat{\tilde{\beta}}\hat{\kappa} + \hat{\tilde{\beta}}^{\ast}\hat{\kappa}^{\ast}\bigr)(\bar{\theta}\gamma_{\mu}\hspace{0.02cm}\theta)\hspace{0.03cm}\hat{\xi}^{\mu}\,-
\]
\[
-\,\frac{1}{2}\bigl(\hspace{0.02cm}\hat{\tilde{\beta}}\rho + \hat{\tilde{\beta}}^{\ast}\rho^{\ast}\bigr)(\bar{\theta}\sigma_{\mu\nu}\gamma_{5\hspace{0.02cm}}\theta)\hspace{0.05cm}\!\,^{\ast\!}\zeta^{\mu\nu}
-\frac{1}{2}\bigl(\hspace{0.02cm}\hat{\tilde{\beta}}\kappa - \hat{\tilde{\beta}}^{\ast}\kappa^{\ast}\bigr)(\bar{\theta}\gamma_{\mu}\gamma_{5\hspace{0.02cm}}\theta)\hspace{0.03cm}\xi^{\mu}
-\frac{1}{2}\bigl(\hspace{0.02cm}\hat{\tilde{\beta}}\alpha + \hat{\tilde{\beta}}^{\ast}\alpha^{\ast}\bigr)(\bar{\theta}\gamma_{5\hspace{0.02cm}}\theta)\hspace{0.03cm}\xi_{5}\Bigr\}.
\]
Comparing the above expression with (\ref{eq:4s}) we lead to the following relations for the coefficients
\[
\begin{split}
&\underline{\mbox{S\hspace{0.02cm}-\hspace{0.02cm}channel:}}\qquad\qquad   \hat{\tilde{\beta}}\hat{\alpha} + \hat{\tilde{\beta}}^{\ast}\hat{\alpha}^{\ast} = \frac{1}{(\bar{\theta}\theta)},\\
&\underline{\mbox{V-\hspace{0.02cm}channel:}}\qquad\qquad  \hat{\tilde{\beta}}\hat{\kappa} + \hat{\tilde{\beta}}^{\ast}\hat{\kappa}^{\ast} = \biggl(\frac{\hat{\tilde{\beta}}}{\hat{\beta}}\biggr)\frac{1}{(\bar{\theta}\theta)},\\
&\underline{\mbox{T-\hspace{0.02cm}channel:}}\qquad\qquad \hat{\tilde{\beta}}\rho + \hat{\tilde{\beta}}^{\ast}\rho^{\ast} =\frac{1}{2} \biggl(\frac{\hat{\tilde{\beta}}}{s}\biggr)\frac{1}{(\bar{\theta}\theta)},\\
&\underline{\mbox{A-\hspace{0.02cm}channel:}}\qquad\qquad \hat{\tilde{\beta}}\kappa - \hat{\tilde{\beta}}^{\ast}\kappa^{\ast} = -\biggl(\frac{\hat{\tilde{\beta}}}{\beta}\biggr)\frac{1}{(\bar{\theta}\theta)},\\
&\underline{\mbox{P-\hspace{0.02cm}channel:}}\qquad\qquad \hat{\tilde{\beta}}\alpha + \hat{\tilde{\beta}}^{\ast}\alpha^{\ast} = \biggl(\frac{\hat{\tilde{\beta}}}{\tilde{\beta}}\biggr)\frac{1}{(\bar{\theta}\theta)}.
\end{split}
\]
By using the relations (\ref{eq:3u}) obtained early, it is not difficult to show that the equation for S-channel is identically satisfied, while the remaining ones are reduced to
\begin{equation}
\begin{split}
&\underline{\mbox{V-\hspace{0.02cm}channel:}}\qquad\qquad  \biggl(\frac{\hat{\tilde{\beta}}^{\ast}}{\hat{\tilde{\beta}}}\biggr)
\hat{\beta}\hat{\kappa}^{\ast} = \frac{1}{2\hspace{0.01cm}(\bar{\theta}\theta)},\\
&\underline{\mbox{T-\hspace{0.02cm}channel:}}\qquad\qquad \biggl(\frac{\hat{\tilde{\beta}}^{\ast}}{\hat{\tilde{\beta}}}\biggr)
s\hspace{0.01cm}\rho^{\ast} = \frac{1}{4\hspace{0.01cm}(\bar{\theta}\theta)},\\
&\underline{\mbox{A-\hspace{0.02cm}channel:}}\qquad\qquad \biggl(\frac{\hat{\tilde{\beta}}^{\ast}}{\hat{\tilde{\beta}}}\biggr)
\beta\!\hspace{0.035cm}\kappa^{\ast} = -\hspace{0.02cm}\frac{1}{2\hspace{0.01cm}(\bar{\theta}\theta)},\\
&\underline{\mbox{P-\hspace{0.02cm}channel:}}\qquad\qquad \biggl(\frac{\hat{\tilde{\beta}}^{\ast}}{\hat{\tilde{\beta}}}\biggr)
\tilde{\beta}\alpha^{\ast} = \frac{1}{2\hspace{0.01cm}(\bar{\theta}\theta)}.
\end{split}
\label{eq:4d}
\end{equation}
The relations of the form
\begin{equation}
\hat{\beta}\hat{\kappa}^{\ast} = -\beta\kappa^{\ast} = 2\hspace{0.01cm}s\hspace{0.01cm}\rho^{\ast} = \tilde{\beta}\alpha^{\ast}
\label{eq:4f}
\end{equation}
will be particular consequence of the expressions obtained. In fact these relations (as however and (\ref{eq:4d})) are relations for phases of the coefficient functions.
However, it is possible to show that the relations (\ref{eq:4f}) contradict both the system of equations for phases (\ref{eq:3p}) and (\ref{eq:3a}). This is not surprising,
since according to (\ref{eq:4y}) the phases of coefficient functions must be identical (accurate within shift over $\pi$), that evident contradict, for example, (\ref{eq:3s}).\\
\indent
Thus in spite of the fact that we do not have contradiction with the main system of algebraic equations (\ref{eq:3y}), which follows from a direct substitution of (\ref{eq:4w}) into
(\ref{eq:4q}), here we again have contradiction in equations for phases of the desired coefficient functions.\\
\indent
In Appendix C the problem of inverse mapping of the bilinear combination $\xi^{\mu}\xi^{\nu}$ is also considered.

\section{\bf Map $(\psi, \bar{\psi}) \rightarrow (\xi_{\mu},\xi_{5})$ subject to contribution of odd pseudo\-tensor $^{\ast\!}\zeta_{\mu \nu}$}
\setcounter{equation}{0}

In section 2 we have considered the simplest map of the form $(\psi, \bar{\psi}) \rightarrow (\xi_{\mu}, \xi_5)$ and shown evident contradiction in algebraic equations
for unknown coefficient function $\kappa$ in (\ref{eq:2q}). Thus, under a mapping of the kinetic term (\ref{eq:2u}) we have obtained the following condition:
\begin{equation}
|\kappa|^2(\bar\theta\theta)=\frac{1}{2}\,,
\label{eq:5q}
\end{equation}
while in the analysis of the mapping of the interaction term, Eq.\,(\ref{eq:2d}), it was derived the condition
\begin{equation}
|\kappa|^2(\bar\theta\theta) = -1.
\label{eq:5w}
\end{equation}
As was mentioned in section\,2, one reason of this contradiction is that such a simple choice of the map (\ref{eq:2q}) is not complete, i.e. we lose
some additional contributions.\\
\indent
Let us consider a more general expansion of the form (\ref{eq:4q}). It is evident that the contribution with odd vector $\hat{\xi}_{\mu}$ (and especially, the contribution
with odd scalar $\hat{\xi}_5$) hardly will solve the problem. At the end of section 3 it was mentioned that the vector $\hat{\xi}_{\mu}$ should be most likely proportional
to the four-velocity $\dot{x}_{\mu}$, where the coefficient of proportionality is some odd scalar. Therefore the interaction terms involving $\hat{\xi}_{\mu}$ can never represent
a purely spin interaction of a particle with a background field. The term in (\ref{eq:4q}) containing the odd pseudotensor $^{\ast}\zeta_{\mu \nu}$ is a different matter. According to
(\ref{eq:3d}) it can be expressed in terms of the pseudovector $\xi_{\mu}$ and there is a good reason to believe that it can generate additional contributions of the required form both in
the kinetic terms (\ref{eq:2i}) and in force term in (\ref{eq:2s}). These new contributions may allow to overcome the above difficulty. Thus we extend the map (\ref{eq:2q}), (\ref{eq:2w})
adding the pseudotensor term
\begin{equation}
\begin{split}
&\psi = \kappa\,\xi_{\mu}(\gamma^{\mu}\gamma_{5\hspace{0.02cm}}\theta)\hspace{0.02cm} +
\hspace{0.02cm}\rho\hspace{0.02cm}\,^{\ast\!}\zeta_{\mu\nu}(\sigma^{\mu\nu}\gamma_{5\hspace{0.02cm}}\theta) +\, \ldots\,,\\
&\bar\psi=-\kappa^{*}(\bar\theta\gamma_{5\hspace{0.02cm}}\gamma^{\mu})\hspace{0.02cm}\xi_{\mu} -
\rho^{\ast}(\bar{\theta}\gamma_{5\hspace{0.02cm}}\sigma^{\mu\nu})\,^{\ast\!}\zeta_{\mu\nu} - \ldots\,,\\
\end{split}
\label{eq:5e}
\end{equation}
where the dots refers to the term with $\xi_5$ which doesn't play any role in this section and therefore it will be omitted.\\
\indent
At first, let us consider the mapping of the kinetic term (\ref{eq:2u}). An additional contribution of the pseudotensor only to this mapping has a form
\[
-\, i\rho^{\ast}(\bar{\theta}\gamma_{5\hspace{0.02cm}}\sigma^{\mu\nu})\,^{\ast\!}\zeta_{\mu\nu}\!
\biggl[\,\frac{d\rho}{d\tau}\,\,^{\ast\!}\zeta_{\lambda\sigma}(\sigma^{\lambda\sigma}\gamma_{5\hspace{0.02cm}}\theta) +
\rho\,\frac{d\,^{\ast\!}\zeta_{\lambda\sigma}}{d\tau}\,(\sigma^{\lambda\sigma}\gamma_{5\hspace{0.02cm}}\theta) +
\rho\,^{\ast\!}\zeta_{\lambda\sigma}\Bigl(\sigma^{\lambda\sigma}\gamma_{5\hspace{0.02cm}}\,\frac{d\theta}{d\tau}\Bigr) \biggr].
\]
Here, our concern is only with the second term in square brackets. Let us written out it separately
\begin{equation}
-\,i\hspace{0.02cm}|\rho|^2(\bar{\theta}\sigma^{\mu\nu}\sigma^{\lambda\sigma}\theta)\,^{\ast\!}\zeta_{\mu\nu}\,\frac{d\,^{\ast\!}\zeta_{\lambda\sigma}}{d\tau}\,.
\label{eq:5r}
\end{equation}
Further, we make use of the formula for the product of two $\sigma$-matrices, Eq.\,(B.3) in Appendix B. Here, we restrict again our consideration to the contribution from only the identity
spinor matrix to the right-hand side of (B.3). This gives us, instead of (\ref{eq:5r})
\[
-\,2\hspace{0.02cm}i\hspace{0.02cm}|\rho|^{2\,}(\bar{\theta}\theta)\biggl(\!\,^{\ast\!}\zeta_{\mu\nu}\,\frac{d\,^{\ast\!}\zeta^{\mu\nu}}{d\tau}\biggr) +\;\ldots\,.
\]
Finally, in the expression obtained we use an explicit form of $^{\ast}\zeta_{\mu \nu}$ presented by the four-velocity $\dot{x}_{\mu}$ and odd
pseudovector $\xi_{\mu}$:
\begin{equation}
\,^{\ast\!}\zeta_{\mu\nu} = \dot{x}_{\mu\hspace{0.02cm}}\xi_{\nu} - \dot{x}_{\nu\hspace{0.02cm}}\xi_{\mu}.
\label{eq:5t}
\end{equation}
Taking into account the above mentioned, we obtain the contribution of interest to us of the pseudotensor to kinetic term
\begin{equation}
-\,2\hspace{0.02cm}i\hspace{0.02cm}|\rho|^{2\,}(\bar{\theta}\theta)\biggl\{2\hspace{0.02cm}\dot{x}^2
\xi_{\mu}\,\frac{d\xi^{\mu}}{d\tau} \,+\, (\dot{x}_{\mu\hspace{0.02cm}}\ddot{x}_{\nu} \,-\, \dot{x}_{\nu\hspace{0.02cm}}\ddot{x}_{\mu})\,\xi^{\mu}\xi^{\nu} \,+\,
(\dot{\xi}_{\mu\hspace{0.02cm}}\xi_{\nu} \,+\, \dot{\xi}_{\nu\hspace{0.02cm}}\xi_{\mu})\,\dot{x}^{\mu\hspace{0.02cm}}\dot{x}^{\nu}
\biggr\} +\;\ldots\,.
\label{eq:5y}
\end{equation}
The first term here has a required form. It is necessary to add the given expression to (\ref{eq:2i}).\\
\indent
Now we turn to analysis of the interaction term of a form
\begin{equation}
- \,\frac{eg}{4}\;Q^aF^{a}_{\mu\nu}(\bar{\psi}\sigma^{\mu\nu}\psi).
\label{eq:5u}
\end{equation}
Let us consider the contribution of pseudotensor $^{\ast}\zeta_{\mu \nu}$ to interaction, i.e. in the expression above we set
\[
\psi \sim \rho\hspace{0.02cm}\,^{\ast\!}\zeta_{\mu\nu}(\sigma^{\mu\nu}\gamma_{5\hspace{0.02cm}}\theta),\qquad
\bar\psi \sim - \rho^{\ast}\hspace{0.02cm}(\bar{\theta}\gamma_{5\hspace{0.02cm}}\sigma^{\mu\nu})\,^{\ast\!}\zeta_{\mu\nu}.
\]
Then the spin tensor in (\ref{eq:5u}) takes a form
\begin{equation}
\frac{1}{2}\,(\bar{\psi}\sigma^{\mu\nu}\psi) = \frac{1}{2}\,|\rho|^{2}(\bar{\theta}\sigma^{\rho\delta}\sigma^{\mu\nu}\sigma^{\lambda\sigma}\theta)
\,^{\ast\!}\zeta_{\rho\delta}\,^{\ast\!}\zeta_{\lambda\sigma}.
\label{eq:5i}
\end{equation}
Here, it is necessary to use an expansion of the product of three $\sigma$-matrices. The general form of such an expansion is given in Appendix B, Eq.\,(B.4).
We will need only take the first term on the right-hand side of (B.4). Substituting this term instead of the product of three $\sigma$-matrices in (\ref{eq:5i}),
taking into account (\ref{eq:5t}), and collecting like terms, we obtain the following simple expression
\begin{equation}
\frac{1}{2}\,(\bar{\psi}\sigma^{\mu\nu}\psi) =
4\hspace{0.02cm}i\hspace{0.02cm}|\rho|^{2\hspace{0.02cm}}(\bar{\theta}\theta)\hspace{0.04cm}\Bigl\{\dot{x}^2 \xi^{\mu}\xi^{\nu} +
(\dot{x}^{\mu\hspace{0.02cm}}\xi^{\nu} - \dot{x}^{\nu\hspace{0.02cm}}\xi^{\mu})\hspace{0.02cm}(\dot{x}\cdot\xi)\Bigr\}
+\;\ldots\,.
\label{eq:5o}
\end{equation}
Here the required contribution proportional to $\xi^{\mu} \xi^{\nu}$ arises. It needs to be added to the first term in (\ref{eq:2s}). With allowance made for the new
additional terms, the algebraic equations (\ref{eq:5q}) and (\ref{eq:5w}) go over into the following equations
\begin{equation}
\begin{split}
&|\kappa|^2(\bar\theta\theta) + 4\hspace{0.02cm}\dot{x}^{2}\hspace{0.02cm}|\rho|^{2\hspace{0.01cm}}(\bar{\theta}\theta) = \frac{1}{2}\,,\\
&|\kappa|^2(\bar\theta\theta) + 4\hspace{0.02cm}\dot{x}^{2}\hspace{0.02cm}|\rho|^{2\hspace{0.01cm}}(\bar{\theta}\theta) = -1.
\end{split}
\label{eq:5p}
\end{equation}
Unfortunately, the system still remains inconsistent. The additional terms have appeared in symmetric fashion, in spite of the fact that they resulted from a much
different terms in the initial Lagrangian (\ref{eq:1r}).\\
\indent
Let us try to consider mixed contributions, i.e. contributions of the product of the pseudovec\-tor term with the pseudotensor one in the map (\ref{eq:5e}). For kinetic term (\ref{eq:2u})
the mixed contribu\-tion has the following form:
\[
i\hspace{0.02cm}\kappa^{*\!}\rho\hspace{0.03cm}(\bar{\theta}\gamma^{\mu}\sigma^{\nu\lambda}\theta)\,\xi_{\mu}\,\frac{d\,^{\ast\!}\zeta_{\nu\lambda}}{d\tau} \;+\,
i\hspace{0.02cm}\kappa\hspace{0.03cm}\rho^{*}\hspace{0.01cm}(\bar{\theta}\sigma^{\nu\lambda}\gamma^{\mu}\theta)\,^{\ast\!}\zeta_{\nu\lambda}\,\frac{d\xi_{\mu}}{d\tau}\,.
\]
Taking into account the expansion formulas (B.2) for the product of $\gamma$- and $\sigma$-matrices and also (\ref{eq:5t}), from the last expression we get
\begin{equation}
\begin{split}
&-2\hspace{0.04cm}(\kappa^{*\hspace{0.01cm}}\!\rho - \kappa\hspace{0.03cm}\rho^{*})\!\left[(\bar{\theta}\gamma^{\nu}\theta)\hspace{0.02cm}\dot{x}_{\nu}\right]\!
\xi_{\mu}\,\frac{d\xi^{\mu}}{d\tau} \;-\,
2\hspace{0.01cm}\kappa^{*\hspace{0.01cm}}\!\rho\!\left[(\bar{\theta}\gamma^{\nu}\theta)\hspace{0.02cm}\xi_{\nu}\right]\!(\xi\cdot\ddot{x})\\
&+2\hspace{0.04cm}\Bigl\{\!\kappa^{*\hspace{0.01cm}}\!\rho\hspace{0.02cm}(\bar{\theta}\gamma^{\nu}\theta)\hspace{0.02cm}\dot{x}^{\mu} \,-\, \kappa\hspace{0.03cm}\rho^{*}(\bar{\theta}\gamma^{\mu}\theta)\hspace{0.02cm}\dot{x}^{\nu}\Bigr\}
\,\xi_{\mu}\,\frac{d\xi_{\nu}}{d\tau}\\
&-2\hspace{0.04cm}i\hspace{0.03cm}\kappa^{*\hspace{0.01cm}}\!\rho\hspace{0.04cm}\epsilon^{\mu\nu\lambda\sigma}\xi_{\mu}\xi_{\nu}
(\bar{\theta}\gamma_{\lambda}\gamma_{5\hspace{0.01cm}}\theta)\hspace{0.03cm}\ddot{x}_{\sigma} -
2\hspace{0.04cm}i\hspace{0.01cm}(\kappa^{*\hspace{0.01cm}}\!\rho - \kappa\hspace{0.03cm}\rho^{*})\hspace{0.04cm}\epsilon^{\mu\nu\lambda\sigma}
\hspace{0.04cm}\xi_{\mu}\,\frac{d\xi_{\nu}}{d\tau}\,(\bar{\theta}\gamma_{\lambda}\gamma_{5\hspace{0.01cm}}\theta)\hspace{0.04cm}\dot{x}_{\sigma}.
\end{split}
\label{eq:5a}
\end{equation}
Here, the first term has the wanted structure: $\xi_{\mu\hspace{0.02cm}}\dot{\xi}^{\mu}$. However, in contrast to the direct contribu\-tions (\ref{eq:2i}) and (\ref{eq:5y}), this term
has more tangled coefficient function proportional to convolution $(\bar{\theta} \gamma^{\nu} \theta)\hspace{0.04cm}\dot{x}_{\nu}$. It hints at that at least phases of the coefficient
functions $\kappa$ and $\varrho$ can depend on four-velocity of particle $\dot{x}_{\mu}$. This is not quite clear from the physical point of view.\\
\indent
Let us define an mixed contribution in the mapping of the tensor of spin $\frac{1}{2}(\bar{\psi} \sigma^{\mu \nu} \psi)$. Here, we have the following expression of the mixed type
\begin{equation}
-\,\Bigl\{\!\hspace{0.03cm}\kappa^{*\!}\rho\hspace{0.05cm}(\bar{\theta}\gamma^{\rho}\sigma^{\mu\nu}\sigma^{\lambda\sigma}\theta) \,-
\kappa\hspace{0.05cm}\rho^{*}\hspace{0.01cm}(\bar{\theta}\sigma^{\lambda\sigma}\sigma^{\mu\nu}\gamma^{\rho}\theta)\Bigr\}
\,\xi_{\rho}\,^{\ast\!}\zeta_{\lambda\sigma}.
\label{eq:5s}
\end{equation}
For the product of spinor matrices one needs to use sequentially expansions (B.2) and (B.3). If we keep only the required terms containing the product $\xi^{\mu}\xi^{\nu}$,
then the calculations lead to the following expression
\begin{equation}
\frac{1}{2}\,(\bar{\psi}\sigma^{\mu\nu}\psi)\,\Bigl|_{\rm mix.} =
2\hspace{0.04cm}(\kappa^{*\hspace{0.01cm}}\!\rho - \kappa\hspace{0.03cm}\rho^{*})\!\left[(\bar{\theta}\gamma^{\lambda}\theta)\hspace{0.02cm}\dot{x}_{\lambda}\right]\!
\xi^{\mu}\xi^{\nu} +\;\ldots\,.
\label{eq:5d}
\end{equation}
If we now add the additional contributions from the mixed terms in (\ref{eq:5a}) and (\ref{eq:5d}) to algebraic system (\ref{eq:5p}), then we again obtain a perfect coincidence
of the left-hand sides of these equations
\begin{equation}
\begin{split}
&|\kappa|^2(\bar\theta\theta) +\hspace{0.01cm} 4\hspace{0.02cm}\dot{x}^{2}\hspace{0.02cm}|\rho|^{2\hspace{0.01cm}}(\bar{\theta}\theta)\hspace{0.01cm}
-2\hspace{0.02cm}i\hspace{0.01cm}(\kappa^{*\hspace{0.01cm}}\!\rho - \kappa\hspace{0.03cm}\rho^{*})\!\left[(\bar{\theta}\gamma^{\lambda}\theta)\hspace{0.02cm}\dot{x}_{\lambda}\right]
= \frac{1}{2}\,,\\
&|\kappa|^2(\bar\theta\theta) +\hspace{0.01cm} 4\hspace{0.02cm}\dot{x}^{2}\hspace{0.02cm}|\rho|^{2\hspace{0.01cm}}(\bar{\theta}\theta)\hspace{0.01cm}
-2\hspace{0.02cm}i\hspace{0.01cm}(\kappa^{*\hspace{0.01cm}}\!\rho - \kappa\hspace{0.03cm}\rho^{*})\!\left[(\bar{\theta}\gamma^{\lambda}\theta)\hspace{0.02cm}\dot{x}_{\lambda}\right]
= -1
\end{split}
\label{eq:5f}
\end{equation}
and the system remains inconsistent.\\
\indent
In this situation we can take the last step. We turn to expression (\ref{eq:5y}) for contribution to the kinetic term $i\bar{\psi}\left(d\psi/d\tau\right)$ from the pseudotensor
$^{\ast}\!\hspace{0.02cm}\zeta_{\mu \nu}$. In the first equation of algebraic system (\ref{eq:5f}) we have taken into account the contribution only from the first term in braces in
(\ref{eq:5y}). However, we can single out another required contribution from the third term of the following form:
\begin{equation}
2\hspace{0.03cm}\dot{x}^{\mu}\dot{x}^{\nu}\dot{\xi}_{\mu\hspace{0.02cm}}\xi_{\nu} = a^{\mu\nu}\xi_{\mu}\hspace{0.02cm}\frac{d\xi_{\nu}}{d\tau},
\label{eq:5g}
\end{equation}
where $a^{\mu\nu}\equiv -2\hspace{0.02cm}\dot{x}^{\mu\hspace{0.02cm}}\dot{x}^{\nu}$. Let us present the function $a^{\mu\nu}$ in the following form:
\begin{equation}
a^{\mu\nu} = \frac{1}{4}\hspace{0.02cm}g^{\mu\nu} a_{\lambda}^{\lambda} + (\mbox{traceless piece})
= -\frac{1}{2}\,g^{\mu\nu}\dot{x}^2 +\, \dots\,.
\label{eq:5h}
\end{equation}
With allowance for the last expression from (\ref{eq:5y}) we single out more accurately the required contribution to kinetic term
\begin{equation}
i\hspace{0.02cm}\bar{\psi}\,\frac{d\psi}{d\tau} \sim
-\,2\hspace{0.02cm}i\hspace{0.02cm}|\rho|^{2\hspace{0.01cm}}(\bar{\theta}\theta)
\biggl\{\!\hspace{0.02cm}2\hspace{0.03cm}\dot{x}^2\xi_{\mu}\,\frac{d\xi^{\mu}}{d\tau} \,+\, \biggl(-\frac{1}{2}\biggr)\dot{x}^2\xi_{\mu}\hspace{0.02cm}\frac{d\xi^{\mu}}{d\tau} +\, \ldots
\biggr\}
\label{eq:5j}
\end{equation}
\[
= -\,3\hspace{0.02cm}i\hspace{0.02cm}|\rho|^{2\hspace{0.02cm}}(\bar{\theta}\theta)\hspace{0.03cm}\dot{x}^2\xi_{\mu}\hspace{0.02cm}\frac{d\xi^{\mu}}{d\tau} +\, \ldots\,.
\]
\indent
Furthermore, let us consider a mixed contribution of $\xi_{\mu}$ and $^{\ast}\zeta_{\mu\nu}$ to the kinetic term, i.e. the expression (\ref{eq:5a}).
In the first equation of system (\ref{eq:5f}) we have also taken into account the contribution only from the first term in (\ref{eq:5a}).
However, it is possible to single out another relevant contribution from the third term in (\ref{eq:5a}) following the same rule (\ref{eq:5g}) and (\ref{eq:5h}),
where we should set
\[
a^{\mu\nu}\equiv
2\hspace{0.04cm}\Bigl\{\!\kappa^{*\hspace{0.01cm}}\!\rho\hspace{0.02cm}(\bar{\theta}\gamma^{\nu}\theta)\hspace{0.02cm}\dot{x}^{\mu} \,-\, \kappa\hspace{0.02cm}\rho^{*}(\bar{\theta}\gamma^{\mu}\theta)\hspace{0.02cm}\dot{x}^{\nu}\Bigr\}
= \frac{1}{2}\,g^{\mu\nu}(\kappa^{*\hspace{0.01cm}}\!\rho - \kappa\hspace{0.03cm}\rho^{*})\!\left[(\bar{\theta}\gamma^{\lambda}\theta)\hspace{0.02cm}\dot{x}_{\lambda}\right]
+\,\ldots\,.
\]
Thus a more accurate analysis for the mixed contribution leads to
\begin{equation}
i\hspace{0.02cm}\bar{\psi}\,\frac{d\psi}{d\tau}\,\bigg|_{\rm mix.} \!\sim\,
-\frac{3}{2}\,(\kappa^{*\hspace{0.01cm}}\!\rho - \kappa\hspace{0.03cm}\rho^{*})\!\left[(\bar{\theta}\gamma^{\lambda}\theta)\hspace{0.02cm}\dot{x}_{\lambda}\right]
\xi_{\mu}\hspace{0.02cm}\frac{d\xi_{\mu}}{d\tau}
\,+\,\ldots\,.
\label{eq:5k}
\end{equation}
\indent
Now we turn our attention to consideration of the spin tensor $\frac{1}{2}(\bar{\psi}\sigma^{\mu\nu}\psi)$. Fist we consider the direct contribution to the spin tensor from
$^{\ast}\!\zeta_{\mu \nu}$, i.e. Eq.\,(\ref{eq:5o}). In the second equation of the system (\ref{eq:5f}) we have taken into account the contribution only from the first term
in braces in (\ref{eq:5l}). However, here there exist a required contribution which is contained in the second term. To single out the relevant contribution we rewritten
(\ref{eq:5o}) in the following form:
\begin{equation}
\frac{1}{2}\,(\bar{\psi}\sigma^{\mu\nu}\psi) = 8\hspace{0.02cm}i\hspace{0.02cm}|\rho|^{2\hspace{0.02cm}}(\bar{\theta}\theta)\hspace{0.03cm}\dot{x}^2 \xi^{\mu}\xi^{\nu}
+ \frac{1}{2}\,b^{\hspace{0.02cm}\mu\nu\lambda\sigma}\xi_{\lambda}\xi_{\sigma},
\label{eq:5l}
\end{equation}
where
\begin{equation}
b^{\hspace{0.02cm}\mu\nu\lambda\sigma}\! = 4\hspace{0.02cm}i\hspace{0.02cm}|\rho|^{2}(\bar{\theta}\theta)\hspace{0.04cm}\Bigl\{\!\hspace{0.02cm}
\bigl(\hspace{0.02cm}\dot{x}^{\mu}\dot{x}^{\sigma}g^{\nu\lambda} - \dot{x}^{\nu}\dot{x}^{\sigma}g^{\mu\lambda}\bigr) -
\bigl(\hspace{0.02cm}\dot{x}^{\mu}\dot{x}^{\lambda}g^{\nu\sigma} - \dot{x}^{\nu}\dot{x}^{\lambda}g^{\mu\sigma}\bigr)\hspace{0.03cm}\!\Bigr\}.
\label{eq:5z}
\end{equation}
From the previous expression we pick out the required term by the following rule:
\begin{equation}
b^{\hspace{0.02cm}\mu\nu\lambda\sigma}\! = b\hspace{0.01cm}
\bigl(\hspace{0.02cm}g^{\mu\lambda}g^{\nu\sigma} - g^{\mu\sigma}g^{\nu\lambda}\bigr) + (\mbox{and all}),
\label{eq:5x}
\end{equation}
where
\[
b = \frac{1}{12}\;g_{\mu\lambda}\hspace{0.01cm}g_{\nu\sigma}\hspace{0.02cm}b^{\hspace{0.02cm}\mu\nu\lambda\sigma}
\]
or for the specific case (\ref{eq:5z}) we take
\[
b = -2\hspace{0.02cm}i\hspace{0.02cm}|\rho|^{2}(\bar{\theta}\theta)\hspace{0.02cm}\dot{x}^{2}.
\]
Thus the analysis performed enables us to obtain instead of (\ref{eq:5o}) the more exact expression
\begin{equation}
\frac{1}{2}\,(\bar{\psi}\sigma^{\mu\nu}\psi) =
i\hspace{0.02cm}|\rho|^{2\hspace{0.02cm}}(\bar{\theta}\theta)\hspace{0.04cm}\Bigl\{\!\hspace{0.03cm}4\hspace{0.04cm}\dot{x}^2 \xi^{\mu}\xi^{\nu} -
\dot{x}^2\bigl(\hspace{0.02cm}g^{\mu\lambda}g^{\nu\sigma} - g^{\mu\sigma}g^{\nu\lambda}\bigr)\xi_{\lambda}\xi_{\sigma}
+\,\ldots\,\Bigr\}
\label{eq:5c}
\end{equation}
\[
= 2\hspace{0.02cm}i\hspace{0.02cm}|\rho|^{2\hspace{0.02cm}}(\bar{\theta}\theta)\hspace{0.02cm}\dot{x}^2\xi^{\mu}\xi^{\nu}
+\,\ldots\,.
\]
\indent
Finally, let us analyze a mixed contribution to the spin tensor, expression (\ref{eq:5s}). A somewhat more cumbersome analysis within the presentation
(\ref{eq:5l})\,--\,(\ref{eq:5x}) allows us to single out other contributions similar to (\ref{eq:5d}). In doing so we obtain instead of (\ref{eq:5d}) the more exact expression
\begin{equation}
\frac{1}{2}\,(\bar{\psi}\sigma^{\mu\nu}\psi)\,\Bigl|_{\rm mix.} =
\frac{1}{2}\hspace{0.04cm}(\kappa^{*\hspace{0.01cm}}\!\rho - \kappa\hspace{0.03cm}\rho^{*})\!\left[(\bar{\theta}\gamma^{\lambda}\theta)\hspace{0.02cm}\dot{x}_{\lambda}\right]\!
\xi^{\mu}\xi^{\nu}+\;\ldots\,.
\label{eq:5v}
\end{equation}
\indent
Taking into account all the above-obtained expression (\ref{eq:5j}), (\ref{eq:5k}), (\ref{eq:5c}) and (\ref{eq:5v}), we derived instead of the system (\ref{eq:5f})
\begin{equation}
\begin{split}
&|\kappa|^2(\bar\theta\theta) +\hspace{0.01cm} 3\hspace{0.02cm}\dot{x}^{2}\hspace{0.02cm}|\rho|^{2\hspace{0.01cm}}(\bar{\theta}\theta)\hspace{0.01cm}
-\frac{3\hspace{0.02cm}i}{2}\hspace{0.02cm}(\kappa^{*\hspace{0.01cm}}\!\rho - \kappa\hspace{0.03cm}\rho^{*})\!\left[(\bar{\theta}\gamma^{\lambda}\theta)\hspace{0.02cm}\dot{x}_{\lambda}\right]
= \frac{1}{2}\,,\\
&|\kappa|^2(\bar\theta\theta) +\hspace{0.01cm} 2\hspace{0.02cm}\dot{x}^{2}\hspace{0.02cm}|\rho|^{2\hspace{0.01cm}}(\bar{\theta}\theta)\hspace{0.01cm}
-\frac{i}{2}\,(\kappa^{*\hspace{0.01cm}}\!\rho - \kappa\hspace{0.03cm}\rho^{*})\!\left[(\bar{\theta}\gamma^{\lambda}\theta)\hspace{0.02cm}\dot{x}_{\lambda}\right]
= -1.
\end{split}
\label{eq:5b}
\end{equation}
The system obtained is consistent at least at a formal level. If we make use of the representation:
$\rho = |\rho|\hspace{0.02cm}{\rm e}^{i\varphi}$ and $\kappa = |\kappa|\hspace{0.02cm}{\rm e}^{i\phi}$,
then for the coefficient function $\kappa^{*\hspace{0.01cm}}\!\rho - \kappa\hspace{0.03cm}\rho^{*}$ we will have
\[
\kappa^{*\hspace{0.01cm}}\!\rho - \kappa\hspace{0.03cm}\rho^{*} = 2\hspace{0.03cm}i|\rho||\kappa|\sin(\varphi -\phi).
\]
Thus the algebraic system (\ref{eq:5b}) relates among themselves three unknown functions: $|\rho|$, $|\kappa|$ and difference of phases $\varphi - \phi$.

\section{\bf Connection with Polyakov's Lagrangian}
\setcounter{equation}{0}

In the previous section we have considered the additional contributions which is generated by the pseudo\-tensor term $^{\ast}\zeta_{\mu \nu}$ in the form of (\ref{eq:5t})
to one-sided map $(\psi, \bar{\psi}) \rightarrow (\xi_{\mu}, \xi_5)$. It was shown that besides the terms of the usual type a number of the rather interesting additional
contributions in the mapped Lagrangian appears in both the kinetic term (\ref{eq:5y}) and interaction term, Eqs.\,(\ref{eq:5u}), (\ref{eq:5o}). In this section
we would like to consider these contributions in more detail and trace rather unusual connection of certain of them with expressions (A.2), (A.3) in the form
presented by A.M. Polyakov \cite{polyakov_book}.\\
\indent
First we consider the kinetic term (\ref{eq:5y}). We write out an initial functional integral for a supersymmetric part in which the action is defined by Lagrangian (A.1)
without regard for the interacti\-on terms with gauge external field
\begin{equation}
\begin{split}
Z = \!\int\!{\cal D}x_{\mu}{\cal D}\xi_{\mu}{\cal D}\chi\hspace{0.02cm}{\cal D}e\hspace{0.02cm}{\cal D}\xi_{5}
\,\exp\hspace{0.02cm}\Biggl\{-\!\int\limits_{0}^{1}d\tau\biggl[
&-\!\displaystyle\frac{1}{2\hspace{0.02cm}e}\,\dot{x}_{\mu\hspace{0.02cm}}\dot{x}^{\mu} - \displaystyle\frac{i}{2}\,\xi_{\mu\hspace{0.035cm}}\dot{\xi}^{\mu} - \displaystyle\frac{e}{2}\,m^2\\
&+ \displaystyle\frac{i}{2\hspace{0.02cm}e}\,\chi\hspace{0.03cm}\dot{x}_{\mu\hspace{0.02cm}}\xi^{\mu} +
\displaystyle\frac{i}{2}\,\xi_{5\,}\dot{\xi_{5}} + \displaystyle\frac{i}{2}\,m\chi\hspace{0.02cm}\xi_{5}\biggr]\Biggr\}.
\end{split}
\label{eq:6q}
\end{equation}
Furthermore, we follow the reasoning by A.M. Polyakov \cite{polyakov_book}. Our first step is the functional integrating over $\xi_5$ according to formula
\[
\!\int{\cal D}\xi_{5}\,\exp\biggl\{-\!\int\limits_{0}^{1}d\tau\biggl[\,\displaystyle\frac{i}{2}\,\xi_{5\hspace{0.02cm}}\dot{\xi_{5}} + \displaystyle\frac{i}{2}\,m\chi\hspace{0.02cm}\xi_{5}\biggr]\biggr\}
=\,
\exp\hspace{0.02cm}\biggl\{-\frac{im^2}{\!16}\!\int\limits_{0}^{1}\!\!\int\limits_{0}^{1}\!d\tau_{1} d\tau_{2}\;{\rm sign}(\tau_{1} - \tau_{2})\hspace{0.02cm}\chi(\tau_{1})\chi(\tau_{2})\biggr\}.
\]
Integral over the gravitino field $\chi$ is also Gaussian one. Performing the $\chi$ integration with allowance for the last equality, we obtain the following expression for
the functional integral, instead of (\ref{eq:6q}),
\[
\begin{split}
Z = \!\!\int\!{\cal D}x_{\mu}{\cal D}\xi_{\mu}{\cal D}e
\,\exp\hspace{0.02cm}\Biggl\{-\!\int\limits_{0}^{1}d\tau\biggl[
&-\!\displaystyle\frac{1}{2\hspace{0.02cm}e}\,\dot{x}_{\mu}\dot{x}^{\mu} - \displaystyle\frac{i}{2}\,\xi_{\mu\hspace{0.035cm}}\dot{\xi}^{\mu} - \displaystyle\frac{e}{2}\,m^2
+ \frac{i}{m^2}
\biggl(\frac{1}{e}\,(\dot{x}\cdot\xi)\biggr)\frac{d}{d\tau\!}\biggl(\frac{1}{e}\,(\dot{x}\cdot\xi)\biggr)\biggr]\\
&-\frac{\!i}{4m^2}\!\int\limits_{0}^{1}\!\!\int\limits_{0}^{1}\!d\tau_{1} d\tau_{2}\;{\rm sign}(\tau_{1} - \tau_{2})\hspace{0.02cm}
\frac{d}{d\tau_{1}\!}\biggl(\frac{1}{e}\,(\dot{x}\cdot\xi)\biggr)\frac{d}{d\tau_{2}\!}\biggl(\frac{1}{e}\,(\dot{x}\cdot\xi)\biggr)\!\Biggr\}
\end{split}
\]
\begin{equation}
\begin{split}
= \!\int\!{\cal D}x_{\mu}{\cal D}\xi_{\mu}{\cal D}e
\,\exp\hspace{0.02cm}\Biggl\{-\!\int\limits_{0}^{1}d\tau\biggl[
&-\!\displaystyle\frac{1}{2\hspace{0.03cm}e}\,\dot{x}_{\mu}\dot{x}^{\mu} - \displaystyle\frac{i}{2}\,\xi_{\mu\hspace{0.035cm}}\dot{\xi}^{\mu} - \displaystyle\frac{e}{2}\,m^2\\
&-\frac{i}{2\hspace{0.03cm}m^{2}e^{2}}\;\omega_{\mu\nu}[\hspace{0.05cm}x(\tau)]\hspace{0.03cm}\xi^{\mu}\xi^{\nu} - \frac{i}{4\hspace{0.03cm}m^{2}e^{2}}\,(\xi^{\mu}\dot{\xi}^{\nu} +
\xi^{\nu}\dot{\xi}^{\mu})\hspace{0.03cm}\dot{x}_{\mu}\dot{x}_{\nu} \biggr]\\
&\qquad\qquad\qquad\quad\, + \frac{i}{2\hspace{0.03cm}m^{2}}\,\biggl[\,\frac{1}{e}\,(\dot{x}\cdot\xi)\biggr]_{\!\tau = 1}\biggl[\,\frac{1}{e}\,(\dot{x}\cdot\xi)\biggr]_{\!\tau = 0}\,
\Biggr\}.
\end{split}
\label{eq:6w}
\end{equation}
Here, the function
\[
\omega_{\mu\nu}[\hspace{0.05cm}x(\tau)]\equiv \frac{1}{2}\,(\dot{x}_{\mu}\ddot{x}_{\nu} - \dot{x}_{\mu}\ddot{x}_{\nu})
\]
has been introduced by A.M. Polyakov. It is the notation of the tangent vector to the trajectory. An expression similar to (\ref{eq:6w}) was also considered in the different context in work \cite{gauntlett_1990} (Eq.\,(36)).\\
\indent
According to the obtained expression (\ref{eq:6w}), if we drop boundary term, we can choose as a Lagran\-gian the following expression, instead of (A.1),
\begin{equation}
\begin{split}
L=
&-\!\displaystyle\frac{1}{2\hspace{0.03cm}e}\,\dot{x}_{\mu\hspace{0.03cm}}\dot{x}^{\mu} - \displaystyle\frac{e}{2}\,m^2\\
&- \displaystyle\frac{i}{2}\;\xi_{\mu\hspace{0.03cm}}\dot{\xi}^{\mu} - \frac{i}{4\hspace{0.03cm}m^{2}e^{2}}\,
\Bigl\{\!\hspace{0.02cm}(\dot{x}_{\mu}\ddot{x}_{\nu} - \dot{x}_{\nu}\ddot{x}_{\mu})\hspace{0.03cm}\xi^{\mu}\xi^{\nu} +\,
(\xi^{\mu}\dot{\xi}^{\nu} +\, \xi^{\nu}\dot{\xi}^{\mu})\hspace{0.03cm}\dot{x}_{\mu}\dot{x}_{\nu}\!\Bigr\} +\, \ldots\,.
\end{split}
\label{eq:6e}
\end{equation}
Formally, it is explicitly independent of the pseudoscalar $\xi_5$ and gravitino $\chi$. On the other hand, according to the results of the previous section we can define a mapping
of initial Lagrangian (\ref{eq:1r}) without consideration of the odd pseudoscalar $\xi_5$ and not to go much on local supersymmetry closely related to the terms with the odd
variable $\chi$. In other words we take map (\ref{eq:5e}) in which the contribution with $\xi_5$ has been dropped and choose the pseudotensor $^{\ast}\zeta_{\mu\nu}$
in the form (\ref{eq:5t}). In accordance with the results of sections 2 and 5 the mapping of kinetic term (\ref{eq:2u}) will have a form
\begin{equation}
i\Bigl(|\kappa|^2 +\, 4\hspace{0.02cm}\dot{x}^2|\rho|^{2\,}\Bigr)\!\hspace{0.02cm}(\bar{\theta}\theta)\hspace{0.03cm}\xi_{\mu\hspace{0.03cm}}\dot{\xi}^{\mu} \,+\,
2\hspace{0.03cm}i\hspace{0.03cm}|\rho|^{2}(\bar{\theta}\theta)\hspace{0.03cm}
\Bigl\{\!\hspace{0.03cm}(\dot{x}_{\mu}\ddot{x}_{\nu} - \dot{x}_{\nu}\ddot{x}_{\mu})\hspace{0.03cm}\xi^{\mu}\xi^{\nu} -\,
(\xi^{\mu}\dot{\xi}^{\nu} +\, \xi^{\nu}\dot{\xi}^{\mu})\hspace{0.03cm}\dot{x}_{\mu}\dot{x}_{\nu}\!\Bigr\}.
\label{eq:6r}
\end{equation}
Here, we have kept only terms proportional to $(\bar{\theta} \theta)$. Comparing this expression with (\ref{eq:6e}) we see practically perfect similarity in the structure!
The only distinction of (\ref{eq:6e}) from (\ref{eq:6r}) is in different signs between two terms in braces. However, it is this difference that gives impossibility
to obtain consistent algebraic system for unknown coefficients $\chi$ and $\rho$. In spite of the similarity in structure they are still different, since they have
different sources. In the case of Lagrangian (\ref{eq:6e}) the terms in braces come from the following derivation:
\[
(\dot{x}\cdot\xi)\,\frac{d}{d\tau} \,(\dot{x}\cdot\xi) = (\dot{x}_{\mu\hspace{0.03cm}}\xi^{\mu})(\ddot{x}_{\nu\hspace{0.03cm}}\xi^{\nu} +\, \dot{x}_{\nu\hspace{0.03cm}}\dot{\xi}^{\nu}) =
\frac{1}{2}\,(\dot{x}_{\mu\hspace{0.03cm}}\ddot{x}_{\nu} - \dot{x}_{\nu\hspace{0.03cm}}\ddot{x}_{\mu})\hspace{0.03cm}\xi^{\mu}\xi^{\nu} +\,
\frac{1}{2}\,(\xi^{\mu}\dot{\xi}^{\nu} +\, \xi^{\nu}\dot{\xi}^{\mu})\hspace{0.03cm}\dot{x}_{\mu}\dot{x}_{\nu},
\]
while in the latter case (\ref{eq:6r}) they are produced by derivative of more complex object
\[
\!\,^{\ast\!}\zeta_{\mu\nu}\,\frac{d\,^{\ast\!}\zeta^{\mu\nu}}{d\tau} =
(\dot{x}_{\mu\hspace{0.02cm}}\xi_{\nu} - \dot{x}_{\nu\hspace{0.02cm}}\xi_{\mu})\hspace{0.03cm}\Bigl\{\!\hspace{0.03cm}
(\ddot{x}^{\mu\hspace{0.02cm}}\xi^{\nu} - \ddot{x}^{\nu\hspace{0.02cm}}\xi^{\mu}) +
(\dot{x}^{\mu\hspace{0.02cm}}\dot{\xi}^{\nu} - \dot{x}^{\nu\hspace{0.02cm}}\dot{\xi}^{\mu})
\Bigr\}
\]
\[
= 2\hspace{0.03cm}\dot{x}^2 \xi_{\mu\hspace{0.03cm}}\frac{d\xi^{\mu}}{d\tau} +
(\dot{x}_{\mu\hspace{0.03cm}}\ddot{x}_{\nu} - \dot{x}_{\nu\hspace{0.03cm}}\ddot{x}_{\mu})\hspace{0.03cm}\xi^{\mu}\xi^{\nu} -\,
(\xi^{\mu}\dot{\xi}^{\nu} +\, \xi^{\nu}\dot{\xi}^{\mu})\hspace{0.03cm}\dot{x}_{\mu}\dot{x}_{\nu}.
\]
The Lagrangian (\ref{eq:6e}) is still SUSY-invariant (up to the total derivative). The residual supersym\-metric transformation is
\[
\delta{x}_{\mu}=i\hspace{0.02cm}\alpha\hspace{0.02cm}\xi_{\mu},
\hspace{2.7cm}
\]
\[
\delta{e}=\hspace{0.02cm}i\hspace{0.02cm}\alpha\,\frac{\!\!2}{m^2}\,\frac{d}{d\tau\!}\biggl(\frac{1}{e}\,(\dot{x}\cdot\xi)\biggr),
\]
\[
\hspace{2.8cm}
\delta{\xi}_{\mu}= -\hspace{0.02cm}\alpha\hspace{0.03cm}\dot{x}_{\mu}\frac{1}{e}
\, +\, i\hspace{0.02cm}\alpha\hspace{0.03cm}\xi_{\mu}\,\frac{\!1}{e\hspace{0.02cm}m^2}\,\frac{d}{d\tau\!}\biggl(\frac{1}{e}\,(\dot{x}\cdot\xi)\biggr).
\]
\indent
We would like to say a few words concerning new interaction terms with a background gauge field that are generated by the odd pseudotensor (\ref{eq:5t}).
By virtue of (\ref{eq:5o}) the most simple term of the interaction has the following form:
\begin{equation}
\sim\, eg\hspace{0.02cm}Q^aF^{a}_{\mu\nu}(\dot{x}^{\mu\hspace{0.02cm}}\xi^{\nu} - \dot{x}^{\nu\hspace{0.02cm}}\xi^{\mu})\hspace{0.02cm}(\dot{x}\cdot\xi).
\label{eq:6t}
\end{equation}
If there exist only the background bosonic field in a system, then by chousing the proper-time gauge $e=1/m$, $\chi=0$, and $\xi_{5}=0$ (see Appendix A) this contribution can be
vanishing by virtue of the constraint
\begin{equation}
\dot{x}\cdot\xi = 0.
\label{eq:6y}
\end{equation}
Recall that this constraint is obtained by varying Lagrangian (A.1) over variable $\chi$. However, in the presence of a background fermionic field, a situation can drastically change.
In our next paper \cite{part_III} we will present a more general approach to construction of interaction terms with the external fermionic field (see Conclusion). The appearance of
new terms in Lagrangian (\ref{eq:1o}) containing the one-dimensional gravitino field $\chi$ as a multiplier is one of nontrivial consequence of this approach. The expression in the form
\begin{equation}
L_{\chi\Psi} = \chi\hspace{0.02cm}\!\left\{\theta^{\dagger{i}}(\bar{\theta}_{\alpha}\Psi^{i}_{\alpha})
- (\bar{\Psi}^{i}_{\alpha}\theta_{\alpha})\hspace{0.02cm}\theta^{i}\right\} + \,\dots
\label{eq:6u}
\end{equation}
is the simplest of them. Taking into account this circumstance instead of the constraint (\ref{eq:6y}) now we will have
\[
\dot{x}\cdot\xi = \frac{2\hspace{0.02cm}i}{m}\left\{\theta^{\dagger{i}}(\bar{\theta}_{\alpha}\Psi^{i}_{\alpha})
- (\bar{\Psi}^{i}_{\alpha}\theta_{\alpha})\hspace{0.02cm}\theta^{i}\right\} \neq 0
\]
and, correspondingly, interaction term (\ref{eq:6t}) takes a form
\[
\sim\, \frac{\!2\hspace{0.02cm}ig}{m^{2}}\,\hspace{0.02cm}Q^aF^{a}_{\mu\nu}(\dot{x}^{\mu\hspace{0.02cm}}\xi^{\nu} - \dot{x}^{\nu\hspace{0.02cm}}\xi^{\mu})
\left\{\theta^{\dagger{i}}(\bar{\theta}_{\alpha}\Psi^{i}_{\alpha})
- (\bar{\Psi}^{i}_{\alpha}\theta_{\alpha})\hspace{0.02cm}\theta^{i}\right\}.
\]
One can make an assumption that the pseudotensor contribution to map (\ref{eq:5e}) will give actually new interaction terms only in the presence of background fermion field in the system.

\section{\bf Mapping into supersymmetric Lagrangian}
\setcounter{equation}{0}

We have spoken repeatedly that our initial Lagrangian (\ref{eq:1r}) written  in terms of the commutative variable $\psi_{\alpha}$ is devoid of any supersymmetry.
Therefore it can be mapped into other nonsuper\-sym\-metric Lagrangian. The terms containing the fermion counterpart $\chi$ to the vierbein field $e$, namely
\begin{equation}
\displaystyle\frac{i}{2\hspace{0.015cm}e}\,\chi\hspace{0.02cm}\dot{x}_{\mu}\hspace{0.02cm}\xi^{\mu},\qquad
\displaystyle\frac{im}{2}\,\chi\hspace{0.02cm}\xi_{5}
\label{eq:7q}
\end{equation}
cannot appear in principle in any map. Counterparts of these two terms a priory must be contained in the initial Lagrangian (\ref{eq:1r}). In this section we would like
to show how the terms of such a kind can really appear in (\ref{eq:1r}).\\
\indent
The basic idea in determining such terms consist in the use of an extended Hamiltonian or superHamiltonian in the construction of the spinning equation (\ref{eq:1r}).
Hamiltonians of such a kind have been considered in a few papers for different reasons. Thus in papers by Borisov, Kulish \cite{borisov_1982} and
Fradkin, Gitman \cite{fradkin_1991} it was used in the construction of the Green's function of a Dirac particle in an background non-Abelian gauge field. Within
operator formalism this superHamiltonian has a form
\begin{equation}
-2\hspace{0.025cm}m\hat{H}_{\rm SUSY} =
\Bigl(\hat{D}_{\mu}\hat{D}^{\mu} + \frac{1}{2}\hspace{0.02cm}g\hspace{0.02cm}\hat{\sigma}_{\mu\nu}F^{a\hspace{0.02cm}\mu\nu}\hat{T}^a  - m^2\Bigr)
\!+\hspace{0.02cm} i\chi\bigl(\hat{D}_{\mu}\hat{\gamma}^{\mu} + m\bigr).
\label{eq:7w}
\end{equation}
All quantities with hats above represent operators acting in appropriate spaces of representa\-tions of the spinor, color and coordinate algebras; $\chi$ is an odd variable.
Analogy of introducing of such a superHamiltonian in the massless limit can be also found in paper by Friedan and Windey \cite{friedan_1984} in the construction of the superheat kernel.
The last one has been used in calculating of the chiral anomaly. In the monograph by Thaller \cite{thaller_book} within the supersymmetric quantum mechanics a notion of {\it Dirac operator with supersymmetry} has been defined in the most general abstract form. The expression (\ref{eq:7w}) is its special case.\\
\indent
Before studying the general case of the Dirac operator with supersymmetry it is necessary to recall briefly the fundamental points of deriving the equation of motion for the commuting spinor $\psi_{\alpha}$, Eq.\,(\ref{eq:1r}). This equation arises when we analyze the connection of the relativistic quantum mechanics with relativistic classical mechanics, first performed by W. Pauli \cite{pauli_1932}
within formalism of the first order for fermions. In book \cite{akhiezer_1969} this analysis has been performed on the basis of the second-order formalism \cite{morgan_1995}. Here, we will
follow the second line.\\
\indent
In the second order formalism the initial QCD Dirac equation for the wave function $\Psi$ is replaced by its quadratic form
\begin{equation}
-2\hspace{0.025cm}m\hspace{0.02cm}\hat{H}\Phi =
\Bigl(D_{\mu}D^{\mu} + \frac{1}{2}\hspace{0.04cm}g\hspace{0.02cm}\sigma_{\mu\nu}F^{\mu\nu} - m^2\Bigr)\Phi = 0,
\label{eq:7e}
\end{equation}
where a new spinor $\Phi$ is connected with initial one by the relation
\[
\Psi = \frac{1}{m}\,\bigl(\gamma_{\mu}D^{\mu} + m\bigr)\Phi.
\]
In this section we restore Planck's constant $\hbar$ in all formulas. Since we are interested in interaction of the spin degree of freedom of a particle with external field most,
then for the sake of simplicity we will consider equation (\ref{eq:7e}) in the case of interaction with an Abelian background field. Presence of the color degree of freedom can
result in qualitatively new features, one of them is appearing a mixed spin-color degree of freedom \cite{arodz_1_1982}. In this respect our initial model Lagrangian (\ref{eq:1r})
is simplified and it corresponds to perfect factorization of spin and color degrees of freedom of a particle. The non-Abelian case also requires appreciable complication of the usual
WKB-method in analysis of Eq.\,(\ref{eq:7e}) that is beyond the scope of this work (see, for example, \cite{arodz_2_1982, belov_1992}).\\
\indent
A solution of equation (\ref{eq:7e}) in the semiclassical limit is defined as a series in powers of $\hbar$
\begin{equation}
\Phi = {\rm e}^{iS/\hbar} (f_{0} + \hbar f_{1} + \hbar^{2\!}f_{2}\,+\,\ldots\,),
\label{eq:7r}
\end{equation}
where $S,\,f_0,\,f_1,\,\ldots$ are some functions of coordinates and time. Substituting this series into (\ref{eq:7e}) and collecting coefficients at the same power of $\hbar$,
we obtain correct to the first order in $\hbar$
\begin{equation}
\hbar^{0}:\hspace{3cm}
\biggl(\frac{\partial S}{\partial x_{\mu}} \,+\, e A_{\mu}\biggr)^{\!2}\! - m^{2} = 0,\hspace{5cm}
\label{eq:7t}
\end{equation}
\begin{equation}
\hbar^{1}:
\biggl[\,\frac{1}{i}\,\frac{\partial}{\partial x_{\mu}}
\biggl(\frac{\partial S}{\partial x^{\mu}} \,+\, e A_{\mu}\biggr)\biggr]\hspace{0.02cm}f_{0} \,+\,
\frac{2}{i}\biggl(\frac{\partial S}{\partial x^{\mu}} \,+\, e A_{\mu}\biggr)\frac{\partial f_{0}}{\partial x_{\mu}} \,+\,
\frac{e}{2}\,\sigma_{\mu\nu}F^{\mu\nu}\! f_{0} =\hspace{0.02cm} 0.
\label{eq:7y}
\end{equation}
Furthermore, we introduce into consideration a flux fermion density
\begin{equation}
s_{\mu} \equiv \bar{\Psi}_{0}\gamma_{\mu}\Psi_{0},
\label{eq:7u}
\end{equation}
where as $\Psi_0$ we take the following expression
\[
\begin{split}
&\Psi_{0} = \frac{1}{m}\,\bigl(\gamma_{\mu}D^{\mu} - m \bigr) f_{0\,}{\rm e}^{iS/\hbar} \simeq \frac{1}{m}\,{\rm e}^{iS/\hbar}\bigl[\pi_{\mu}\gamma^{\mu} -
m \bigr]\hspace{0.01cm}f_{0},\\
&\pi_{\mu}\equiv\frac{\partial S(x,\boldsymbol{\alpha})}{\partial x^{\mu}} \,+\, e A_{\mu}(x).
\end{split}
\]
Here, $\boldsymbol{\alpha}$ designates three arbitrary constants defining solution for $S$, Eq.\,(\ref{eq:7t}). In terms of the spinor $f_{0}$ the flux density (\ref{eq:7u}) has a form
\[
s_{\mu} = \frac{2\,}{m^2}\,\pi_{\mu}\bigl[\hspace{0.03cm}\bar{f}_{0}(\gamma_{\nu}\pi^{\nu} - m)f_{0}\bigr]
\]
and by virtue of Eqs.\,(\ref{eq:7t}) and (\ref{eq:7y}) it satisfies the equation of continuity
\[
\frac{\partial s_{\mu}}{\partial x_{\mu}} = 0.
\]
\indent
The equation (\ref{eq:1q}) arises from an analysis of equation for the spinor $f_{0}$ (\ref{eq:7y}). This equation in terms of the function $\pi_{\mu}$ can be written in the compact form
\begin{equation}
\frac{\partial \pi_{\mu}}{\partial x_{\mu}}\,f_{0} \,+\, 2\hspace{0.02cm}\pi_{\mu}\,\frac{\partial f_{0}}{\partial x_{\mu}} \,+\,
\frac{ie}{2}\,\sigma_{\mu\nu}F^{\mu\nu}\! f_{0} =\hspace{0.02cm} 0.
\label{eq:7i}
\end{equation}
At this point we introduce a new variable
\[
\eta\equiv \frac{2\,}{m^2} \bigl[\hspace{0.03cm}\bar{f}_{0}(\gamma_{\nu}\pi^{\nu} - m)f_{0}\bigr],
\]
such that $s_{\mu} = \pi_{\mu} \eta$. Owing to the continuity equation we have an important relation for this function
\begin{equation}
\frac{\partial \pi_{\mu}}{\partial x_{\mu}}\,\eta = -\pi_{\mu}\,\frac{\partial\eta}{\partial x_{\mu}}.
\label{eq:7o}
\end{equation}
At the final step we substitute $f_{0}=\sqrt{\eta}\hspace{0.03cm}\varphi_{0}$ into Eq.\,(\ref{eq:7y}). With allowance for (\ref{eq:7o}) this equation has a following form for a new spinor
function $\varphi_{0}$
\[
\pi_{\mu}\frac{\partial \varphi_{0}}{\partial x_{\mu}} = -\frac{ie}{4}\,\sigma_{\mu\nu}F^{\mu\nu}\! \varphi_{0}.
\]
In the book \cite{akhiezer_1969} a solution of the equation obtained just above is expressed by means of a solution of Schrodinger's equation for the wave function $\psi_{\alpha}(\tau)$,
Eq.\,(\ref{eq:1q}). The latter describes a motion of spin in a given field $F_{\mu \nu}(x)$. This field is defined along the trajectory of the particle
$x_{\mu} = x_{\mu}(\tau,\boldsymbol{\alpha},\boldsymbol{\beta})$ which in turn is defined from a solution of the equation
\[
m\,\frac{d x_{\mu}}{d\tau} = \pi_{\mu}(x,\boldsymbol{\alpha})
\]
with the initial value given by a vector $\boldsymbol{\beta}$.\\
\indent
Let us now consider the question of modification of the expressions above if instead of usual Hamilton operator in equation (\ref{eq:7e}) we take its supersymmetric extension, i.e.
consider the equation in the form
\begin{equation}
-2\hspace{0.025cm}m\hspace{0.02cm}\hat{H}_{\rm SUSY}\Phi\equiv
\Bigl\{\!\Bigl(D_{\mu}D^{\mu} + \frac{e\hbar}{2}\,\sigma_{\mu\nu}F^{\mu\nu} - m^2\Bigr)\! +\hspace{0.02cm}
i\chi\bigl(\gamma_{\mu}\gamma_{5}D^{\mu} +\, m\gamma_{5}\bigr)\!\Bigr\}\Phi = 0.
\label{eq:7p}
\end{equation}
Here, in the second expression following \cite{fradkin_1991} in parentheses we have introduced the $\gamma_5$ matrix into the definition of linear Dirac's operator. This operator
$\bigl(\gamma_{\mu}\gamma_5 D^{\mu} + m \gamma_5\bigr)$ should be considered as an odd function. We will seek also a solution of equation (\ref{eq:7p}) in the formal series (\ref{eq:7r})
with the only condition that the function $S$ is considered as an usual commuting function, and the spinor functions $f_0, f_1,\ldots$ are considered as  containing both Grassmann
even and odd parts. The equations (\ref{eq:7t}) and (\ref{eq:7y}) are modified as follows
\[
\begin{split}
\hbar^{0}:\hspace{0.5cm} \bigl(\hspace{0.03cm}\pi^2 - m^2\bigr)f_{0} +\, i\chi\bigl(\hspace{0.03cm}&\pi_{\mu}\gamma^{\mu}\gamma_{5} +\, m\gamma_{5}\bigr)f_{0} = 0,\\
\hbar^{1}:\hspace{0.5cm} \bigl(\hspace{0.03cm}\pi^2 - m^2\bigr)f_{1} +\, i\chi\bigl(\hspace{0.03cm}&\pi_{\mu}\gamma^{\mu}\gamma_{5} +\, m\gamma_{5}\bigr)f_{1}\; + \\
&+ \biggl[\frac{1}{i}\,\frac{\partial \pi_{\mu}}{\partial x_{\mu}}\,f_{0} \,+\, \frac{2}{i}\,\pi_{\mu}\,\frac{\partial f_{0}}{\partial x_{\mu}} \,+\,
\frac{e}{2}\,\sigma_{\mu\nu}F^{\mu\nu}\! f_{0}\biggr] +\,
\chi\hspace{0.02cm}\gamma_{\mu}\gamma_{5}\,\frac{\partial f_{0}}{\partial x_{\mu}} = 0.
\end{split}
\]
\indent
The next step is to present the spinors $f_{0}$ and $f_{1}$ as a sum of even and odd parts
\begin{equation}
\begin{cases}
f_{0} = f_{0}^{(0)} + \chi\hspace{0.03cm}f_{0}^{(1)},\\
f_{1} = f_{1}^{(0)} + \chi\hspace{0.03cm}f_{1}^{(1)}.
\end{cases}
\label{eq:7a}
\end{equation}
In the decomposition of (\ref{eq:7a}) we believe the functions $(f_{0}^{(0)}, f_{1}^{(0)})$ are even ones, and $(f_0^{(1)}, f_1^{(1)})$ are odd ones. Another variant of
partition in Grassmann evenness will be mentioned at the end of this section. By the use of (\ref{eq:7a}) the equation of zeroth order in $\hbar$ is decomposed into two equations
\[
\begin{split}
&\bigl(\hspace{0.03cm}\pi^2 - m^2\bigr)f^{(0)}_{0} = 0,\\
&\bigl(\hspace{0.03cm}\pi^2 - m^2\bigr)f_{0}^{(1)} +\, i\bigl(\hspace{0.03cm}\pi_{\mu}\gamma^{\mu}\gamma_{5} +\, m\gamma_{5}\bigr)f_{0}^{(0)} =\hspace{0.02cm} 0,
\end{split}
\]
the first of which defines the Hamilton-Jacobi equation for the function $S$, Eq.\,(\ref{eq:7t}), and the second one is reduced to
\[
\bigl(\hspace{0.03cm}\pi_{\mu}\gamma^{\mu}\gamma_{5} +\, m\gamma_{5}\bigr)f_{0}^{(0)} =\hspace{0.02cm} 0.
\]
\indent
Furthermore, the equation of the first order in $\hbar$ is also decomposed into two equations which with the use of (\ref{eq:7t}) take the form
\[
\frac{1}{i}\biggl(\frac{\partial \pi_{\mu}}{\partial x_{\mu}}\biggr)\!f_{0}^{(0)} \,+\, \frac{2}{i}\,\pi_{\mu}\,\frac{\partial f_{0}^{(0)}}{\partial x_{\mu}} \,+\,
\frac{e}{2}\,\sigma_{\mu\nu}F^{\mu\nu}\! f_{0}^{(0)} = 0,
\hspace{7.2cm}
\]
\begin{equation}
\frac{1}{i}\biggl(\frac{\partial \pi_{\mu}}{\partial x_{\mu}}\biggr)\!f_{0}^{(1)} \,+\, \frac{2}{i}\,\pi_{\mu}\,\frac{\partial f_{0}^{(1)}}{\partial x_{\mu}} \,+\,
\frac{e}{2}\,\sigma_{\mu\nu}F^{\mu\nu}\! f_{0}^{(1)} +\hspace{0.03cm}
\gamma_{\mu}\gamma_{5}\,\frac{\partial f_{0}^{(0)}}{\partial x_{\mu}} =
\bigl(\hspace{0.03cm}\pi_{\mu}\gamma^{\mu}\gamma_{5} +\, m\gamma_{5}\bigr)f_{1}^{(0)}.
\label{eq:7s}
\end{equation}
Notice that the term on the right-hand side of Eq.\,(\ref{eq:7s}) represents the contribution of quantum correction in contrast to the other terms. The first equation
for the even spinor $f_0^{(0)}$ is analyzed similar to previous one by the replacement
\begin{equation}
f_{0}^{(0)} \!= \sqrt{\eta}\hspace{0.03cm}\varphi_{0}^{(0)},\quad \eta\equiv\frac{2\,}{m^2} \bigl[\hspace{0.03cm}\bar{f}_{0}^{(0)}(\gamma_{\mu}\pi^{\mu} - m)f_{0}^{(0)}\bigr].
\label{eq:7d}
\end{equation}
For the odd spinor $f_{0}^{(1)}$ we define a similar replacement introducing a new odd spinor $\theta_{0}^{(1)}$ by the rule
\begin{equation}
f_{0}^{(1)} \!= \sqrt{\eta}\,\theta_{0}^{(1)},
\label{eq:7f}
\end{equation}
with the same scalar function $\eta$ as it was defined in (\ref{eq:7d}). Taking into account the continuity equation in the form (\ref{eq:7o}) and replacement (\ref{eq:7f}),
we obtain instead of (\ref{eq:7s})
\begin{equation}
\pi_{\mu}\,\frac{\partial \theta_{0}^{(1)}}{\partial x_{\mu}} \,+\,
\frac{ie}{4}\,\sigma_{\mu\nu}F^{\mu\nu}\theta_{0}^{(1)} +
\hspace{0.02cm}i\hspace{0.02cm}\gamma_{\mu}\gamma_{5}\,\frac{1}{2\sqrt{\eta}}\frac{\partial \sqrt{\eta}\hspace{0.02cm}\varphi_{0}^{(0)}}{\partial x_{\mu}} =
\frac{1}{2}\,\bigl(\pi_{\mu}\gamma^{\mu}\gamma_{5} +\, m\gamma_{5}\bigr)\varphi_{1}^{(0)}\!,
\label{eq:7g}
\end{equation}
where on the right-hand side we also have set $f_{1}^{(0)}\equiv\sqrt{\eta}\hspace{0.03cm}\varphi_1^{(0)}$. The equation obtained can be connected with the equation of motion of spin in external
field in form (\ref{eq:1q}), but instead of the even spinor $\psi_{\alpha}(\tau)$, here we have the odd spinor $\theta_{\alpha}(\tau)\equiv \theta_{0\hspace{0.02cm}\alpha}^{(1)}(\tau)$.
The latter can be identified with the auxiliary Grassmann spinor which we use throughout this work. Further, spinor $\varphi_1^{(0)}$ in the left-hand side of (\ref{eq:7g}) is even one
and it can be compared with our commuting spinor $\psi_{\alpha}$ setting
\[
\varphi_{1}^{(0)}\equiv m\hspace{0.02cm}\psi.
\]
\indent
The expression in parentheses on the right-hand side of (\ref{eq:7g}) should be considered as Grassmann odd one by virtue of oddness of the initial operator expression which correlates with it
(see the text after formula (\ref{eq:7p})). The oddness of this expression can be made explicitly if we again enter the Grassmann scalar $\chi$ as a multiplier. Taking into account all the
above mentioned and also the relation $\dot{x}_{\mu} = \pi_{\mu}/m$, we obtain the final expression of equation for the odd spinor $\theta_{\alpha}$:
\begin{equation}
\frac{1}{i}\,\frac{d\theta}{d\tau} \,+\,
\frac{e}{4m}\,\sigma_{\mu\nu}F^{\mu\nu}\theta\, +\, \dotsb =
\frac{m}{2\hspace{0.02cm}i}\,\chi\hspace{0.02cm}\dot{x}_{\mu}\bigl(\gamma^{\mu}\gamma_{5}\psi\bigr) +\, \frac{m}{2\hspace{0.03cm}i}\,\chi\bigl(\gamma_{5}\hspace{0.02cm}\psi\bigr).
\label{eq:7h}
\end{equation}
Here, the dots denotes contribution of the last term on the left-hand side of Eq.\,(\ref{eq:7g}). Its physical meaning is not clear. The terms on the right-hand side of
(\ref{eq:7h}) can be obtained by varying with respect to $\bar{\theta}$ from the following terms, which have to be added in Lagrangian (\ref{eq:1r})
\[
L =\,\ldots\,+\,\left\{\left(\frac{im}{2}\,\chi\hspace{0.03cm}\dot{x}_{\mu}\bigl(\bar{\theta}\gamma^{\mu}\gamma_{5}\psi\bigr) +\, \frac{im}{2}\,\chi\bigl(\bar{\theta}\gamma_{5}\psi\bigr)\!\right)
+ (\mbox{conj.\,part})\right\}.
\]
Finally, in turn, in the mapping of Lagrangian (\ref{eq:1r}) into (A.1) the expressions in braces should be identified with the Grassmann pseudovector $\xi_{\mu}$ and pseudoscalar $\xi_5$
introduced in section\,2 by the rule
\[
\begin{split}
(\bar{\theta}\theta)\hspace{0.03cm}\xi_{\mu} &\,\sim\, \bigl(\bar{\theta}\gamma_{\mu}\gamma_{5}\psi\bigr) \,+\, (\mbox{conj.\,part}),\\
(\bar{\theta}\theta)\hspace{0.03cm}\xi_{\hspace{0.02cm}5} &\,\sim\, \bigl(\bar{\theta}\gamma_{5}\psi\bigr) \,+\, (\mbox{conj.\,part}),
\end{split}
\]
and thereby we can obtain the missing terms (\ref{eq:7q}) in our map. Although we have obtained here, the equation of motion for the odd spinor, Eq.\,(\ref{eq:7h}), such an equation can be
obtained for the even spinor $\psi_{\alpha}$ by changing Grassmann evenness of the spinors $(f_{0}^{(0)}\!,\hspace{0.02cm} f_{1}^{(0)})$ and $(f_0^{(1)}\!,\hspace{0.02cm}f_1^{(1)})$ in the decomposition of (\ref{eq:7a}) to the opposite ones.

\section{\bf Explicit form of the odd spinor $\theta_{\alpha}$}
\setcounter{equation}{0}

In this section we would like to consider a question of construction in an explicit form the auxiliary odd spinor $\theta_{\alpha}$ in terms of known functions
of the problem under consideration. It is clear that this spinor has to contain by all means the dependence of background field $\Psi_{\alpha}^i(x)$ defined along
the particle world line $x_{\mu}=x_{\mu}(\tau)$. This spinor field is the only function, besides the commuting spinor $\psi_{\alpha}$, that possess the spinor index.\\
\indent
We will seek the required dependence in the following simple form
\begin{equation}
\theta_{\alpha} = \theta_{\alpha}(\theta^{i},\theta^{\dagger{}i},\vartheta^a,\Psi^{i}_{\alpha},\bar\Psi^{i}_{\alpha}).
\label{eq:8q}
\end{equation}
The only feature required of this function is that it must be gauge invariant. Performing the infinitesimal gauge transformations:
\[
\begin{split}
\theta^{i}\,&\rightarrow\;\theta^{i} +\hspace{0.02cm} ig\Lambda^a(t^a)^{ij}\theta^{j},\\
\vartheta^{a}&\rightarrow\,{\vartheta}^{a} -\hspace{0.02cm} gf^{abc}\Lambda^{b}\vartheta^{c},\\
\Psi^{i}_{\alpha}&\rightarrow\Psi^{i}_{\alpha} +\hspace{0.02cm} ig\Lambda^{a}(t^{a})^{ij}\Psi^{j}_{\alpha},\\
\end{split}
\]
where $\Lambda^{a}$ is a parameter of the transformations, the requirement of gauge invariance of the expression (\ref{eq:8q}) is reduced to the condition
\begin{equation}
\delta_{\Lambda}{\theta_{\alpha}}=
\label{eq:8w}
\end{equation}
\[
=ig\Lambda^{a}\!\left(\frac{\overrightarrow{\partial}\!{\theta_{\alpha}}}{\partial{\theta^{i}}}(t^{a})^{ij}\theta^{j}
-\theta^{\dagger{j}}(t^a)^{ji}\frac{\overleftarrow{\partial}\!{\theta_{\alpha}}}{\partial{\theta^{\dagger{i}}}}
-\frac{\overrightarrow{\partial}\!{\theta_{\alpha}}}{\partial{\vartheta^{b}}}(T^b)^{ac}\vartheta^{c}
+\frac{\overrightarrow{\partial}\!{\theta_{\alpha}}}{\partial{\Psi^{i}_{\alpha}}}(t^a)^{ij}\Psi^{j}_{\alpha}
-\bar{\Psi}^{j}_{\alpha}(t^a)^{ji}\frac{\overleftarrow{\partial}\!{\theta_{\alpha}}}{\partial{\bar{\Psi}^{i}_{\alpha}}}\right)=0.
\]
Here, the right (left) arrow above the partial derivative with respect to the corresponding Grassmann variables indicates that the derivative acts from the right (left)
on the $\theta_{\alpha}$. We restrict ourselves to the case of the linear dependence of the odd spinor on the background Dirac fermion field $\Psi_{\alpha}^i(x)$,
i.e. we can set\footnote{\,\label{foot_8}If we wish that $\theta_{\alpha}$ represents a Majorana spinor, then a choose of the representation $\theta_{\alpha}$ by $\Psi_{\alpha}^i(x)$
will be already ambiguous
\begin{equation}
\begin{cases}
\theta_{\alpha}^{(1)} = \hspace{0.03cm}
\bigl({\cal F}^{\dagger i}\Psi_{\alpha}^{i} + (\Psi^{c})_{\alpha}^{i}{\cal F}^{i}\bigr)\!/\hspace{0.03cm}2,\\
\theta_{\alpha}^{(2)} = \hspace{0.03cm}
\bigl({\cal F}^{\dagger i}\Psi_{\alpha}^{i} - (\Psi^{c})_{\alpha}^{i}{\cal F}^{i}\bigr)\!/\hspace{0.03cm}2i.
\end{cases}
\label{eq:8e}
\end{equation}
}
\begin{equation}
\theta_{\alpha}={\cal F}^{\dagger i}(\theta^{\dagger}\!,\theta,\vartheta)\Psi_{\alpha}^{i}.
\label{eq:8r}
\end{equation}
If we substitute the last expression into (\ref{eq:8w}) then the condition of gauge invariance is reduced to an equation for unknown even function ${\cal F}^{{\dagger}i}$
\begin{equation}
-\hspace{0.02cm}\frac{\overrightarrow{\partial}\!{\cal F}^{\dagger i}}{\partial{\theta}^{j}}\,(t^a)^{jk}\theta^k
\,+\,\theta^{\dagger{k}}(t^a)^{kj}\,\frac{\overleftarrow{\partial}\!{\cal F}^{\dagger i}}{\partial{\theta}^{\dagger{j}}}
\,+\,{\cal F}^{\dagger j}(t^a)^{ji}
\,+\,\frac{\overrightarrow{\partial}\!{\cal F}^{\dagger i}}{\partial{\vartheta}^{b}}\,(T^b)^{ac}\vartheta^c=0.
\label{eq:8t}
\end{equation}
It is easy to verify that the following simple combination of color charges:
\[
\theta^{\dagger{j}}\vartheta^a(t^a)^{ji}
\]
satisfies the equation (\ref{eq:8t}). We point out here, the principle importance of introducing the Grassmann real color charge $\vartheta^{\alpha}$ belonging to the adjoint
representation of the color $SU(N_c)$ group. A particular consequence of this is appearance of additional contribution (\ref{eq:1i}) in the total Lagrangian (\ref{eq:1r}).\\
\indent
By using the solution obtained above we transform equation (\ref{eq:8t}) to a slightly different form considering a new unknown {\it matrix} function $f$ by the replacement
\begin{equation}
{\cal F}^{\dagger i} =\hspace{0.03cm} \theta^{\dagger{j}}\vartheta^a(t^{a\!}f)^{ji}.
\label{eq:8y}
\end{equation}
Then, instead of (\ref{eq:8t}) we will have
\begin{equation}
\frac{\overrightarrow{\partial}\negmedspace f^{ij}}{\partial{\theta}^{k}}\hspace{0.02cm}(t^a)^{ks}\theta^s
-\hspace{0.04cm} \theta^{\dagger{s}}(t^a)^{sk}\hspace{0.02cm}\frac{\overleftarrow{\partial}\negmedspace f^{ij}}{\partial{\theta}^{\dagger{k}}}
\,-\hspace{0.03cm}\frac{\overrightarrow{\partial}\negmedspace f^{ij}}{\partial{\vartheta}^{b}}\,(T^b)^{ac}\vartheta^c
+\, (t^a)^{ik}\negmedspace f^{kj} -\, \negmedspace f^{ik}(t^a)^{kj} = 0.
\label{eq:8u}
\end{equation}
This equation exactly coincides with the equation obtained earlier in \cite{markov_J_Phys_G_2010}, Eq.\,(\ref{eq:5w}). According to results of the above-mentioned paper we can
immediately give a rich class of solutions of equation (\ref{eq:8u})
\[
f^{ij}\, \sim\;\; \delta^{ij},\quad \theta^i\theta^{\dagger j},\quad
(t^a\theta)^i(\theta^{\dagger} t^a)^{j},\quad Q^a(t^a)^{ij},\quad
{\cal Q}^a(t^a)^{ij},\,\ldots\,.
\]
Taking into account relations (\ref{eq:8y}) and (\ref{eq:8r}), we find a broad class of possible expressions that can be taken as the required odd spinor $\theta_{\alpha}$
\begin{equation}
\theta_{\alpha}\;\sim \;\; \vartheta^{a}(\theta^{\dagger}t^{a}\Psi_{\alpha}),\quad \vartheta^{a}Q^{a}(\theta^{\dagger}\Psi_{\alpha}),\quad
\vartheta^{a}(\theta^{\dagger}t^{a}t^{b}\theta)(\theta^{\dagger}t^{b}\Psi_{\alpha}),\quad
\vartheta^{a}Q^{b}(\theta^{\dagger}t^{a}t^{b}\Psi_{\alpha}),
\label{eq:8i}
\end{equation}
etc. Let us recall that here, $Q^{a}\equiv(\theta^{\dagger}t^{a}\theta)$ or $(\vartheta^{b}(T^{a})^{bc}\vartheta^{c})/2$.\\
\indent
One can rise a question about the dependence of the spinor $\theta_{\alpha}$ (\ref{eq:8q}) on derivative of background fermionic field, i.e. on the functions of type
$\partial_{\mu}\Psi^i_{\alpha}(\,\equiv\!\!{\Psi}^i_{\alpha,\,\mu})$. In the case of the linear dependence we set
\begin{equation}
\theta_{\alpha}={\cal F}^{\dagger i}_{\mu}(\theta^{\dagger}\!,\theta,\vartheta)\partial^{\mu}\Psi_{\alpha}^{i}(x).
\label{eq:8o}
\end{equation}
It is clear that the derivative of the $\Psi$-field can enter the spinor in question in the form of the covariant derivative only, therefore instead of the previous
we can write down at once
\[
\theta_{\alpha}={\cal F}^{\dagger i}_{\mu}(\theta^{\dagger}\!,\theta,\vartheta)\overrightarrow{D}^{ij}_{\mu}(x)\Psi_{\alpha}^{j}(x).
\]
This relation already includes the simplest dependence on an background gauge field. The dependence of the spinor on the $\Psi$-field derivatives leads to the fact that
the terms of the following form
\[
\frac{\overrightarrow{\partial}\!{\theta_{\alpha}}}{\partial{\Psi^{i}_{\beta,\,\mu}}}\,(t^a)^{ij}\delta\Psi^{j}_{\beta,\,\mu}
-\,\delta\bar{\Psi}^{j}_{\beta,\,\mu}(t^a)^{ji}\frac{\overleftarrow{\partial}\!{\theta_{\alpha}}}{\partial{\bar{\Psi}^{i}_{\beta,\,\mu}}}
\]
must be added to the total variation (\ref{eq:8w}), where in a particular case of variations induced by the infinitesimal gauge transformation, we have
\[
\begin{split}
\delta\Psi^{i}_{\alpha,\,\mu} &= ig(\partial_{\mu}\Lambda^{a})(t^{a})^{ij}\Psi^{j}_{\alpha} +\, ig\Lambda^{a}(t^{a})^{ij}(\partial_{\mu}\Psi^{j}_{\alpha}),\\
\delta\bar{\Psi}^{i}_{\alpha,\,\mu} &= -\hspace{0.02cm}ig(\partial_{\mu}\Lambda^{a})\bar{\Psi}^{j}_{\alpha}(t^{a})^{ji} - ig\Lambda^{a}(\partial_{\mu}\bar{\Psi}^{j}_{\alpha})(t^{a})^{ji}.
\end{split}
\]
It is easy to convince ourselves that the basic requirement of gauge independence of the spinor (\ref{eq:8o}) leads to the fulfilment of equations (\ref{eq:8t}),
where we need to substitute ${\cal F}^{\dagger i}_{\mu}$ in place of ${\cal F}^{\dagger i}$. The vector index enters these equations in the parametrical manner and
therefore without loss of generality, these functions can be taken in the factored form
\[
{\cal F}_{\mu}^{\dagger i} = {\cal F}^{\dagger i} (\theta^{\dagger}\!,\theta,\vartheta)\,{\cal F}_{\mu}^{\ast}(\dot{x}),
\]
where as ${\cal F}_{\mu}^{\ast}$ the function $\dot{x}_{\mu}$ can be taken. In this way by using an explicit form of solutions for ${\cal F}^{\dagger i}$ we can write out more nontrivial
expressions for the spinor $\theta_{\alpha}$ involving the background fermionic field:
\[
\theta_{\alpha} \,\sim\; \theta^{\dagger{j}}\vartheta^a(t^a)^{ji} \left(\hspace{0.02cm}\dot{x}^{\mu}D^{jk}_{\mu}(A)\Psi^{k}_{\alpha}(x)\right),\quad
\theta^{\dagger{j}}\vartheta^{a} Q^{b}(t^{a}t^{b})^{ji}  \left(\hspace{0.02cm}\dot{x}^{\mu}D^{jk}_{\mu}(A)\Psi^{k}_{\alpha}(x)\right),
\]
and so on.\\
\indent
In the mapping of spinors $(\psi_{\alpha}, \bar{\psi}_{\alpha})$ into Grassmann pseudovector and pseudoscalar in kinetic term (\ref{eq:2i}) we have obtained the terms with derivatives of
the odd spinor $\theta_{\alpha}$ in a form
\[
\left(\frac{d\bar\theta}{d\tau}\,\sigma^{\mu\nu}\theta\right) \,-\, \left(\bar\theta\hspace{0.03cm}\sigma^{\mu\nu}\frac{d\theta}{d\tau}\right)
\quad \mbox{and}\quad
\left(\frac{d\bar\theta}{d\tau}\,\gamma_{\mu}\hspace{0.03cm}\theta\right) \,-\, \left(\bar\theta\,\gamma_{\mu}\hspace{0.02cm}\frac{d\theta}{d\tau}\right)\!.
\]
In view of the line of this paper these terms can be considered as some additional interaction terms. Their explicit form can be found by the substitution of any one of expressions
(\ref{eq:8i}) into the above expressions. Unfortunately, such a direct approach results in very cumbersome expressions not interpreted even for the simplest functions in (\ref{eq:8i}).
It can be a hint that such an awkwardness of the suggested method is generated by some preference of the function we used for the description of spin degree of freedom of a particle.
In this case we deal with the even spinor $\psi_{\alpha}$. Here, more balanced and symmetric approach is needed. Such a possibility will be discussed just below in Conclusion. At the end
of this section for completeness we would like to consider an interesting question: how does the composite spinor $\theta_{\alpha}$ transform under local SUSY transformations (A.5).
To be specific, we consider the simplest combinations of a form
\begin{equation}
\theta_{\alpha} =\hspace{0.02cm} \theta^{\dagger{j}}\vartheta^{a}(t^{a})^{ji}\Psi_{\alpha}^{i}(x).
\label{eq:8p}
\end{equation}
In addition to (A.5) we believe that the odd color charge $\vartheta^{\alpha}$ is transformed under infinitesimal sypersymmetric transformations by the rule
\[
\vartheta^{a} \rightarrow\,  \vartheta^{a} +\hspace{0.02cm} g\alpha\,\xi^{\mu}A_{\mu}^{c}(x)(T^{c})^{ab}\vartheta^{b}.
\]
Subject to this fact it is not difficult to see that the function (\ref{eq:8p}) is transformed as follows:
\[
\theta_{\alpha} \rightarrow\,  \theta_{\alpha} +\hspace{0.02cm} i\hspace{0.02cm}\alpha\hspace{0.03cm}\theta^{\dagger{j}}\vartheta^{a}(t^{a})^{jk\!}
\bigl(\hspace{0.02cm}\xi^{\mu}D_{\mu}^{ki}(x)\Psi_{\alpha}^{i}(x)\bigr).
\]
We notice that in obtaining this formula we have taken into account that the coordinate $x_{\mu}$ entering argument of the background fermion field $\Psi_{\alpha}^i$, changes under the SUSY transformation by the rule
\[
x_{\mu} \rightarrow\, x_{\mu} +\, i\hspace{0.02cm}\alpha\hspace{0.02cm}\xi_{\mu}.
\]
The most nontrivial moment here is the fact that if we will require that the background fermion field transforms under the supersymmetric transformation as follows
\[
\Psi_{\alpha}^{i}(x) \rightarrow\, \Psi_{\alpha}^{i}(x) -\, i\hspace{0.02cm}\alpha \bigl(\hspace{0.02cm}\xi^{\mu}D_{\mu}^{ij}(x)\Psi_{\alpha}^{j}(x)\bigr),
\]
then the spinor $\theta_{\alpha}$ (\ref{eq:8p}) is SUSY invariant.

\section{Conclusion}
\setcounter{equation}{0}

In this paper we have presented further analysis of interaction of a classical color spinning particle with background non-Abelian fermionic field. We confined
close attention to spin interaction sector. We have used the Grassmann odd pseudovector $\xi_{\mu}$ and pseudoscalar $\xi_5$ variables as the main dynamical variables for
the description of the spin degree of freedom of the particle. The simplest explicit form of interaction terms with the fermionic field $\Psi_{\alpha}^{i}(x)$
in terms of these odd variables can be obtained by substituting expressions (\ref{eq:2q}) and (\ref{eq:2w}) with the coefficients (\ref{eq:2g}) into the interaction
Lagrangian (\ref{eq:1o}). A somewhat more complicated variant of the interaction terms can be obtained by substituting the expression (\ref{eq:5e}) containing
the odd pseudotensor $^{\ast}\zeta_{\mu \nu}$ into (\ref{eq:1o}). This pseudotensor was defined in the form (\ref{eq:5t}). Finally, the most complete and exact expression for the
interaction terms is derived by the substitution of (\ref{eq:4q}) (and even though by the substitution of (\ref{eq:4o}), (\ref{eq:4p})) into (\ref{eq:2q}),
although a physical interpretation of certain terms in these maps remains unclear.\\
\indent
Furthermore, it was shown that we can formally take any one of expressions (\ref{eq:8i}) as the auxiliary odd spinor $\theta_{\alpha}$.
In such a choice of the spinor $\theta_{\alpha}$ with linear dependence on $\Psi_{\alpha}^i(x)$, the Lagrangian of interaction (\ref{eq:1o})
becomes quadratically depending on the background fermion field. In general it is quite natural if one recalls that in the effective one-loop QCD action
$i\hspace{0.03cm}\Gamma [A,\bar{\psi},\psi]$ in the presence of both classical external bosonic and fermionic fields (Eq.\,(\ref{eq:7w}) in \cite{markov_J_Phys_G_2010})
the latter enter only in the quadratic combination $\bar{\Psi}_{\alpha}^{i}(x) \Psi_{\beta}^{j}(x)$. The interaction terms in (\ref{eq:1o}) represent in fact fragments
of a total Lagrangian that would enter into the worldline path integral representation of the effective action $i\hspace{0.03cm}\Gamma [A,\bar{\psi},\psi]$.\\
\indent
However, as was mentioned in the previous section, though the approach suggested in this paper results in entirely concrete expressions for the interaction terms, but
nevertheless it is not quite clear and obvious. Eventually, the interaction terms are sufficiently tangled and cumbersome. The analysis in section\,7 suggests the way
of ameliorating the situation. In just mentioned section it was shown that to construct a map into complete supersymmetric Lagrangian (A.1), the initial Lagrangian
(\ref{eq:1r}) must be also supersymmetric. To accomplish these ends we must add the terms in an explicit form containing auxiliary anticommuting classical spinor
$\theta_{\alpha}$ to the initial expression (\ref{eq:1r}). Furthermore, the equation obtained (\ref{eq:7h}) for the odd spinor serves as a hint at that the spinor
should be considered as an {\it independent} dynamical variable subject to own dynamical equation. And finally, this odd spinor $\theta_{\alpha}$ should be related
to its superpartner: the even spinor $\psi_{\alpha}$, and thus one must consider a single {\it superspinor}
\[
\Theta_{\alpha} = \theta_{\alpha} + \eta\hspace{0.04cm}\psi_{\alpha},
\]
as was done in the paper \cite{sorokin_1989}. Here, $\eta$ is a real odd scalar. On such a view at the problem under consideration there is no need to define an explicit
form of $\theta_{\alpha}$ as was presented in section\,8. This procedure on its own is ambiguous and rather artificial.\\
\indent
The next step forward in this direction is to use at the outset all considered variables $(\psi_{\alpha},\theta_{\alpha},\xi_{\mu})$ for the description of spin degree
of freedom, assuming that they are equivalent. This approach is known in literature as the construction of Lagrangians with {\it double supersymmetry}, i.e. possessing
both local (world-line) and global (space-time) SUSY \cite{kowalski_1988}. The circumstances that a description of such a purely quantum property of a particle
as a spin in terms of a certain classical commuting or anticommuting, spinor or vector variable can be not quite complete, at least in the description of interaction of
the spinning particle with background gauge and all the more with fermionic field, offers the justification of such a hybrid description.
Similar reasoning can be applied to the color charges $\theta^i,\theta^{\dagger{i}}$ and $\vartheta^a$ considering them as elements of appropriate superfields
\begin{equation}
\begin{split}
&N^{i} = \theta^i + \eta\hspace{0.02cm}\rho^i,\\
&N^{a\!} = \vartheta^a + \eta\hspace{0.04cm}Q^a.
\end{split}
\label{eq:9q}
\end{equation}
\indent
In the next paper \cite{part_III} we will suggest an approach for construction of the most general Lagrangian of interaction of a color spinning particle with external fields of
different statistics. Appearance of new type of interaction terms with the fermion field (one of such terms has mentioned in section\,5, Eq.\,(\ref{eq:6u})) and nontrivial relations
between components of color super\-char\-ges (\ref{eq:9q}) involving the background $\Psi$-field and so on, will be the consequen\-ce of such an approach. This provides a rather strong
argument that the presence background fermion field in the system introduces qualitatively new features into dynamics of a particle. They have no analog in the presence of only
background bosonic field in the system.


\section*{\bf Acknowledgments}
This work was supported by the Russian Foundation for Basic Research (project No. 09-02-00749), by the grant of the President of Russian Federation
for the support of the Leading Scientific Schools (NSh-1027.2008.2), in part by the Federal Target Programs "Development of Scientific Potential in
Higher Schools" (project 2.2.1.1/1483, 2.1.1/1539), and "Research and Training Specialists in Innovative Russia, 2009-2013", contract 02.740.11.5154.


\newpage
\section*{\bf Appendix A}
\setcounter{equation}{0}

Here, for convenience of references we give the Lagrangian for a spin massive particle in external non-Abelian gauge field written out in the paper \cite{balachandran_1977}.
We also written out the local super-transformation under which this Lagrangian is invariant, the constrain equations and the equations of motion for dynamical variables.\\
\indent
The most general Lagrangian for a classical relativistic spin\,$\frac{1}{2}$ particle moving in a background non-Abelian gauge field is
$$
L=L_{0}+L_{m}+L_{\theta},
\eqno{\rm (A.1)}
$$
where
$$
L_{0} = -\displaystyle\frac{1}{2\hspace{0.02cm}e}\,\dot{x}_{\mu\hspace{0.02cm}}\dot{x}^{\mu} - \displaystyle\frac{i}{2}\,\xi_{\mu\hspace{0.02cm}}\dot{\xi}^{\mu} + \displaystyle\frac{i}{2\hspace{0.02cm}e}\,\chi\hspace{0.02cm}\dot{x_{\mu}}\hspace{0.02cm}\xi^{\mu},
\eqno{\rm (A.2)}
$$
$$
L_{m} = -\displaystyle\frac{e}{2}\,m^2 + \displaystyle\frac{i}{2}\,\xi_{5\,}\dot{\xi_{5}} + \displaystyle\frac{i}{2}\,m\chi\hspace{0.02cm}\xi_{5},
\hspace{0.75cm}
\eqno{\rm (A.3)}
$$
$$
L_{\theta} = i\theta^{\dagger{i}}D^{ij}\theta^{j}
+ \displaystyle\frac{i}{2}\,eg\,Q^aF^{a}_{\mu\nu}\,\xi^{\mu}\xi^{\nu}.
\hspace{0.5cm}
\eqno{\rm (A.4)}
$$
Here, $\xi_{\mu},\,\mu=0,1,2,3$, and $\xi_{5}$ are the dynamical variables\footnote{\,Here, in contrast to \cite{balachandran_1977}, as the notation of spin variable we use
Greek letter $\xi$ instead of generally accepted $\psi$, since the latter is used throughout the present work for the notation of the bispinor $\psi_{\alpha}$.}, describing
the relativistic spin dynamics of the massive particle. These variables are elements of the Grassmann algebra \cite{berezin_1975}. The Lagrangian is invariant up to a total
derivative under the following infinitesimal supersymmetry transformation
\[
\begin{split}
&\delta{x}_{\mu}=i\alpha\hspace{0.02cm}\xi_{\mu},\\
&\delta{\xi}_{\mu}=-\alpha\Bigl(\dot{x}_{\mu}-\frac{1}{2}\,i\chi\hspace{0.02cm}\xi_{\mu}\Bigr)\!\Big/e,\\
&\delta{e}=-i\alpha\chi,\\
&\delta{\chi}=2\hspace{0.025cm}\dot{\alpha},\\
&\delta{\xi}_{5}=m\hspace{0.02cm}\alpha,\\
&\delta{\theta}^{i}=g\alpha\,\xi^{\mu}A_{\mu}^{a}(t^{a})^{ij}\theta^{j},
\end{split}
\tag{\rm A.5}
\]
where $\alpha=\alpha(\tau)$ is an arbitrary Grassmann-valued function.\\
\indent
Varying the variables $e$, $\chi$ и $\xi_{5}$, we obtain the constraint equations
\[
\begin{split}
&(\hspace{0.02cm}\dot{x}^2-i\chi\hspace{0.02cm}{\dot{x}}_{\mu\hspace{0.02cm}}\xi^{\mu})/e^{2}-m^{2\!} +\hspace{0.02cm} ig\hspace{0.025cm}Q^aF^{a}_{\mu\nu}\,\xi^{\mu}\xi^{\nu}=0,\\
&\dot{x}_{\mu\hspace{0.02cm}}\xi^{\mu} + m\hspace{0.02cm}e\hspace{0.02cm}\xi_{5}=0,\\
&2\hspace{0.04cm}\dot{\xi}_{5}-m\chi=0,
\end{split}
\tag{\rm A.6}
\]
which for a special choice of the proper time gauge $e=1/m,\,\chi=0$ and $\xi_{5}=0$ are reduced to the following equations
\[
\begin{split}
&m^2\dot{x}^{2}-m^2\!+ig\hspace{0.02cm}Q^aF^{a}_{\mu\nu}\,\xi^{\mu}\xi^{\nu}=0,\\
&\dot{x}_{\mu\hspace{0.02cm}}\xi^{\mu}=0.
\end{split}
\tag{\rm A.7}
\]
Finally, variation over the remaining dynamical variables gives the equations of motion
$$
\dot{\xi}_{\mu}-\frac{g}{m}\,Q^aF^{a}_{\mu\nu\,}\xi^{\nu}=0,
\hspace{2.6cm}
\eqno{\rm (A.8)}
$$
$$
\hspace{1.1cm}
\dot{\theta}^{i}+ig\Bigl(A^{a}_{\mu\,}\dot{x}^{\mu}-\frac{i}{2m}\,F^{a}_{\mu\nu}\,\xi^{\mu}\xi^{\nu}\Bigr)(t^{a})^{ij}\theta^j=0,
\eqno{\rm (A.9)}
$$
$$
\hspace{1.8cm}
m\hspace{0.02cm}\ddot{x}_{\mu} - g\hspace{0.02cm}Q^a\Bigl(F^{a}_{\mu\nu\hspace{0.03cm}}\dot{x}^{\nu}-\frac{i}{2m}\,D^{ab}_{\mu}(x)F^{b}_{\nu\lambda}\hspace{0.02cm}\xi^{\nu}\xi^{\lambda}\Bigr)=0.
\eqno{\rm (A.10)}
$$
Here, ${D}^{ab}_{\mu}(x)=\delta^{ab}\partial/\partial x^{\mu}+igA^{c}_{\mu}(x)(T^{c})^{ab},$ where $(T^{c})^{ab}\equiv -if^{cab}$ is the covariant derivative in the adjoint representation.
In deriving (A.10), we have used the equation of motion for the commuting color charge $Q^a(\equiv\theta^{\dagger}t^a\theta)$. This equation follows from the equation of motion for $\theta^i$
$$
\hspace{1cm}
\dot{Q}^{a}+ig\Bigl(A^{b}_{\mu\hspace{0.02cm}}\dot{x}^{\mu}-\frac{i}{2m}\hspace{0.02cm}F^{b}_{\mu\nu}\,\xi^{\mu}\xi^{\nu}\Bigr)(T^b)^{ac}Q^{c}=0.
\eqno{\rm (A.11)}
$$
\indent
The color current of the particle which enters as the source into the equation of motion for the gauge field,
\[
D^{ab}_{\mu}(x)F^{b\,\mu\nu}(x)=j^{a\nu}(x),
\]
is
$$
j^{a\mu}(x)=g\!\!\int\!\!{d}\tau\Bigl(Q^{a}\dot{x}^{\mu}
-i\hspace{0.03cm}\xi^{\mu}\xi^{\nu}\frac{1}{m}\,{D}^{ab}_{\nu}(x)\,Q^b\Bigr)
\delta^{(4)}(x-x(\tau)).
\eqno{\rm (A.12)}
$$


\section*{\bf Appendix B}
\setcounter{equation}{0}

In this Appendix we give some needed formulas for the $\gamma^{\mu}$ spinor matrix algebra. The first basic formula of them is
$$
\gamma^{\mu}\gamma^{\nu} = {\rm I}\cdot g^{\mu\nu} + i\sigma^{\mu\nu}, \quad \sigma^{\mu\nu}\equiv\frac{1}{2i}\,[\gamma^{\mu},\gamma^{\nu}],
$$
where ${\rm I}$ is the identity spinor matrix. We use the metric $g^{\mu\nu}={\rm diag}(1,-1,-1,-1)$. The useful identity is also
$$
\sigma^{\mu\nu}\gamma_5 = \frac{1}{2\hspace{0.02cm}i}\,\epsilon^{\mu\nu\lambda\sigma}\sigma_{\lambda\sigma},
\eqno{\rm (B.1)}
$$
where $\gamma_5\equiv i\gamma^{0}\gamma^{1}\gamma^{2}\gamma^{3}$; $\epsilon^{\mu\nu\lambda\sigma}$ is the totaly antisymmetric tensor so that $\epsilon^{0123}= +1$.\\
\indent
The expansion of product for the $\gamma$- and $\sigma$-matrices reads
\[
\begin{split}
&\sigma^{\mu\nu}\gamma^{\lambda} = \frac{1}{i}\left(g^{\nu\lambda}\gamma^{\mu} - g^{\mu\lambda}\gamma^{\nu}\right)\cdot {\rm I} - \epsilon^{\mu\nu\lambda\sigma}\gamma_{\sigma}\gamma_5,\\
&\gamma^{\lambda}\sigma^{\mu\nu} = \frac{1}{i}\left(g^{\mu\lambda}\gamma^{\nu} - g^{\mu\lambda}\gamma^{\nu}\right)\cdot {\rm I} - \epsilon^{\mu\nu\lambda\sigma}\gamma_{\sigma}\gamma_5.
\end{split}
\tag{\rm B.2}
\]
The formula of expansion for product of two $\sigma$-matrices has the following form
$$
\sigma^{\mu\nu}\sigma^{\lambda\sigma} = {\rm I}\cdot(\hspace{0.03cm}g^{\mu\lambda}g^{\nu\sigma} - g^{\mu\sigma}g^{\nu\lambda}) +
\frac{1}{i}\left(\hspace{0.03cm}g^{\nu\lambda}\sigma^{\mu\sigma} + g^{\mu\lambda}\sigma^{\nu\sigma} - g^{\nu\sigma}\sigma^{\mu\lambda} - g^{\mu\sigma}\sigma^{\nu\lambda}\right)
+ \frac{1}{i}\,\epsilon^{\mu\nu\lambda\sigma}\gamma_5.
\eqno{\rm (B.3)}
$$
Finally, for a product of three $\sigma$-matrices we have
$$
\sigma^{\rho\delta}\sigma^{\mu\nu}\sigma^{\lambda\sigma} =
\eqno{\rm (B.4)}
$$
\[
=\frac{\,1}{\,i}\,\left\{
g^{\lambda\nu}\!\!\left(g^{\rho\mu}g^{\delta\sigma}\!-\!g^{\rho\sigma}g^{\delta\mu}\right)\!-\!
g^{\mu\lambda}\!\!\left(g^{\rho\nu}g^{\delta\sigma}\!-\!g^{\rho\sigma}g^{\delta\nu}\right)\!-\!
g^{\nu\sigma}\!\!\left(g^{\rho\mu}g^{\delta\lambda}\!-\!g^{\rho\lambda}g^{\delta\mu}\right)\!+\!
g^{\mu\sigma}\!\!\left(g^{\rho\nu}g^{\delta\lambda}\!-\!g^{\rho\lambda}g^{\delta\nu}\right)
\right\}\cdot{\rm I}
\]
\[
+ \left(g^{\mu\lambda}g^{\nu\sigma} - g^{\mu\sigma}g^{\lambda\nu}\right)\sigma^{\rho\delta} - \frac{1}{2}\,\epsilon^{\mu\nu\lambda\sigma}\epsilon^{\rho\delta\alpha\beta}\sigma_{\alpha\beta}
\hspace{9.5cm}
\]
\[
-\, g^{\lambda\nu}\left(g^{\mu\delta}\sigma^{\rho\sigma} - g^{\rho\mu}\sigma^{\delta\sigma} - g^{\delta\sigma}\sigma^{\rho\mu} + g^{\rho\sigma}\sigma^{\delta\mu}\right)
+\, g^{\nu\sigma}\left(g^{\mu\delta}\sigma^{\rho\lambda} - g^{\rho\mu}\sigma^{\delta\lambda} - g^{\delta\lambda}\sigma^{\rho\mu} + g^{\rho\lambda}\sigma^{\delta\mu}\right)
\hspace{1.8cm}
\]
\[
+\, g^{\mu\lambda}\left(g^{\mu\delta}\sigma^{\rho\sigma} - g^{\rho\nu}\sigma^{\delta\sigma} - g^{\delta\sigma}\sigma^{\rho\nu} + g^{\rho\sigma}\sigma^{\delta\nu}\right)
- g^{\mu\sigma}\left(g^{\nu\lambda}\sigma^{\rho\lambda(?)} - g^{\rho\nu}\sigma^{\delta\lambda} - g^{\delta\lambda}\sigma^{\rho\nu} + g^{\rho\lambda}\sigma^{\delta\nu}\right)
\hspace{1.8cm}
\]
\[
- \left\{g^{\lambda\nu}\epsilon^{\rho\delta\mu\sigma} - g^{\mu\lambda}\epsilon^{\rho\delta\nu\sigma} - g^{\nu\sigma}\epsilon^{\rho\delta\mu\lambda} + g^{\mu\sigma}\epsilon^{\rho\delta\nu\lambda}
\right\}\gamma_5.
\hspace{8.4cm}
\]

\newpage

\section*{\bf Appendix C}
\setcounter{equation}{0}

The Appendix is concerned with a problem of inverse map for the quadratic combination of the odd pseudovectors $\xi^{\mu}\xi^{\nu}$ in the representation in terms of spinors
$\psi$ and $\theta$. Taking into account (\ref{eq:2e}) we have
$$
\xi^{\mu}\xi^{\nu}\! = \frac{1}{4}\,\beta^{2}
(\bar{\theta}\hspace{0.01cm}\gamma^{\mu}\gamma_{5}\hspace{0.01cm}\psi)(\bar{\theta}\hspace{0.01cm}\gamma^{\nu}\gamma_{5}\hspace{0.01cm}\psi)
+
\frac{1}{4}\,(\beta^{\ast})^{2}
(\bar{\psi}\hspace{0.01cm}\gamma_{5}\hspace{0.01cm}\gamma^{\mu}\theta)(\bar{\psi}\hspace{0.01cm}\gamma_{5}\hspace{0.01cm}\gamma^{\nu}\theta)
\eqno{\rm (C.1)}
$$
$$
\hspace{0.6cm}
-\hspace{0.04cm}\frac{1}{4}\,|\beta|^{2}\Bigl\{(\bar{\theta}\hspace{0.01cm}\gamma^{\mu}\gamma_{5}\hspace{0.01cm}\psi)(\bar{\psi}\hspace{0.01cm}\gamma_{5}\hspace{0.01cm}\gamma^{\nu}\theta) +
(\bar{\psi}\hspace{0.01cm}\gamma_{5}\hspace{0.01cm}\gamma^{\mu}\hspace{0.01cm}\theta)(\bar{\theta}\hspace{0.01cm}\gamma^{\nu}\gamma_{5}\hspace{0.01cm}\psi)\Bigr\}.
$$
For an analysis of the terms in the right-hand side of this equation we use the following general formula (Appendix\,3.5 in the book \cite{okun_book}). Let F and G be some matrix expressions,
then the following expression is true
$$
{\rm F}_{\alpha\gamma}{\rm G}_{\beta\delta} = \frac{1}{4}\sum\limits_{A}\Delta_{A}({\rm F}\Gamma_{A}{\rm G})_{\alpha\delta}(\Gamma)_{\beta\gamma},
\eqno{\rm (C.2)}
$$
where index $A$ runs over five channels: $S,\,V,\,T,\,A$ and $P$; $\Delta_{A}$ is some sign factor. In our case it is necessary to set ${\rm F} = \gamma^{\mu}\gamma_{5}$ and
${\rm G} = \gamma^{\nu}\gamma_{5}$. We are essentially interested in the last term in the expression (C.1). Straightforward, but some cumbersome calculations with the use of
formula (C.2), enables us to introduce this term as a sum of the following contributions:
\[
\begin{split}
&\underline{\mbox{S\hspace{0.02cm}-\hspace{0.02cm}channel:}}\qquad\qquad  -\!\frac{i}{8}\,|\beta|^{2}(\bar{\theta}\hspace{0.01cm}\sigma^{\mu\nu}\hspace{0.02cm}\theta)(\bar{\psi}\hspace{0.01cm}\psi),\\
&\underline{\mbox{V-\hspace{0.02cm}channel:}}\qquad\qquad  +\hspace{0.01cm}\! \frac{i}{8}\,|\beta|^{2}\hspace{0.01cm}\epsilon^{\mu\nu\lambda\sigma}(\bar{\theta}\hspace{0.01cm}\gamma_{\sigma}\gamma_{5}\hspace{0.01cm}\theta)(\bar{\psi}\hspace{0.01cm}\gamma_{\lambda}\psi),\\
&\underline{\mbox{T-\hspace{0.02cm}channel:}}\qquad\qquad -\!\frac{i}{8}\,|\beta|^{2}(\bar{\theta}\hspace{0.01cm}\theta)(\bar{\psi}\hspace{0.01cm}\sigma^{\mu\nu}\hspace{0.02cm}\psi)
\hspace{0.01cm}+\hspace{0.01cm} \frac{1}{16}\,|\beta|^{2}\epsilon^{\mu\nu\lambda\sigma}(\bar{\theta}\gamma_{5}\hspace{0.01cm}\theta)(\bar{\psi}\hspace{0.01cm}\sigma_{\lambda\sigma}\hspace{0.02cm}\psi) ,\\
&\underline{\mbox{A-\hspace{0.02cm}channel:}}\qquad\qquad -\! \frac{i}{8}\,|\beta|^{2}\hspace{0.01cm}\epsilon^{\mu\nu\lambda\sigma}(\bar{\theta}\hspace{0.01cm}\gamma_{\sigma}\theta)(\bar{\psi}\gamma_{\lambda}\gamma_{5}\hspace{0.01cm}\psi),\\
&\underline{\mbox{P-\hspace{0.02cm}channel:}}\qquad\qquad +\hspace{0.01cm}\! \frac{1}{16}\,|\beta|^{2}\epsilon^{\mu\nu\lambda\sigma}(\bar{\theta}\hspace{0.01cm}\sigma_{\lambda\sigma}\hspace{0.02cm}\theta)(\bar{\psi}\hspace{0.01cm}\gamma_{5}\hspace{0.01cm}\psi).
\end{split}
\tag{\rm C.3}
\]
The required spin tensor quantity $\frac{1}{2}\hspace{0.02cm}(\bar{\psi}\hspace{0.01cm}\sigma_{\mu\nu}\hspace{0.01cm}\psi)$ has appeared in the T-channel, and the combination
$\frac{1}{2}\hspace{0.02cm}(\bar{\theta} \sigma_{\mu \nu} \theta)$ symmet\-ric to it has appeared in the S-channel. Let us especially note that appearance of these spin tensors
has turned out possible due to {\it Grassmann} character of the auxiliary spinor $\theta$ only!\\
\indent
If we now consider a comparison of the force terms in the different representations inverse to (\ref{eq:2d})
$$
\frac{ieg}{2}\,Q^aF^{a}_{\mu\nu\,}\xi^{\mu}\xi^{\nu} \sim \hspace{0.02cm}
- \frac{eg}{4}\,Q^aF^{a}_{\mu\nu}(\bar{\psi}\sigma^{\mu\nu}\psi) + \ldots,
$$
then in the choice of $\xi^{\mu}\xi^{\nu}$ on the left-hand side according to (C.3)
$$
\xi^{\mu}\xi^{\nu}\sim
-\frac{i}{8}\,|\beta|^{2}(\bar{\theta}\hspace{0.01cm}\theta)(\bar{\psi}\hspace{0.02cm}\sigma^{\mu\nu}\hspace{0.02cm}\psi)
$$
we obtain the following equation for the coefficient $\beta$
$$
|\beta|^{2} = -\frac{4}{(\bar{\theta}\hspace{0.01cm}\theta)}.
\eqno{\rm (C.4)}
$$
On the other hand from (\ref{eq:2g}) we have
$$
|\beta|^{2} = +\hspace{0.02cm}\frac{2}{(\bar{\theta}\hspace{0.01cm}\theta)}
\eqno{\rm (C.5)}
$$
and there is a distinction in factor $(-1/2)$ as it already took place for equations (\ref{eq:2o}) and (\ref{eq:2f}).\\
\indent
By using the formula (C.2) the first term in the right-hand side of (C.1) can be also presented as a sum of the following contributions:
\[
\begin{split}
&\underline{\mbox{S\hspace{0.02cm}-\hspace{0.02cm}channel:}}\qquad\qquad  -\!\frac{i}{16}\,\beta^{2}(\bar{\theta}\hspace{0.01cm}\sigma^{\mu\nu}\hspace{0.01cm}\psi)(\bar{\theta}\hspace{0.01cm}\psi),\\
&\underline{\mbox{V-\hspace{0.02cm}channel:}}\qquad\qquad  \!+\!\hspace{0.01cm} \frac{i}{16}\,\beta^{2}\hspace{0.01cm}\epsilon^{\mu\nu\lambda\sigma}(\bar{\theta}\hspace{0.01cm}\gamma_{\sigma}\gamma_{5}\hspace{0.01cm}\psi)(\bar{\theta}\hspace{0.01cm}\gamma_{\lambda}\psi),\,\\
&\underline{\mbox{T-\hspace{0.02cm}channel:}}\!\qquad\qquad +\frac{i}{16}\,\beta^{2}(\bar{\theta}\hspace{0.01cm}\psi)(\bar{\theta}\hspace{0.01cm}\sigma^{\mu\nu}\hspace{0.02cm}\psi)
\hspace{0.01cm}-\hspace{0.01cm} \frac{1}{32}\,\beta^{2}\epsilon^{\mu\nu\lambda\sigma}(\bar{\theta}\gamma_{5}\hspace{0.01cm}\psi)(\bar{\theta}\hspace{0.01cm}\sigma_{\lambda\sigma}\hspace{0.02cm}\psi) ,\\
&\underline{\mbox{A-\hspace{0.02cm}channel:}}\qquad\qquad - \frac{i}{16}\,\beta^{2}\hspace{0.01cm}\epsilon^{\mu\nu\lambda\sigma}(\bar{\theta}\hspace{0.01cm}\gamma_{\sigma}\psi)(\bar{\theta}\gamma_{\lambda}\gamma_{5}\hspace{0.01cm}\psi),\\
&\underline{\mbox{P-\hspace{0.02cm}channel:}}\qquad\qquad +\hspace{0.01cm}\! \frac{1}{32}\,\beta^{2}\epsilon^{\mu\nu\lambda\sigma}(\bar{\theta}\hspace{0.01cm}\sigma_{\lambda\sigma}\hspace{0.02cm}\psi)(\bar{\theta}\hspace{0.01cm}\gamma_{5}\hspace{0.01cm}\psi).
\end{split}
\]
The fact that terms of vector and pseudovector contributions in exact cancel each other is rather remarkable feature, while scalar and pseudoscalar contributions double tensor one.
Thus we remain here with contributions containing only the antisymmetric spinor matrix $\sigma^{\mu\nu}$.


\end{document}